\documentclass[a4paper,12pt]{article}
\usepackage{amsfonts,amsmath,amssymb,color}
\usepackage[paper=letterpaper,margin=1.0in]{geometry}
\usepackage{graphicx}
\usepackage{cite}
\usepackage{hyperref}
\usepackage{mathabx} 
\usepackage{youngtab} 
\usepackage{comment}

\newcommand{\be}{\begin{equation}}
\newcommand{\bea}{\begin{eqnarray}\displaystyle}
\newcommand{\eea}{\end{eqnarray}}

\newcommand{\ba}{\begin{array}}
\newcommand{\ea}{\end{array}}
\newcommand{\ben}{\begin{enumerate}}
\newcommand{\een}{\end{enumerate}}
\newcommand{\bi}{\begin{itemize}}
\newcommand{\ei}{\end{itemize}}
\newcommand{\bc}{\begin{center}}
\newcommand{\ec}{\end{center}}
\newcommand{\bfig}{\begin{figure}}
\newcommand{\efig}{\end{figure}}
\newcommand{\bq}{\begin{quotation}}
\newcommand{\eq}{\end{quotation}}
\newcommand{\bt}{\begin{table}}
\newcommand{\et}{\end{table}}
\newcommand{\btab}{\begin{tabular}}
\newcommand{\etab}{\end{tabular}}
\newcommand{\bmi}{\begin{minipage}}
\newcommand{\emi}{\end{minipage}}
\newcommand{\bs}{\begin{slide}}
\newcommand{\es}{\end{slide}}

\def\Tr{ {\rm Tr } } 
\def\Dim{ {\rm Dim} }

\newcommand{\Ust}{ { \boldmath   U_{ \star } }}

\def\cA{{\cal A}}  
  \def\cF{{\cal F}}
  
  \def\cL{{\cal L}}
 \def\cN{{\cal N}} \def\cO{{\cal O}}
  
\def\cS{{\cal S}}  \def\cU{{\cal U}}
\def\cV{{\mathcal V}} \def\cW{{\mathcal W}} 
 
\def\bU{ {\bold{U} } } 
 
\def\Ust{ { \bold   U_{ \star } }} 
\def\Vst{ { \bold V^{ \star } } } 
\def\tV{ { \widetilde V }} 

\def\Dim{{\rm{Dim}}} 
\newcommand{\Ustt}{ { \bold U_{ \star , 2 } }}

\newcommand{\mC}{{\mathbb C}}

\newcommand{\bfd}{ { \bf { d } } }

\def\theequation{\thesection.\arabic{equation}}

\begin{document}

\makeatletter
\@addtoreset{equation}{section}
\makeatother
\renewcommand{\theequation}{\thesection.\arabic{equation}}

\rightline{QMUL-PH-20-03}

\vspace{10pt}


{\LARGE{ 
\centerline{\bf Perturbative 4D  conformal  field theories and  } 
\centerline{ \bf   representation theory of diagram  algebras  } 
}}  

\vskip.5cm 

\thispagestyle{empty} \centerline{
    {\large \bf Robert de Mello Koch${}^{a,b,}$\footnote{ {\tt robert@neo.phys.wits.ac.za}}}
   {\large \bf and Sanjaye Ramgoolam
               ${}^{c,b,}$\footnote{ {\tt s.ramgoolam@qmul.ac.uk}}   }
                                                       }

\vspace{.2cm}
\centerline{{\it ${}^a$ Guangdong Provincial Key Laboratory of Nuclear Science, Institute of Quantum Matter},}
\centerline{{ \it South China Normal University, Guangzhou 510006, China}}

\vspace{.2cm}
\centerline{{\it ${}^b$ National Institute for Theoretical Physics,}}
\centerline{{\it School of Physics and Mandelstam Institute for Theoretical Physics,}}
\centerline{{\it University of the Witwatersrand, Wits, 2050, South Africa } }

\vspace{.2cm}
\centerline{{\it ${}^c$ Centre for Research in String Theory, School of Physics and Astronomy},}
\centerline{ {\it Queen Mary University of London, Mile End Road, London E1 4NS, UK}}

\vskip.4cm

\thispagestyle{empty}

\centerline{\bf ABSTRACT}

\vskip.1cm 

The correlators of free four dimensional conformal field theories (CFT4)  have been shown to 
be given by amplitudes in  two-dimensional $so(4,2)$ equivariant topological field theories (TFT2), by using a vertex operator formalism for the correlators. We show that this can be extended to perturbative interacting conformal field theories, using two representation theoretic constructions. A  co-product deformation for the conformal algebra accommodates the equivariant construction of composite operators in the presence of non-additive anomalous dimensions. Explicit expressions for the co-product deformation are given within a sector of $ \cN =4 $ SYM and for the Wilson-Fischer fixed point near four dimensions. The extension of conformal equivariance beyond integer dimensions (relevant for the Wilson-Fischer fixed point) leads to the definition of an  associative diagram algebra $ \Ust $,  abstracted from $ Uso(d)$ in the limit of large integer  $d$,  which admits extension of $ Uso(d)$ representation  theory to general  real (or complex)   $d$. The algebra is related, via oscillator realisations, to $so(d)$ equivariant maps and Brauer category diagrams.  Tensor representations are constructed where  the diagram algebra acts on tensor products of a fundamental diagram representation. A similar diagrammatic algebra ${\bf U}_{\star ,2}$, related  to a  general $d$ extension for  $ Uso(d,2)$ is defined,  and some of its lowest weight representations relevant to the Wilson-Fischer fixed point are described.

\setcounter{page}{0}
\setcounter{tocdepth}{2}

\newpage

\tableofcontents

\setcounter{footnote}{0}
\linespread{1.1}
\parskip 4pt

{}~
{}~

\section{ Introduction} 

Conformal field theories in dimensions higher than two have attracted a lot of interest motivated by the 
AdS/CFT correspondence \cite{Malda}. The calculation of  loop-corrected dilatation operators in $ \cN=4$ SYM \cite{MZ0212,BKS0303} revealed a link to integrable spin chains which has launched an active and successful programme \cite{AdsCFTInteg}.  Integrable spin chains  were also found   in high energy scattering in QCD \cite{L9311,FK9404}. Loop-corrected dilatation operators provide an elegant algebraic way of thinking about anomalous dimensions and loop-corrected primary states. 

In recent years, the  bootstrap programme \cite{FGG73,P74,BPZ84} has been revived \cite{RRTV0812}  and achieved notable results on the operator spectrum of the 3D Ising model \cite{EPPRSV1403}.  The epsilon-expansion near four dimensions \cite{Wilson2} has been treated with  the  perspective of the bootstrap programme \cite{RT1505}: a set of axioms are used to obtain results on anomalous dimensions without the use of the usual methods of perturbative Feynman integrals. The loop-corrected dilatation operator for the 
Wilson-Fischer fixed point near four dimensions has been studied in \cite{Liendo}.

The study of supersymmetric sectors of the AdS/CFT correspondence for four-dimensional 
$ \cN =4$ SYM has led to the recognition that much of the colour-combinatorics of general gauge invariant operators \cite{CJR,BHR1,BHR2,BCD08,BDS08} in these sectors can be captured by two-dimensional topological field theories  (TFT2) based on group algebras \cite{FeynString,QuivCalc1301,Kimura1403}. A natural question is whether the TFT2 perspective can be adapted to capture, not just the colour-combinatorics, but also the space-time dependence in CFTs. This question was addressed in the context of the free scalar field theory in \cite{CFT4TFT2}. 

In  free 4D scalar quantum field theory, the states corresponding to the field  $ \phi $ and its derivatives span the representation $V_+$ which contains a lowest weight  state $|v^+ \rangle  $ of dimension $1$, annihilated by the special conformal generators $ K_{ \mu}$. There is a dual representation with a highest weight state of dimension $ -1$, annihilated by $ P_{ \mu}$. There is an $ so(4,2)$  invariant map
 $ \eta :  V_+ \otimes V_- \rightarrow \mC $. The CFT4/TFT2 construction of free scalar field correlators starts with the  formula 
\bea 
\Phi = { 1 \over \sqrt {2} } 
\left ( e^{ i P. x } | v^+ \rangle  + (x')^2 e^{ - i K. x' }  | v^- \rangle \right )   
\eea
which expresses the basic field as a linear combination of states in $ V = V_+ \oplus V_-$, and exploits the $so(4,2)$ invariant map $ \eta : V_+ \otimes V_- \rightarrow \mC$. Correlators of composite fields made from $n$ copies of $ \phi$ and  derivatives, are constructed by extending the invariant map $ \eta $ to an invariant map 
\bea 
\eta : Sym^n ( V ) \otimes Sym^n ( V ) \rightarrow \mC 
\eea
for general $n$, using the combinatorics of Wick contractions. The space $ Sym^n  ( V )$ is the subspace of 
$ V^{ \otimes n } $ which is invariant under $S_n$ permutations of the $n$ tensor factors. This construction of correlators of composite fields using standard tensor products of $so(4,2)$ representations is possible because the classical dimension of $ \phi^n$ is $n$  times the dimension of $ \phi$. 
This representation theoretic  construction of correlators was also described for vector and matrix scalar fields in \cite{CFT4TFT2}. 

Further investigation of the algebraic structure of the space of states in free field theory 
led to the development of a many-body perspective on the description of primary fields made from $n$ scalars in $ d$ dimensions, where it was found that the primaries are given by a simple system of linear equations, and symmetry constraints,  for functions of $nd$ variables \cite{DRRR170506,DRRR1705,deMelloKoch:2018fze,DR1806}. An interesting  corollary is that the primary fields at fixed $n$ form a ring \cite{HLMM1706,DR1806}, which also has applications in the classification of effective actions modulo equations of motion and integration by parts \cite{HLMM1706}.

The first steps in  generalizing the CFT4-TFT2 programme 
to perturbatively interacting theories were given in \cite{II1512}.
We developed  the equivariant interpretation of Feynman integrals given in \cite{FrenkLib} 
using  the harmonic expansion method for Feynman integrals \cite{CKT80}. One of  the observations was the role of  indecomposable representations of $ so(4,2)$. 

In this paper,  we will set aside Feynman integrals. In the present discussion, we will be content to 
take the input of operator dimensions and OPE coefficients from Feynman integrals. Instead we will address
and resolve some basic issues with the notion that CFT4-TFT2 can be extended beyond free fields. 
In CFT4-TFT2, composite  operators (e.g. $\phi^2$)of the CFT  correspond to states in tensor products of the basic representation of the conformal algebra corresponding to single field operator $ \phi$ and its derivatives. 

One immediate objection is that  using tensor products of $so(4,2)$ representations to describe composite operators  in free CFT makes sense, but is bound to fail for perturbative interacting theories, and the usual treatment of tensor products implies that dimensions are additive, whereas the anomalous dimension corrections to the dimension of  $ \phi^2$ in WF theory for example, it is no longer true that this dimension is twice the dimension of $ \phi$.    On the other hand,  given two representations $ V , W $ of $so(4,2)$, the dilatation operator  $D$ acts on $ V \otimes W$ in the simplest representation theoretic constructions as $ D \otimes 1 + 1 \otimes D$. In the presence of 
non-additive anomalous dimensions, the description of the  compositeness structure of quantum fields 
 in an $ so(4,2)$ equivariant manner requires that the action of $so(4,2)$ on the tensor product of two 
 representation spaces involves a deformation of the standard action via $ \cL \otimes 1 + 1 \otimes \cL$ for the 
 Lie algebra element $ \cL \in so(4,2)$.  We refer to the deformation as a ``co-product deformation''. 
 We build on loop dilatation operators and describe the deformed co-product, which we illustrate with a discussion on $ \cN=4$ SYM and the WF fixed point. 

The example of the WF fixed point, which is defined in  $ d=  4 - \epsilon $ dimensions, raises 
another important issue for the possibility of giving a conformally equivariant description of   perturbative scalar field correlators. Calculations at generic $d$ are done using an algorithm for $so(d)$ tensors which includes a  rule for evaluating  Kronecker deltas as $ \delta^{\mu}_{\mu}  = d $ in index notation. The second challenge is to make sense of this rule as a construction within an appropriate extension of conformal representation theory. To address this challenge, we define two  algebras $ \Ust$ and $ \Ustt $, based on diagrams and  develop elements of  their representation theory,  also based on diagrams. This approach starts from  the fact that the Brauer algebras $ B_d ( n)$  of operators commuting with orthogonal groups in dimension $d$  acting on tensor spaces $V_d^{ \otimes n }$ (the $n$-fold tensor product of the fundamental $V_d$) are known to have a diagrammatic formulation which allows the extension of the definition of Brauer algebras beyond integer dimensions \cite{Brauer}. 
 Brauer algebras have applications in statistical physics  \cite{Martin,Saleur} and have also been used to 
 solve enumeration and combinatorics problems of multi-matrix gauge invariants in $\cN=4$ SYM \cite{BAB,YBQ}. The key is that the dimension $d$ arises from the Brauer algebra point of view, as the evaluation of a loop occurring in the composition of Brauer algebra diagrams. The definition of the algebra $ \Ust  $, an infinite dimensional associative algebra which can be viewed as a generalization  of $ Uso(d)$ beyond integer $d$, along with its representation $ \Vst  $ (the analog of $V_d$ at integer $d$)  allows us to show that the action of $ \Ust   $ on $ \Vst  \otimes \Vst $ commutes with the generators of  the Brauer algebra  $ B_d (2)$ for  generic $d$ (section \ref{sec:ObsComm}).

The paper is organised as follows. 
In section \ref{sec:indecomp} we discuss the role of indecomposable representations in the free CFT.
This point of view on the free CFT is needed to properly understand the process of turning on interactions in the free 
theory, during which short representations that include null states are replaced by long representations.
The basic indecomposable representation of interest for W.F scalar field theory is denoted $ \widetilde V$. 
To account for the fact that the dimension of a composite operator is not in general the sum of the
dimensions of its constituents, section \ref{sec:defcoprod} introduces a deformed co-product.
An example illustrating how the deformed co-product correctly reproduces non-trivial anomalous dimensions at one loop in ${\cal N}=4$ SYM is developed in section \ref{sec:Nis4CoProd}.
We then turn to an analogous treatment of the Wilson-Fisher fixed point in section \ref{sec:ACRTDC}.
To develop the construction of primary fields, we adapt, in the framework of vertex operators for perturbative CFT4-TFT2 the usual physics rules for  tensor calculus beyond integer $d$. 
This rule is described as the ACRTDC (analytically continued representation theory with deformed co-product)  algorithm allowing concrete calculations to be carried out in section \ref{sec:ACRTDC}, in particular reproducing the conserved symmetric traceless stress tensor  at first order in $ \epsilon$ 
as a state in $ \tV \otimes \tV$. 
The question of how to understand the meaning of $so(d)$ for non-integer $d$ is answered in
section \ref{sec:diagalg} in terms of a diagram algebra $\Ust$.
We discuss a representation $\Vst$ of this diagram algebra that is the natural generalization of the vector
representation of $so(d)$ in section \ref{sec:DefVstar}. The algebra $ \Ust$ and its representations provide  a representation theoretic framework for ACRTDC. We present evidence that the structure of 
 $ \Vst^{ \otimes n }$ as a representation is similar to that of $ V_d^{ \otimes n }$ as a representation 
 of $ Uso(d)$ in a large $d$ limit.    In section \ref{sec:TensRepsHd},
 we present concrete conjectures on the representation theory of $ \Ust$.  
In section \ref{sec:UsoStar}, we explain the key elements of the diagram algebra that implements
conformal symmetry for non-integer $d$, i.e. we define an algebra $ \Ustt$ and discuss elements of its representation theory, which mirrors $ Uso(d,2)$ rep theory and extends it to non-integer $d$.    
Our conclusions and some avenues for further exploration are outlined in section \ref{sec:discussion}.

An interesting  recent paper \cite{BR1911} also takes up the
 question of continuing $O(d)$ symmetries to non-integer dimensions in physical models.  
Our results are aligned in one important respect: to make sense of conformal symmetry 
in general dimensions, we  interpret $d$ as the evaluation of a loop as in Brauer algebras and 
Brauer categories. This point of view is then developed in \cite{BR1911} in terms of diagrams and further, symmetries are expressed in terms of Deligne categories. This makes contact with general perspectives from 
category theory, which have inspired a rich literature exploring many aspects of 
complex $d$ in mathematics (see for example \cite{Deligne,Etingof,Ent}). 
The perspective in focus in the present paper is to work within the framework 
of physical symmetry operations being realised as elements of an algebra and quantum states 
as forming linear representations of the algebra. Finally, a clarification regarding the diagrams we draw, which form part of the bases for the algebras and state spaces we define:  while we will often, for clarity,  draw diagrams (e.g. \ref{fig:H21diag}) on the page with over-crossings or under-crossings to avoid intersections, there is no distinction between over- and under-crossings as in knot theory and quantum groups. As in classical (without quantum group  $q$-deformation) Brauer algebras, the key information in the diagrams is in  the start and end-points of the lines.  

\section{  Indecomposable representations  in the interacting theory }\label{sec:indecomp}

Perturbative  interacting CFTs start from  the Hilbert space of the free theory and switches on the interaction, e.g. we can start with the free scalar field and then turn on the $ \phi^4$ interactions. 
When the interaction is turned on, the generators of the conformal group are corrected and the state space is modified
because there is some multiplet recombination.
The dimension of the free field saturates the unitarity bound, and consequently, it has null states and belongs to a short
representation of the conformal group.
The null state corresponds to the equation of motion
\bea
\partial^\mu\partial_\mu\phi =0
\eea
and its descendants.
The interaction contributes an anomalous dimension to $\phi$, the unitarity bound is no longer saturated
and the null state is no longer null.
For the CFT we consider
\bea
{\cal L}={1\over 2}\partial_\mu\phi\partial^\mu\phi-{g\over 4!}\phi^4
\eea
so the equation of motion is 
\bea
\partial_\mu\partial^\mu\phi =-{g\over 3!}\phi^3
\eea
We see that in the interacting theory the null state is related to $\phi^3$, which itself was a primary in the free theory.
This is simply reflecting the fact that the short $\Delta=1$ spin zero multiplet and the $\Delta=3$ spin zero multiplet combine to give a single long multiplet in the interacting theory.

In the CFT4-TFT2 construction of free field correlators, the free scalar field corresponds to 
a state labelled by the space-time position  $x$ living in a direct sum of representations 
$ V_+ \oplus V_-$.  In the following, we will rename $V_+$ to $V$. This representation contains 
a lowest weight state $v$ obeying 
\bea 
&& D v = v \cr 
&& K_{ \mu} v = 0 \cr 
&& M_{ \mu \nu } v = 0 
\eea
and states of dimension $ 1+n$ are spanned by $ P_{ \mu_1} \cdots P_{ \mu_n } v $. 
The free equation of motion corresponds to setting to zero $ P^2v$. 

In the interacting theory,  it is helpful to consider the full state space $ \tV$
spanned all polynomials $P_{ \mu_1} \cdots P_{ \mu_n} v$, without setting $P^2v$ to 
zero from the beginning. 
The representation $\tV$ contains a state $v$, defined by the conditions
\bea\label{defv}  
&& K_{ \mu } v = 0 \cr 
&& M_{ \mu \nu } v = 0 \cr 
&& D v = v
\eea
The representation is spanned by states of the form
\bea 
P_{ \mu_1 } \cdots P_{ \mu_k } v 
\eea
At level $k$ these states span the space of symmetric   tensors of degree $k$  in $4$ variables.
Since we do not impose $ P_{\mu} P_{\mu} v = 0$, there is no restriction to symmetric traceless tensors.  
$\tV$  is a representation of $ so(4,2)$. $ P_{ \mu } $ acts by multiplication. 
The action of $K_{\mu},M_{\mu \nu}$ is defined by using commutators defining the $so(4,2)$ algebra. 

This representation also admits the linear operators 
\bea 
{\partial \over \partial P_{\nu}}
\eea
which annihilate $v$, act as 
\bea 
{\partial\over\partial P_{\nu}} P_{\mu} v = \delta_{\mu\nu} v 
\eea
and on higher polynomials by using the Leibniz rule. 
Together with the $P_{\mu}$, these form a Heisenberg algebra
\bea 
[  { \partial \over \partial P_{ \nu } }  , P_{ \mu } ] = \delta_{ \mu \nu } 
\eea
This demonstrates that $\tV$ is an irreducible representation of this Heisenberg algebra. 
The Heisenberg algebra structure will be a necessary ingredient below when we consider how
the generators of the conformal group are corrected when interactions are turned on. The operators 
$ { \partial \over \partial P_{ \nu } }  $ can be interpreted as position operators, which are well-defined as operators on $ \tV$. 

The representation $\tV$ is an indecomposable representation of $so(4,2)$.
To demonstrate this point, one checks that 
\bea 
K_{ \mu } P^2 v = 0 
\eea
This implies that the states spanned by 
\bea 
P_{ \mu_1} \cdots P_{ \mu_k } P^2 v
\eea
form a sub-representation of $\tV$, which we denote as $ V^{ (p^2)}  $. 
There is an exact sequence 
\bea 
0 \rightarrow V^{ (p^2)} \rightarrow \widetilde V \rightarrow V \rightarrow 0 
\eea
The quotient representation $V$  is not a sub-representation, i.e. we are not able to write $\tV$ as a direct sum of $V$
with another subspace.

 The equation of motion of the interacting theory sets $P^2v$ to be proportional 
  $v\otimes v\otimes v \in V^{ \otimes 3}$. This is compatible with $so(4,2)$ equivariance since 
  $ V^{(p^2)} $ and $ V^{ \otimes 3}$ are isomorphic representations of $ so(4,2)$. 
Thus the  quantum equation of motion is naturally described as a mixing between $\tV$ and $V\otimes V\otimes V$.

\section{ Deformation of co-product for $so(4,2)$ from anomalous dimensions    } \label{sec:defcoprod} 

 A key feature of interacting theories is that anomalous dimensions of composite operators 
 are typically not additive. 
This feature is visible already at first order in $\epsilon$ for the WF theory.  This means that the 
compositeness structure of the state space, when described in an $so(4,2)$ covariant way, involves a deformed co-product. The deformed co-product is given by a formula involving a position operator. 

The field $\phi^2$ picks up an anomalous dimension of  ${\epsilon\over 3}$ at first order in $\epsilon$. 
We want to describe the structure of the state space using representation theory constructions. 
The state corresponding to $\phi$ is  $v\in V$ with the properties in (\ref{defv}). 
The state corresponding to $\phi^2$ is  $v\otimes v\in  V\otimes V$ - the multiplet of states spanned 
by all the derivatives of $ \phi^2$ forms a representation we will call $V_0$ here, and the projector for 
$V_0$ in $ V \otimes V$ is denoted $P_0$.  This means that
\bea 
D (v \otimes v) = (D \otimes 1 + 1 \otimes D) (v \otimes v)  + {\epsilon\over 3}P_0(v\otimes v) 
\eea
The anomalous dimensions of the WF theory thus motivate the study of the deformed co-product 
\bea\label{coprodD}  
\Delta ( D ) = D \otimes 1 + 1 \otimes D +  { \epsilon  \over 3 } P_0 
\eea
compatible with the fact  that the one-loop dilatation operator is 
proportional to $P_0$ \cite{Liendo,KWP93,KW94}. The deformed co-product
means that we have an  action of the algebra generators in a tensor product space $V\otimes W$, which is 
a deformation of the standard action. 
So we have, for each generator $\cL_a$ of $so(4,2)$, linear operators 
\bea 
&& \Delta ( \cL_a ) \in End ( V \otimes W ) ~~ , ~~  \Delta_{ \epsilon } ( \cL_a ) \in End (  V \otimes W )  
\eea
with 
\bea 
&& \Delta ( \cL_a )  = \Delta_0 ( \cL_a ) + \epsilon \Delta_{ \epsilon } ( \cL_a ) \cr  
&& \Delta_{0 } ( \cL_a ) = \cL_a \otimes 1 + 1 \otimes \cL_a  
\eea 
such that 
\bea 
&& [ \cL_a , \cL_b ] = f_{ ab}^c \cL_c  \cr 
&& [ \Delta ( \cL_a ) , \Delta ( \cL_b ) ] = f_{ ab}^c  \Delta ( \cL_c  ) 
\eea
In other words, the $ \Delta ( \cL_a ) $ give the tensor product space $ V \otimes W$ the structure of 
a representation of the Lie algebra $ g  = so(4,2)$.  We will shortly extend the formula (\ref{coprodD}) to all the generators of $ so(4,2)$. 

The structure of deformed co-products also arises in quantum groups for $q = 1 + h  + \cO ( h^2)$,
 where the correction from the standard co-product is given in terms of $ U ( g ) \otimes U ( g )$, the tensor product of 
the universal enveloping algebras of $g$.  For example, in $ U_q(sl(2)$ we have 
$ \Delta ( J_+ ) = ( J_+ \otimes 1 + 1 \otimes J_+ ) + h ( J_+ \otimes J_3 - J_3 \otimes J_+)$.  In the case at hand, the correction is not constructed from 
$ U ( g )$. 

 A comment of the definition of $P_0$, when we are thinking about states in the tensor product 
 $\tV \otimes \tV$,  is in order.  To discuss representation theory at order $\epsilon$, we need  to think about $\tV$ and not $V$.  In $V \otimes V$, which has an orthogonal decomposition into irreps, it is clear what we mean by $P_0$. 
\bea 
V_0 \rightarrow V \otimes V \rightarrow \tV \otimes \tV 
\eea
$V_0$ is a subspace of $V \otimes V$ and hence a subspace of $ \tV \otimes \tV$. 
When we are thinking about $P_0$ as an operator in $ V \otimes V$, then we can exploit the fact that  the free field inner product allows an orthogonal decomposition  of $ V \otimes V$, which allows us to identify 
the orthogonal complement of $ V_0$ as a subspace of $ V \otimes V$. 
 The states of vanishing norm (under the free field inner product) 
 in $ \tV \otimes \tV$ are orthogonal to everything, so in particular to $ V_0$. 
So when thinking about $ P_0$ as an operator in $ \tV \otimes \tV$, 
 we may define it  to be $1$ on all states in $V_0$, and $0$ on all states which are orthogonal 
to $ V_0 \rightarrow \tV \otimes  \tV $ using the free field inner product.  

The co-products for the complete set of generators are 
\bea\label{defcop} 
&& \Delta ( D ) = D \otimes 1 + 1 \otimes D +  { \epsilon \over 3 }  P_0 \cr  
&& \Delta ( P_{ \mu } ) = P_{ \mu } \otimes 1 + 1 \otimes P_{ \mu  } \cr 
&& \Delta ( K_{ \mu } ) = K_{ \mu } \otimes 1 + 1 \otimes K_{ \mu }  - { \epsilon  \over 3 } P_0 
\left( {  \partial \over \partial P_{ \mu } }  \otimes 1 + 1 \otimes {  \partial \over \partial P_{ \mu } } \right)  P_0 \cr 
&& \Delta ( M_{ \mu \nu } ) = M_{ \mu \nu } \otimes 1 + 1 \otimes M_{ \mu \nu } 
\eea
It is a straightforward exercise to verify that the above co-products are consistent with the commutation relations of $ so(4,2)$. 
We check, for example, that 
\bea
[ \Delta ( K_{ \mu } )  , \Delta ( P_{ \nu } )  ] = 2 \Delta ( M_{ \mu \nu } ) 
 - 2  \delta_{ \mu \nu } \Delta ( D )   \, . 
\eea
It is useful to note $ \Delta_0 ( \cL_a ) P_0 = P_0 \Delta_0 ( \cL_a ) $ and $ P_0^2 =P_0 $.

\subsection{ Co-product for  $ so(4,2)$ and the Heisenberg algebra } 

Consider the indecomposable rep $\widetilde V$, which has the same states as an irreducible representation
with lowest weight not equal to $1$. 
The states are of the form 
\bea\label{basis}  
P_{\mu_1}\cdots P_{\mu_n}\,v 
\eea
where 
\bea 
&&D\, v = \delta\,v\cr 
&&M_{\mu\nu}\,v=0 
\eea
This vector space is isomorphic to the space  of polynomials in the variables $ P_{ \mu }$. 
Each monomial 
\bea
P_{ \mu_1} \cdots P_{ \mu_n } 
\eea
maps to a  state 
\bea 
P_{ \mu_1} \cdots P_{ \mu_n }  v 
\eea

On this space, $ P_{\mu } $ acts by multiplication on the vectors in (\ref{basis}), equivalently 
 $P_{\mu}$ acts on the polynomials by multiplication. 
The other generators act by using commutation relations. 
These actions can all be expressed in terms of differential operators. 
\bea\label{so42diffops}  
&& M_{ \mu \nu } =  P_{ \mu }  { \partial \over \partial P_{ \nu } } -  P_{ \nu }  {  \partial \over \partial P_{ \mu } }  \cr 
&& D =  P_{ \mu }  { \partial \over \partial P_{ \mu } } + \delta \cr 
&& K_{\mu} = 2 P_{\mu}{\partial^2\over\partial P_{\alpha}\partial P_{\alpha}} 
   - 2{\partial\over\partial P_{\mu}}(P_{\alpha}{\partial \over \partial P_{ \alpha}}+\delta - 1) 
\eea

We defined a deformed co-product to account for the anomalous dimensions which included the 
operator $ { \partial \over \partial P_{ \mu } } $. 
With the above differential expressions of the generators, it is easy to work out the commutators 
\bea 
&& [ { \partial \over \partial P_{ \mu } } , P_{ \nu } ] = \delta_{ \mu \nu } \cr 
&& [ M_{ \mu \nu } , { \partial \over \partial P_{ \alpha  } } ]  = - 
\delta_{ \mu \alpha } { \partial \over \partial P_{ \nu   } } + \delta_{ \nu \alpha } { \partial \over \partial P_{ \mu  } }  \cr 
&& [ K_{ \mu } , { \partial \over \partial P_{ \alpha } } ] 
=  - 2 \delta_{ \mu \alpha } { \partial^2 \over \partial P_{ \beta } \partial  P_{ \beta } } 
  + 2 { \partial \over \partial P_{ \mu }}  {  \partial \over \partial P_{ \alpha } }
  \equiv M_{ \mu \alpha }^{ (-2)}  
\eea
Note that $ M^{ (-2)}_{ \mu \alpha } = M^{ (-2)}_{ \alpha \mu  }$. 
\bea 
&& [ M_{ \mu \nu } , M^{ (-2)}_{ \alpha \beta } ] 
= [ M_{ \mu \nu } ,  2 { \partial \over \partial P_{ \alpha }}  {  \partial \over \partial P_{ \beta } } ]  \cr 
&& = \delta_{ \nu \alpha }  {  \partial \over \partial P_{ \mu } } {  \partial \over \partial P_{ \beta } } - 
 \delta_{ \mu \alpha} {  \partial \over \partial P_{ \nu } }  {  \partial \over \partial P_{ \beta } }
 + \delta_{ \nu \beta }  {  \partial \over \partial P_{ \mu } } {  \partial \over \partial P_{ \alpha } }
 - \delta_{ \mu \beta} {  \partial \over \partial P_{ \nu } }  {  \partial \over \partial P_{ \alpha } }
\eea

The form of $K_\mu$ might not be entirely obvious.
Since it is second order in derivatives with respect to $P_\mu$, it is useful to consider 
\bea 
&& K_{ \alpha } P_{ \mu_1} P_{ \mu_2} | v > 
= ( 2 M_{ \alpha \mu_1} - 2 D \delta_{ \alpha \mu_1}  ) P_{ \mu_2} | v  > 
   +  P_{ \mu_1} ( 2 M_{ \alpha \mu_2} - 2 D \delta_{ \alpha \mu_2} ) | v >  \cr 
   &&  =  \left( 2 \delta_{ \mu_1 \mu_2} P_{ \alpha } - 2 \delta_{ \alpha \mu_2} P_{ \mu_1} 
          -2 ( \delta +1)\delta_{ \alpha \mu_1} P_{ \mu_2} - 2\delta\, \delta_{ \alpha \mu_2} P_{ \mu_1} \right) | v >  \cr 
          && = \left( 2 \delta_{ \mu_1 \mu_2} P_{ \alpha }  
              - 2 ( \delta+1) (  \delta_{ \alpha \mu_1} P_{ \mu_2} + \delta_{ \alpha \mu_2} P_{ \mu_1} ) \right) | v > 
\eea
This result was obtained using the commutators of the algebra and nothing else.
It is clear that the differential operator for $K_{\alpha}$ in (\ref{so42diffops}) reproduces this. 
To prove that this expression is correct, apply the differential operator for $ K_{\alpha}$ to a general state
\bea 
P_{ \mu_1} \cdots P_{ \mu_n }\, | v > 
\eea 
to get 
\bea 
&& 2 P_{ \alpha } \sum_{ i < j =1}^n \delta_{ \mu_i \mu_j } P_{ \mu_1 \cdots \mu_n \setminus \{  \mu_i , \mu_j \} } | v > - 
2 ( n +\delta -1)  \sum_{i=1}^n \delta_{ \alpha \mu_i }  P_{ \mu_1 \cdots \mu_n \setminus  \{ \mu_i \} }  | v > \cr 
&& = 2 P_{ \alpha } \sum_{ i < j =1}^n \prod_{ k \notin \{ i , j \}} P_{ \mu_k } | v > 
-2 (n+\delta -1)\sum_{ i =1}^n \delta_{ \alpha \mu_i } \prod_{ k \ne i } P_{ \mu_k } | v > 
\eea
where $P_{ \mu_1 \cdots \mu_n \setminus \{  \mu_i , \mu_j \} }$ stands for the product
$P_{ \mu_1} \cdots P_{ \mu_n }$ with $P_{\mu_i}$ and $P_{\mu_j}$ removed and
similarly the obvious interpretation for $P_{ \mu_1 \cdots \mu_n \setminus  \{ \mu_i \} }$.
On the other hand, we have 
\bea 
&& K_{ \alpha } \prod_{ i =1}^n P_{ \mu_i } | v > 
= \sum_{ i =1}^{ n } P_{ \mu_1} \cdots P_{ \mu_{ i-1} } [ K_{ \alpha } , P_{ \mu_i } ] P_{ \mu_{ i+1} } \cdots P_{ \mu_n } | v >  \cr 
&& = \sum_{ i =1}^n P_{ \mu_1} \cdots P_{ \mu_{ i-1} }  
( 2M_{ \alpha \mu_i } - 2 D \delta_{ \alpha \mu_i } )  P_{ \mu_{ i+1} } \cdots P_{ \mu_n } | v >   \cr 
&& =\sum_{ i =1}^n P_{ \mu_1} \cdots P_{ \mu_{ i-1} }   \left ( 
2 \sum_{ j = i +1} ( \delta_{ \mu_i \mu_j } P_{ \alpha } - \delta_{ \alpha \mu_j } P_{ \mu_i } ) 
\prod_{\substack  { k \ne j \\ k  = i+1 }}^{ n } P_{ \mu_k } 
 - 2 \delta_{ \alpha \mu_i } ( n + \delta -1 )  \prod_{ k = i+1}^m 
 P_{ \mu_{ i} } \right ) | v >   \cr 
 &&  = 2 P_{ \alpha } \sum_{ i < j } \prod_{ k \notin \{ i , j \} } P_{ \mu_k } | v >  
    - 2 \sum_{ i =1}^n \sum_{ j =  i +1}^n \delta_{ \alpha \mu_j } \prod_{\substack  { k \ne j  \\ k  = 1 }}^{ n } P_{ \mu_k }  | v >   - 2 \sum_{ i =1}^n ( n - i + \delta ) 
\delta_{ \alpha \mu_i } \prod_{ \substack { k \ne i \\ k =1 } }^n  P_{ \mu_k } | v >  \cr 
    && = 2 P_{ \alpha } \sum_{ i < j } \prod_{ k \notin \{ i , j \} } P_{ \mu_k } | v >   
     - 2 \sum_{ j =1}^n \delta_{ \alpha \mu_j } ( j -1) \prod_{ k \ne j } P_{ \mu_k } | v > 
      - 2 \sum_{ j =1}^n ( n - j + \delta ) \delta_{ \alpha \mu_j } \prod_{ k \ne j } P_{ \mu_k }  | v > \cr 
      &&  = 2 P_{ \alpha } \sum_{ i < j } \prod_{ k \notin \{ i , j \} } P_{ \mu_k } | v > 
-2 ( n + \delta -1 ) \sum_{ i } \delta_{ \alpha \mu_i } \prod_{ k \ne i } P_{ \mu_k } | v > 
\eea
This proves the form of the differential operator for $ K$. 
The first order operator for $M$, follows from the basic $ [ M , P ] $ commutation relation 
and  the Leibniz rule of commutators. Likewise for $ D$. 

The above formulae clarify the proof of one aspect of the homomorphism property of the co-product. 
According to (\ref{defcop}), we consider co-products of the form 
\bea 
\Delta ( K_{ \mu } ) = \Delta_0 ( K_{ \mu } ) + \lambda \epsilon P_0  \Delta_0 ( { \partial \over \partial P_{ \mu } }  ) P_0 
\eea
The commutator of any two components of $K_\mu$ must vanish.
It is straightforward to see that
\bea
[ \Delta ( K_{ \mu } )  , \Delta ( K_{ \nu  } ) ]  
&=& \lambda \epsilon P_0 \Delta_0 ( [ { \partial \over \partial P_{ \mu } }  , K_{ \nu } ] ) P_0 
   + \lambda \epsilon P_0  \Delta_0 ( [ K_{ \mu } ,  { \partial \over \partial P_{ \nu } }   ] ) P_0  \cr 
  &=&  -  \lambda \epsilon P_0  M^{ (-2)}_{ \nu \mu } P_0  + \lambda \epsilon P_0  M^{ (-2)}_{ \mu \nu } P_0 = 0 
\eea
In the last step, we used the symmetry of $ M^{ (-2)}_{ \mu \nu }$. 
Note that this is exactly zero, even though our primary interest is in terms up to order $ \epsilon$. 
The commutation relations for $ so(4,2)$ are satisfied exactly. 

\subsection{ Deformed co-product and a pair of algebras  } 

We can give the co-product an algebraic characterization, but it involves two algebras not one. 
We have an algebra $A$ (the universal enveloping algebra of the Lie algebra $\mathcal{L}$ - in our case this is $so(4,2)$)
and a representation $W$ of $A$. $W$ also happens to be a representation of $B$ and 
$A$ is a sub-algebra of $B$ - in fact the representation structure of $W$ as a representation of 
$A$ follows from that of $B$ : in other words we have expressions for the action of  $ A$ in terms
of the action of $B$ as in (\ref{so42diffops}). 

\begin{figure}[ht]%
\begin{center}
\includegraphics[width=0.25\columnwidth]{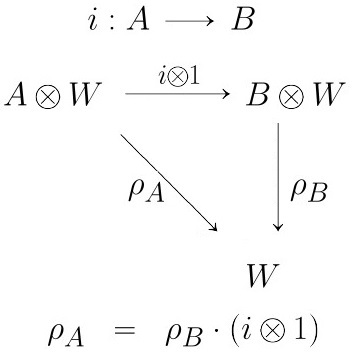}%
\caption{Sub-algebra representation consistency }%
\label{fig:subalgrep}%
\end{center}
\end{figure}

We can express this in terms of diagrams as in Figure \ref{fig:subalgrep}.
The map $ i $ is the embedding of $A$ as a sub-algebra of $ B$. The maps $ \rho_A , \rho_B$ give the linear representation of $ A , B $ on $ W$. The commutativity of the diagram  expresses  the consistency of  the sub-algebra embedding with the representation maps.  

We define a co-product for $A$, which deforms the Lie algebra co-product, but is expressed 
in terms of the algebra $B$. The co-product still obeys the commutation relations of 
$A$ but is constructed from the larger set of  generators of $B$. 
Consider  a quadruple of $ A , \cL ,   B $ as above.  For $ a \in \cL \subset A$ we have the standard co-product 
\bea 
\Delta_0 ( a ) = a \otimes 1 + 1 \otimes a 
\eea
obeying 
\bea 
[\Delta_0 ( a ) , \Delta_0 ( b ) ]  = f_{ ab }^c \Delta_0 ( c ) \, . 
\eea
For any $ a$ outside $ \cL$, the co-product follows as usual using  the product in $ A$. 
We can  ask for  co-products  $\Delta : A \rightarrow B$ such that 
\bea 
\Delta : A \rightarrow  B \otimes B 
\eea
of the form  
\bea 
\Delta ( a ) = \Delta_0 ( a ) +  \epsilon \Delta_1 ( a )  
\eea
such that 
\bea 
[ \Delta ( a ) , \Delta ( b )  ] = f_{ ab}^c \Delta ( c  ) 
\eea
Given such a deformed co-product $\Delta $ for $A$,  whenever $W$ is a representation of $A$ obtained from a representation $W$ of $ B$, as in Figure \ref{fig:subalgrep},
we can use this co-product to turn $W\otimes W$ into a representation of $A$. 
The deformed co-product needed to give an $ so(4,2)$ equivariant description of composite operators 
in perturbative CFT is an instance of this construction, with $ A = U ( so(4,2)) $ and $ B$ is the
Heisenberg algebra generated by $ P_{ \mu} , { \partial \over \partial P_{ \nu}}$.

\section{ Deformed co-product in $ N=4 $ SYM } \label{sec:Nis4CoProd}

Another  simple example of deformed co-products associated with anomalous dimensions 
in a perturbative CFT  comes from the   planar $SU(2)$ sector in ${\cal N}=4$ SYM. 
Consider the three operators 
\bea 
&& \cO_{ z} = {1\over \sqrt{2}N}\Tr (Z^2) \qquad \cO_{ y } = {1\over \sqrt{2}N}\Tr (Y^2)  \cr \cr
&& \cO_{ zy } =  {1\over\sqrt{3}N^2}\left( \Tr(YZYZ)  - \Tr (Y^2Z^2)\right) 
\eea
These operators are all eigenstates of the one loop dilatation operator. 
$\cO_{ zy}$ has a non-zero anomalous dimension $\delta={3\lambda\over 4\pi^2}$ in the planar limit \cite{BKS0303}. 
The anomalous dimensions of both $\cO_{z}$ and $\cO_{y}$ vanish.
The normalization of the operators is chosen so that they have normalized two point functions
\bea
\langle \cO_{ z}(x_1)\cO_{ z}^\dagger (x_2)\rangle = {1\over |x_1-x_2|^4}
=\langle \cO_{ y}(x_1)\cO_{ y}^\dagger (x_2)\rangle
\eea
\bea
\langle \cO_{ zy}(x_1)\cO_{ zy}^\dagger (x_2)\rangle = {1\over |x_1-x_2|^8}
\eea
Consider the correlator 
\bea 
\langle \cO_{ z} ( x_1 ) \cO_{ y} ( x_2 ) \cO_{ zy}^{ \dagger}  ( x_3 ) \rangle
= c ( x_1 - x_3 )^{ - 4  -  2 \delta }  ( x_2 - x_3 )^{ - 4  -  2 \delta } 
\eea
$c$ is a spatial constant and a function of $\lambda$, given by the OPE coefficient
\bea
\cO_{ z} (x) \cO_{ y} (0)= c \cO_{zy}(0)+\cdots
\eea
At tree level $c=-{2\over \sqrt{3}N}$.
The order $\lambda$ correction is scheme dependent.
$\delta$ is the anomalous dimension of $\cO_{ zy}$, proportional to the 't Hooft $\lambda=g_{ YM}^2N$. 
At zero coupling, 
\bea 
\Dim ( \cO_z ) + \Dim ( \cO_y ) = \Dim ( \cO_{ zy } ) 
\eea
At first order in the interaction 
\bea 
\Dim ( \cO_z ) + \Dim ( \cO_y ) = \Dim ( \cO_{ zy } ) - \delta 
\eea

In the operator-state correspondence, the operator $ \cO_{ z} $ corresponds to a tower of operators
\bea 
 \cO_{ z} ( 0 ) & \rightarrow & v_z \cr 
 \partial_{ \mu_1} \cO_{ z} ( 0 ) & \rightarrow &  P_{ \mu_1} v_z  \cr 
   \partial_{ \mu_1} \partial_{ \mu_2} \cO_{ z} ( 0 ) & \rightarrow &  P_{ \mu_1} P_{ \mu_2}  v_z  \cr 
   & \vdots  & 
\eea
The states live in a representation $V_z$ of $so(4,2)$. The lowest weight state $v_z$ has 
the properties
\bea 
&& D v_z = 2 v_z \cr 
&& M_{ \mu \nu} v_z = 0 \cr 
&& K_{ \mu} v_z = 0 
\eea
At dimension $(2+k)$ we have states 
\bea  
 V_{ k } = { \rm Span }  \{ P_{ \mu_1} \cdots P_{ \mu_k } v_z \} 
 \eea
 The direct sum forms the $so(4,2)$ irrep $V^{(z)} $
 \bea 
 V = \bigoplus_{ k =0  }^{ \infty }  V_k 
\eea 
There is a similar representation $V^{(y)}$ built on the primary $\cO_y$ and it is an isomorphic representation of $so(4,2)$.  Let us denote this $so(4,2)$ irrep as $V_2$. 
We can also represent the states as polynomials in the variables $x_\mu$ appearing in the vertex operator as
explained in \cite{CFT4TFT2}.
The generators are then given by the following differential operators acting on the polynomials
\bea
   P_\mu &=& 2x_\mu\left( 2 +x_\rho\partial_\rho\right)-x\cdot x\, \partial_\mu\qquad
  K_\mu =\partial_\mu\cr\cr
  D&=& 2 + x_\rho{\partial\over\partial x_\rho}\qquad
  M_{\mu\nu} = x_\mu{\partial\over\partial x_\nu} - x_\nu{\partial\over\partial x_\mu}
\eea

Now,  $V_{zy}$ is a representation containing a vector 
\bea\label{StandardRep}  
&& D v_{ zy} = ( 4 + \delta ) v_{ zy } \cr
&& K_{ \mu } v_{ zy} = 0 \cr 
&& M_{ \mu \nu } v_{ zy} = 0   
\eea
States at $ D = 4 + \delta +k $ are obtained by acting with $k$ $P$'s. 
The action of all  the generators  on a generic state follow by action of the commutators. 
We can again also represent the states as polynomials in the variables $x_\mu$\cite{CFT4TFT2}.
The generators are now given by the following differential operators acting on the polynomials
\bea
   P_\mu &=& 2x_\mu\left( 4+\delta +x_\rho\partial_\rho\right)-x\cdot x\, \partial_\mu\qquad
  K_\mu =\partial_\mu\cr\cr
  D&=& 4+\delta+x_\rho{\partial\over\partial x_\rho}\qquad
  M_{\mu\nu} = x_\mu{\partial\over\partial x_\nu} - x_\nu{\partial\over\partial x_\mu}
\eea

Given the non-additivity of the anomalous dimensions, we cannot model 
the 3-point correlator with the standard action of the Lie algebra on 
$  V_2 \otimes V_2 $. If we use the standard action, we would have 
\bea 
\Delta_0 ( D ) ( v_z \otimes v_y ) 
= ( D \otimes 1 + 1 \otimes D ) ( v_z \otimes v_y ) = 4 v_z \otimes v_y 
\eea
whereas the dimension of $ v_{ zy}$ is $ 4 + \delta $. The map 
$ f :  v_{ zy } \rightarrow v_z \otimes v_y $ 
 \bea 
\Delta_0 ( D ) f ( v_{zy}   ) = f \Delta_0 ( D ) ( v_{ zy } ) 
\eea
can be extended to an equivariant map $ V_{zy} \rightarrow V_z \otimes V_y$
at zero coupling, but cannot be so extended when we turn on $ \delta $ at non-zero coupling. 
Let $P_4$ be the projector to $ V_4 $ - the $so(4,2)$ representation with scalar 
 lowest weight of dimension $4$  -   in the standard tensor product decomposition of $ V_2 \otimes V_2$. 
 We can define a deformed co-product 
\bea 
&& \Delta ( D ) = \Delta_0 ( D  ) + \delta ~  P_{ 4}  \cr 
&& \Delta ( P_{ \mu } ) = \Delta_0 (  P_{ \mu } ) \cr 
&&  \Delta ( M_{ \mu \nu } ) = \Delta_0 (  M_{ \mu  \nu } ) \cr 
&& \Delta ( K_{ \mu } ) = \Delta_0 ( K_{ \mu } ) -  { \delta \over 2 }  P_{ 4} \Delta_{ 0 } ( { \partial \over \partial P_{ \mu } }  )  P_4  
\eea

With $ so(4,2)$  action  on $ V_{ z} \otimes V_{ y } $  given by 
\bea 
\cL_{ a} : v_1 \otimes v_2  \rightarrow \Delta ( \cL_a ) ( v_1 \otimes v_2 )  
\eea
and the $ so(4,2)$ action on $V_{zy}$ which we will refer to as $ \rho_{ zy} $, we can extend $f$ 
\bea 
f : V_{ zy } \rightarrow V_z \otimes V_y 
\eea
such that 
\bea 
f \rho_{ zy} (  \cL_a ) =  \Delta ( \cL_a  )  f 
\eea

Now we can construct the correlator as follows 
\bea 
g  ( ( e^{ - i P. x_1 } v^+ \otimes e^{ - i P. x_2 } v^+  )  , 
( x_3')^2 f  ( e^{ -  i P. x_3 } v^+_{ zy} )  )  
\eea
The inner product $ g$ on $ V_z \otimes V_y$ is related by using the an anti-automorphism on 
$ so(4,2)$ to the invariant pairing on $ \eta :  ( V_+ \otimes V_+ ) \otimes ( V_- \otimes V_- ) \rightarrow \mC $.

\section{ ACRTDC  and the construction of primaries of WF using tensor products  } 
\label{sec:ACRTDC} 

In this section we will consider the Wilson-Fisher fixed point theory.
Conformal invariance is realized in this theory by balancing the  growth with scale of  the coupling 
due to the classical dimension of the relevant operator $\phi^4$
in $4-\epsilon$ dimensions against loop corrections which decrease the coupling, to obtain a vanishing $\beta$ function. There is a recipe that dictates how calculations are carried out, which we name {\it analytically continued representation 
theory with a deformed co-product}  (ACRTDC). 
As explained above, the correct setting to describe how to deform the free theory to obtain the interacting theory is the representation $\tV$. 
ACRTDC is needed to construct the stress tensor with the right properties as a state in $\tV\otimes\tV$. 

In $4-\epsilon$ dimensions with ACRTDC we have
\bea
D\, v=(1-{\epsilon\over 2})\, v\qquad\qquad M_{\alpha\beta}\, v=0\qquad\qquad \delta_{\mu\mu}=4-\epsilon
\label{inepsilondim}
\eea
The first equation gives the dimension of the free scalar field in $d=4-\epsilon$ dimensions, so the dimension of the scalar field still
saturates the unitarity bound and there is still a null state. 
Indeed, its simple to verify that
\bea
K_\alpha P_\mu P_\mu v&=&(2M_{\alpha\mu}-2D\delta_{\alpha\mu})P_\mu v-P_\mu (-2D\delta_{\alpha\mu})v\cr
&=&(2\delta_{\mu\mu}P_\alpha-2P_\alpha-2DP_\alpha-2P_\alpha D)v=0
\eea
Notice that both the first and last of the relations in (\ref{inepsilondim}) are needed to get this zero: the $\epsilon$
dependence in the dimension of $v$ cancels against the $\epsilon$ dependence in $\delta_{\mu\mu}$. 
The state corresponding to the energy momentum tensor of the free theory is given by
\bea
T_{\mu\nu}&=&{1\over 2}\left( P_\mu v\otimes P_\nu v + P_\nu v\otimes P_\mu v 
-\delta_{\mu\nu}P_\tau v\otimes P_\tau v\right)
-{\alpha\over 6}\Delta(P_\mu P_\nu-P^2 \delta_{\mu\nu})v\otimes v
\eea
where
\bea
\alpha = {1-{\epsilon\over 2}\over 1-{\epsilon\over 3}}
\eea
The above $\alpha$ is needed so that the trace vanishes after setting $P^2 v=0$.
Indeed with the above choice for $\alpha$ we have
\bea
\delta_{\mu\nu}T_{\mu\nu}={1-{\epsilon\over 2}\over 2}(P^2 v\otimes v+v\otimes P^2 v)
\eea
It is easy to verify that 
\bea
\Delta (K_\alpha)T_{\mu\nu}=0
\eea
is identically obeyed, and that
\bea
\Delta (P_\mu)T_{\mu\nu}={1\over 2}(P^2 v\otimes P_\nu v+P_\nu v\otimes P^2 v)
\eea
Thus the stress tensor is conserved after we set $P^2v=0$.
This demonstrates that we have a sensible stress tensor for the free theory in $4-\epsilon$ dimensions.
The advantage of continuing away from 4 dimensions is that we can now add an interaction with coupling
$g^*$ of order $\epsilon$, without spoiling conformal invariance.
When interactions are turned on, the generators of the conformal group are modified. 
There is a non-trivial correction to $K_\mu$
\bea 
\Delta (K_\mu) \to \Delta(K_\mu) 
+{\epsilon\over 3}\sum_{i<j}\rho_{ij}\left(P_0\Delta_0\left({\partial\over\partial P^\mu}\right)P_0\right)
\eea
and a non-trivial correction to the dilatation operator
\bea
\Delta(D)\to \Delta(D)+{\epsilon\over 3}\sum_{i<j}\rho_{ij}\left(P_0\right)
\eea
$P_0$ projects onto the representation built on top of the $v\otimes v$ primary.
Also, $\rho_{ij}(\cdot)$ acts as the identity on all factors in the tensor product, and with its argument on the 
tensor product of the $i$th and $j$th factors.
All of the generators of spacetime symmetries can be constructed from the energy momentum tensor.
The above corrections to $D$ and $K_\mu$ are both accounted for by correcting $T_{\mu\nu}$ as follows
\bea
T_{\mu\nu}&=&{1\over 2}\left( P_\mu v\otimes P_\nu v + P_\nu v\otimes P_\mu v 
-\delta_{\mu\nu}P_\tau v\otimes P_\tau v\right)
-{\alpha\over 6}\Delta(P_\mu P_\nu-P^2 \delta_{\mu\nu})v\otimes v\cr\cr
&+&\delta_{\mu\nu}g^* v\otimes v\otimes v\otimes v
\eea
where $\alpha$ is what it was above and $g^*$ is order $\epsilon$.
The trace of the energy momentum tensor is given by
\bea
\delta_{\mu\nu}T_{\mu\nu}
&=&{1-{\epsilon\over 2}\over 2}(P^2 v\otimes v+v\otimes P^2 v)+(4-\epsilon)g^* v\otimes v\otimes v\otimes v\cr
&=&{1\over 2}(P^2 v\otimes v+v\otimes P^2 v)+4 g^* v\otimes v\otimes v\otimes v
\eea
which vanishes upon using the equation of motion
\bea
  P^2 v = -4g^* v\otimes v\otimes v
\eea
demonstrating that conformal invariance is preserved to this order in $\epsilon$. 

To verify that $T_{\mu\nu}$ is a primary operator is straight forwards: the order 1 pieces are as above.
The order $\epsilon$ piece has $P_0$ hitting the old $T_{\mu\nu}$ which vanishes, and the old $K_\mu$ hitting
$v\otimes v\otimes v\otimes v$ which also vanishes.
To  check that $T_{\mu\nu}$ remains conserved, we verify that 
\bea
\Delta (P_\mu)T_{\mu\nu}={1\over 2}(P^2 v\otimes_{ s}  P_\nu v+P_\nu v\otimes_{ s}  P^2 v)
+\Delta (P_\mu)(g^* v\otimes v\otimes v\otimes v)
\eea
vanishes. We have  used $ \otimes_{ s} $ to denote the operation of  taking the tensor product and symmetrising the factors
after the equation of motion has been used.

\section{ QFT algorithm (ACRTDC)  and diagram algebras} \label{sec:diagalg}

Formulating the construction of composite primary fields of Wilson-Fischer theory in 
general dimension $d=(4-\epsilon)$ requires a deformed co-product explained in Section \ref{sec:defcoprod} 
as well as ``the rule $\delta_{\mu\mu}=d= 4-\epsilon$''. 
In Section \ref{sec:ACRTDC} we explained this algorithm using an analytic continuation from general $d$, 
which we have described as ACRTDC (analytically continued representation theory with deformed co-product). 
While ACRTDC is a well-defined algorithm based on the usual physics extension  of 
tensor calculus beyond integer $d$, it naturally raises the question: What algebra are we representing 
for general $d$ ? What do we mean by $ U so (d , 2 )$ for non-integer $d$ ? In fact since primary fields in CFT are labelled by a scaling dimension along with representation labels for $ U so(d)$, we need to define an extension  of $ Uso(d)$ representation theory to non-integer $d$ in order to have a TFT2 construction for CFT$d$.  It has been observed that going beyond integer $d$ can be done in terms of diagrams and further that symmetries should be expressed in terms of Deligne catagories \cite{BR1911}.
We take a similar point of view regarding the need for a diagrammatic formulation. However, we take a more conservative point of view on how the symmetries are realized. We propose to formulate 
the CFT$d$/TFT2 correspondence by defining an associative algebra $\Ust$, generically infinite dimensional,  and its linear representation theory. 

\begin{figure}[h]
\begin{center}
\includegraphics[width=0.1\columnwidth]{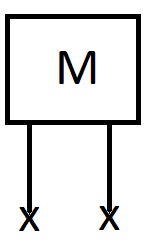}%
\caption{Diagram representing the generator $M_{ij}$.}
\label{fig:DiagM}
\end{center}
\end{figure}

We will introduce an associative algebra of diagrams $ \cF  $ and a quotient of this algebra 
$\Ust $.  We will specify a set of diagrams and define a vector space  $\cF  $ consisting of linear combinations of these diagrams, with coefficients in the field of  complex numbers $ \mC$. 
An associative product on this space,   a map $\cF \otimes \cF \rightarrow \cF $ is defined  by juxtaposition of the diagrams. The product leads to a well-defined product on the quotient so we have 
a map $ \Ust \otimes \Ust \rightarrow \Ust$. The definition is motivated  by a process of associating diagrams to tensorial expressions we encounter when working with $Uso(d)$ Lie algebras for generic $d$. For example the generators $M_{ij}$ are associated, in either $ \cF  $ or $\Ust$
with the diagram in Figure \ref{fig:DiagM}. 
Note that the diagram does not keep track of the labels $i , j $ : in a sense, we may think of the diagram as being obtained by forgetting the labels and consequently forgetting the range of values they take - which is necessary if we want to have a framework that makes sense for non-integer $d$.

If we depict the product  $M_{ ij} M_{ kl}$ in the universal enveloping algebra $ Uso(d)$ by 
juxtaposing  two boxes side to side, we can express 
\bea\label{commutator}  
M_{ ij} M_{ kl} - M_{ kl} M_{ ij} = \delta_{ jk} M_{ il} + \delta_{il} M_{ jk} - \delta_{ jl} M_{ ik} - \delta_{ ik} M_{ jl}  
\eea 
as a relation between diagrams as follows
\bea
\begin{gathered}\includegraphics[scale=0.4]{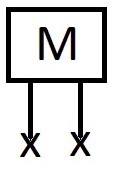}\end{gathered}
\begin{gathered}\includegraphics[scale=0.4]{M}\end{gathered}
-\begin{gathered}\includegraphics[scale=0.4]{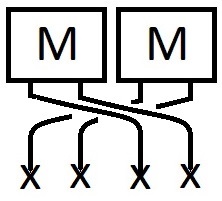}\end{gathered}=
\begin{gathered}\includegraphics[scale=0.4]{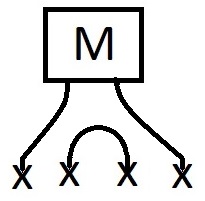}\end{gathered}
+\begin{gathered}\includegraphics[scale=0.4]{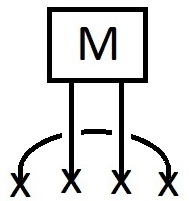}\end{gathered}
-\begin{gathered}\includegraphics[scale=0.4]{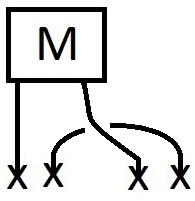}\end{gathered}
-\begin{gathered}\includegraphics[scale=0.4]{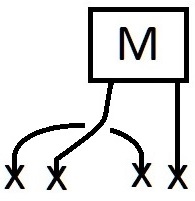}\end{gathered}\label{fig:Mcomm}
\eea
To go from the diagrammatic relation to the equation in $ Uso(d)$,
we attach the labels $ i , j , k , l $ to the crosses 
starting with $i $ for the left-most cross and proceeding with $j , k , l $ as we go to the crosses towards the right. 

The antisymmetry can be expressed diagrammatically as follows
\bea
\begin{gathered}\includegraphics[scale=0.4]{M}\end{gathered}=
-\begin{gathered}\includegraphics[scale=0.4]{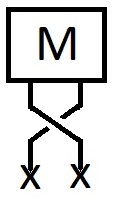}\end{gathered}\label{fig:antisymmM}
\eea
The quadratic Casimir  $ M_{ij} M_{ ij}$ is associated to the diagram shown below
\bea
\begin{gathered}\includegraphics[scale=0.4]{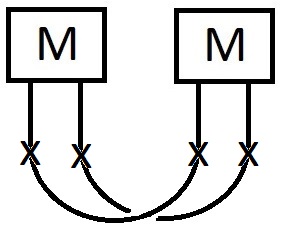}\end{gathered}=
\begin{gathered}\includegraphics[scale=0.4]{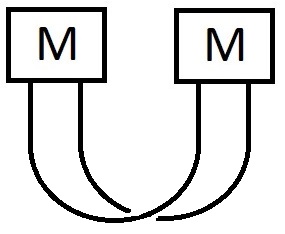}\end{gathered}\label{CasDiag}
\eea

The examples above illustrate the key features appearing in the diagrams defining  $\cF$ and $\Ust$. The diagrams include a number of crosses arranged horizontally in a line, a number of $M$-boxes each of which has two incident lines. In addition a diagram 
can have some cups as in the Casimir diagram, or caps as in the diagram describing the commutation relation.

As in Brauer algebras, a quantity $ \bfd$ can be introduced to keep track of 
loops arising when manipulating algebra elements in  $ \cF$ and $\Ust$. This quantity $ \bfd$  should not be thought as a specific  integer, but rather as an indeterminate. 
Loops arise when $\Ust$ acts on itself (through an action we will define shortly, generalizing 
the action of $ Uso(d)$ on itself by commutators) or on a representation $\Vst$ (as well as 
its tensor powers)  which we will define. The quotienting construction that takes us 
from $ \cF $ to $\Ust$ is modelled on the fact that the enveloping algebra $Uso(d)$ can be defined as a  quotient of  the free algebra generated by $M_{ i j } $( for $ i , j \in \{ 1, \cdots , d \}$) by the commutation relation. We will subsequently define a specialization 
where we go from treating $ \bfd $ as a formal variable to a real number  $d$. In this step important issues of unitarity arise for generic real $d$ - here we make contact  with discussion 
of non-unitarity in  \cite{HRV1512}.

The general diagrams  in $ \cF  $ and $\Ust$ will have a number $k$ of $M$-boxes, a number $l_1 $ of caps and $l_2$ of cups and the diagrams can be described by taking  
\bea 
 d \gg  k , l_1 , l_2
 \eea
and considering $Uso(d)$  equivariant maps inspired by the oscillator realization of $so(d)$. 
We will describe this in section \ref{sec:DiagsAsEquivs}. Since $k, l_1, l_2$ can be as large as we like,  $\Ust$ is obtained, in a sense, from a  $d \rightarrow \infty $ limit of $ U so(d)$. 

We will define a vector space $\Vst$ of diagrams. We will give an action of $\Ust$ on this vector space. 
We conjecture that the decomposition of $ ( \Vst )^{ \otimes n } $ in terms of irreps of $\Ust$ will be isomorphic to the decomposition of $ V_d^{ \otimes n }$   as a representation of the orthogonal Lie algebra 
 $ so(d)$ for $ d \gg n$. Results on the stable (large $d$)  limit of $O(d)$ tensor representations are given in \cite{HoweOdstab}. 
In section \ref{sec:HVVdecomp}, we set up the foundation for the conjecture through explicit calculations at  $ n =2$.

$\Ust$ acts on itself by commutators. We conjecture that  
this action decomposes into irreducible representations 
in the same way that $ Uso (d )$ decomposes under  the adjoint action of $so (d)$, in an appropriate large $d$ limit. The large $d$ limit is defined by considering polynomials of degree $k$ and $ d \gg k$. 
The Poincare-Birkhoff-Witt theorem states that $ Uso(d)$ transforms in the same was as symmetric polynomials of the generators $M_{ ij}$ (see e.g. \cite{Humphreys}). Explicit results can be obtained from 
Theorem 2.1.2 of \cite{HoweOdstab}. 

These constructions and conjectures are motivated by the insight that tensor manipulations at generic $d$ can be given a mathematical formulation using diagrams, and the expectation in the physics literature\cite{Wilson,Collins} that generic $d$ should involve infinite dimensional  spaces.

\subsection{ A free associative diagram algebra $\cF$ and a quotient $\Ust $  }\label{sec:DiagsAsEquivs}

We will define an infinite dimensional  associative algebra over $ \mC$, denoted  $ \cF $, abstracted 
from the generators $M_{ ij}$ of $Uso (d) $. An associative algebra is a vector space equipped with a product $m$ 
\bea 
m : \cF \otimes \cF  \rightarrow \cF 
\eea
The vector space $ \cF$   is  
\bea\label{firstpiece}  
\cF  = \mC \oplus Span_{ \mC } ( M ) \oplus \cdots 
\eea
The $ \cdots $ refers to subspaces to be specified later, which will be  done efficiently 
after we introduce the oscillator construction of $ U so(d)$ and its interpretation in terms of 
equivariant maps and diagrams. Focusing on the subspaces made explicit in (\ref{firstpiece}) 
a general vector is specified by two complex numbers $a_0,a_1$ and can be written as
\bea
\left(a_0+a_1\begin{gathered}\includegraphics[scale=0.4]{M}\end{gathered}\right)\label{fig:vecDiagM}
\eea
The multiplication map can be written as 
\bea 
m ( a_0 + a_1 M , a_0' + a_1' M ) = a_0 a_0' + ( a_0 a_1' + a_0' a_1 ) M   + a_1 a_1' ( M \otimes M )  \cdots  
\eea
where $M \otimes M$ represents the diagram with the two $M$-boxes juxtaposed next to each other. 
The product with $ a_1 =  a_1' = 1 ; a_0 = a_0' = 0 $ is given in terms of diagrams as
\bea
m\left(\begin{gathered}\includegraphics[scale=0.4]{M}\end{gathered}\, ,\,
\begin{gathered}\includegraphics[scale=0.4]{M}\end{gathered}\right)=
\begin{gathered}\includegraphics[scale=0.4]{M}\end{gathered}\,
\begin{gathered}\includegraphics[scale=0.4]{M}\end{gathered}\label{fig:MMprod}
\eea
Note that the vector space spanned by the $M$-box diagram is just one-dimensional. This is despite the fact that the $M$-box diagram is inspired by $ M_{ i j }$, of which we have $ d ( d-1)/2$ in $ so(d)$. 
We wish to  define algebras $\cF  $ and $\Ust$ which can be specialised to 
$ \bfd \rightarrow d $ where $d$ is a generic real number. So it will certainly not do to have vector subspaces of $ \cF$ which have dimension $ d ( d -1)/2$.

In order to give the general definition of the infinite dimensional vector space $ \cF $, 
we will use the oscillator construction of $ so(d)$ in order to interpret the 
$M$-box diagram above, and its generalizations which span $ \cF  $ 
in terms of tensor algebras and  $ so(d)$ equivariant maps.

The $d$-dimensional oscillator relations are 
\bea\label{OscAlg}  
[ a^{ \dagger}_{ i} ,  a_{ j } ]  = - \delta_{ ij } 
\eea
and the Lie algebra generators of $ so(d)$ can be written as 
\bea\label{Mosc} 
M_{ ij } = a^{ \dagger}_i a_j - a^{ \dagger}_j a_i 
\eea
Taking the $ i,j$ indices to range over $ \{ 1, \cdots , d \}$, the $ a^{ \dagger}_i$ span a 
$d$-dimensional vector space, which is the vector representation of $ so(d)$.  
It is useful to rethink these equations in terms of tensor products : We will think of the 
$d$-dimensional space spanned by $ a^{ \dagger}_i$ as a tensor product  
$V_+ \otimes W$ of a one-dimensional vector space $V_+$ with a $d$-dimensional vector space $W$. 
Likewise the annihilation operators span $ V_- \otimes W$. 
The action of $ M_{ij}$ on $ W$ is given by the commutator.

Consider a 1-dimensional vector space spanned by $a $, which we call $V_-$. 
And a 1-dimensional vector space spanned by $a^{ \dagger } $ which we call $V_+$. 
We can also consider a direct sum $ V_+ \oplus V_- = V$. 
Let us define an anti-symmetric map
\bea 
COM : V  \otimes V \rightarrow \mC 
\eea
where 
\bea\label{commutatorCOM}  
COM( a , a^{ \dagger} ) = - COM ( a^{ \dagger} , a ) = 1
\eea
This is the commutator expressed as a linear map between vector spaces. 

Let $ e_i$ form a basis for the vector space $W$.  
We will write 
\bea 
&& a \otimes e_i  \in V \otimes W  \cr 
&& a^{ \dagger} \otimes e_i \in V \otimes W
\eea
and express our calculations with $a_i , a_j^{\dagger} $ in terms of the COM map. 
The commutator is giving a map 
\bea 
COM : V \otimes W \otimes V \otimes W \rightarrow \mC 
\eea
which acts as 
\bea 
COM ( a \otimes e_i \otimes a^{ \dagger} \otimes  e_j )  = - COM ( a^{ \dagger} 
 \otimes e_i \otimes a \otimes  e_j ) 
= g ( e_i , e_j ) 
\eea
$g$ is an inner product, for which the $e_i$ form an orthonormal basis. 
In fact 
\bea\label{comdef}  
COM ( a \otimes w_1  \otimes a^{ \dagger} \otimes  w_2  )  
= g ( w_1 , w_2 ) 
\eea

Now the formula 
\bea 
M_{ ij} = a^{ \dagger}_i a_{ j } -  a^{ \dagger}_j a_{ i }
\eea
specifies a state (which we can also call $M_{ij}$) 
\bea\label{Mijtensor}  
M_{ ij}  =  a^{ \dagger} \otimes e_i \otimes a \otimes e_j 
 - a^{ \dagger} \otimes e_j \otimes a \otimes e_i  \in V \otimes  W \otimes V \otimes W  
 && 
\eea
It is useful to write this as 
\bea 
 M_{ ij} = P_{ A }^{ W \otimes W }   ( a^{ \dagger} \otimes e_i \otimes a \otimes e_j  )
\eea
$P^{W \otimes W}_{ A } $ is the anti-symmetrizer acting on the $ W \otimes W $ factor of  $V \otimes  W \otimes V \otimes W $.
The number of these $M_{ ij}$ is $ d ( d-1)/2$. But consider the  space of equivariant maps 
\bea 
P_{ A } ( W \otimes W )  \rightarrow 
P_{ A}^{ W \otimes W } ( V_+ \otimes W \otimes V_- \otimes W ) \, . 
\eea
This is a one-dimensional vector space (for $ d >4$) \footnote{ for $ d=4$ we can also use $ \epsilon_{ i_i i_2 i_3 i_4 } $  which gives another map : so we will use large $d$ in the appropriate places in our definitions to keep things as simple as possible}  spanned by the map $M$ acting as  
\bea\label{Mmap}  
M : ( e_{ i_1} \otimes e_{i_2}  - e_{ i_2 } \otimes e_{ i_1} ) 
 \rightarrow  ( a^{ \dagger} \otimes e_{ i_1 }  \otimes a \otimes e_{ i_2} 
 - a^{ \dagger} \otimes e_{ i_2 }  \otimes a \otimes e_{ i_1}  ) 
\eea
This map is associated to  the diagram in Figure \ref{fig:DiagM}. 
General juxtapositions of this diagram, associated with tensor products involving 
$ V \otimes W$ and $W$ will be used to describe an infinite dimensional associative algebra.

\noindent 
{\bf Definition of $ \cF   $ and $\Ust$ } 

Our goal is therefore  to use the above observations to give a definition of  infinite dimensional algebras 
 $\Ust$ defined in terms of diagrams, such that $\Ust$ has 
a representation theory which is similar to that of $ so (d)$ in a large $d$ limit. The formal parameter 
$ \bfd $ can be specialised to general complex numbers. We first give the proposal for the definition of the algebras $\Ust$ as a quotient of an algebra $ \cF$. This will be followed 
in section \ref{sec:DefVstar} by a definition of an infinite dimensional representation $ \Vst$ of $\Ust$. In section \ref{sec:HVVdecomp} we prove that the $ \Vst \otimes \Vst $ decomposes 
as a representation of $\Ust$ into a direct sum of three irreducible representations in the same way   $ V_d \otimes V_d$ decomposes into irreducible representations of $ so(d)$. We describe the conjecture for 
the decomposition of $ ( \Vst )^{ \otimes n }$ in terms of irreducible representations of $\Ust$, developing further the connection to the representation theory of $ so(d)$ at large $d$. 

Let $W$ be the $d$-dimensional vector representation of $so(d)$. 
\bea 
 W_2 & = &  P_{ A }  ( W  \otimes W )   \cr 
 (VW)_2 & = &  P^{ W \otimes W}_{  A }  ( ( V_+ \otimes W ) \otimes  ( V_- \otimes W) ) \cr 
T(W_2)  & =  & \bigoplus_{ n=0}^{ \infty } W_2^{ \otimes n } \cr 
T(VW)_2 &  = &  \bigoplus_{ n =0}^{ \infty } ( VW)_2^{ \otimes n } 
\eea
For any vector space $ V$, the tensor algebra is the vector space 
\bea 
T ( V ) &  =  & \mC \oplus V \oplus ( V \otimes V ) \oplus ( V \otimes V \otimes V ) \oplus \cdots \cr 
& = & \bigoplus_{ n =0}^{ \infty } V^{ \otimes n } 
\eea
with $ \otimes $ as an associative product. 
$ P_{A}$ denotes the projection to the anti-symmetric part so that 
\bea 
W_2 & = &  P_{ A} ( W \otimes W ) = \hbox{ Span }  \{ e_i \otimes e_j - e_j \otimes e_i \}   \cr 
(VW)_2 & = &  P^{ W \otimes W}_{  A }  ( ( V_+ \otimes W ) \otimes  ( V_- \otimes W) )  \cr 
&=& \hbox{ Span } \{  a^{  \dagger} \otimes e_i \otimes a \otimes e_j - 
 a^{  \dagger} \otimes e_j  \otimes a \otimes e_i  \}  \cr 
 && 
\eea
where $ e_i $ for $ i \in \{ 1, 2 , \cdots , d \} $ span the vector representation 
of $ so(d)$. 

The space of $so(d)$ equivariant maps   $  W_2 \rightarrow (VW)_2$ 
is a one-dimensional vector space for $ d > 8$. One such  map takes  
every  anti-symmetric vector in $ W \otimes W $  to the corresponding anti-symmetric vector 
in $ V^+ \otimes W \otimes V^+ \otimes W$. We can scale by an arbitrary scalar in $ \mC$. 
We will define $M$ to be the map 
\bea 
M ( e_i \otimes e_j  - e_j \otimes e_i ) 
= a^{  \dagger} \otimes e_i \otimes a \otimes e_j - 
 a^{  \dagger} \otimes e_j  \otimes a \otimes e_i 
\eea
It is a basis vector  for the one-dimensional vector space of equivariant maps $ W_2 \rightarrow ( VW)_2$.  

As a space which corresponds to  diagrams with any number of $M$-boxes, we propose to consider  
\bea 
\cF  = { \rm Hom}_{ so( \infty )  } ( T ( W_2)  , T  ( ( VW)_2 )  ) 
\eea
which will be defined in terms of $so(d)$ equivariant maps for $ d$ sufficiently large. 
The space of linear maps from 
the tensor algebra $ T ( W_2)$ to the tensor algebra $ T ( ( VW)_2) $, 
is  therefore graded by two non-negative  integers 
\bea 
Hom_{ so( \infty )  }  ( T ( W_2) , T ( VW)_2)  = \bigoplus_{ m ,n =0}^{ \infty }  Hom_{ so( \infty )  }   ( W_2^{ \otimes m } , ( VW)_2^{ \otimes n } ) 
\eea
When we consider $so(d)$ equivariant maps $ W_2^{ \otimes m } \rightarrow (VW)_2^{ \otimes n } $, 
the space of these maps simplifies for large $d$ ($ d > 2m + 2n $).  When this restriction 
on $d$ is not met, we can use $\epsilon_{i_1,\cdots,i_d}$ to contract indices from $W_2^{\otimes m}\otimes (VW)_2^{\otimes n}$ - our use of large $d$ avoids this complication. 
 We define 
\bea 
Hom_{ so(\infty) }  ( T ( W_2) , T ( VW)_2)   = \bigoplus_{ m ,  n =0}^{ \infty } 
   Hom_{ so(d ) : d > 2m + 2n }   ( W_2^{ \otimes m } , ( VW)_2^{ \otimes n } ) 
\eea
We denote 
\bea 
\cF_{ m , n } =  Hom_{ so(d ) : d > 2m + 2n }   ( W_2^{ \otimes m } , ( VW)_2^{ \otimes n } )
\eea
$ \cF_{ 1,1} $ is spanned by $M$, which is associated with the diagram in Figure \ref{fig:DiagM}. 
$\cF_{ 2,1} $ is spanned by the linear combination of diagrams shown below
\bea
\begin{gathered}\includegraphics[scale=0.4]{RHSComm1}\end{gathered}
+\begin{gathered}\includegraphics[scale=0.4]{RHSComm2}\end{gathered}
-\begin{gathered}\includegraphics[scale=0.4]{RHSComm3}\end{gathered}
-\begin{gathered}\includegraphics[scale=0.4]{RHSComm4}\end{gathered}\label{fig:H21diag}
\eea
%
This linear combination of diagrams corresponds to the map 
\bea 
 &&  ( e_{ i_1} \otimes e_{ i_2} \otimes e_{ i_3} \otimes e_{ i_4 } ) -  ( e_{ i_2} \otimes e_{ i_1} \otimes e_{ i_3} \otimes e_{ i_4 } )  -  ( e_{ i_1} \otimes e_{ i_2} \otimes e_{ i_4} \otimes e_{ i_3 } )  + 
  ( e_{ i_2 } \otimes e_{ i_1 } \otimes e_{ i_4} \otimes e_{ i_3 } )  \cr 
  &&  \rightarrow 
 \delta_{ i_2 i_3 } M_{ i_1 i_4 } +  \delta_{ i_1 i_4 } M_{i_2 i_3 } - \delta_{ i_1 i_3 } M_{ i_2 i_4 } 
- \delta_{ i_2 i_4 } M_{ i_1 i_3 }
\eea
There is an associative  product on this space of equivariant maps, which is obtained by taking the tensor product operation, or by juxtaposing diagrams. 
An element $ a \in \cF_{ m_1 , n_1 }$ multiplies an element $b$ in $ \cF_{ m_2 , n_2 }$ 
to give $ m( a , b ) = a \otimes  b \in \cF_{ m_1 + m_2 , n_1 + n_2 }$. We are using 
the associative tensor product in $ T ( W_2)$ and $ T (  (VW)_2 )$ to define 
a product on the equivariant maps. To make this more explicit, suppose 
\bea 
a : ( W_2)^{ \otimes m_1}  \rightarrow ( VW)_2^{ \otimes m_2 }  \cr 
b : ( W_2)^{ \otimes n_1}  \rightarrow ( VW)_2^{ \otimes n_2 } 
\eea
with the equivariance condition 
\bea 
&& g a = a g \qquad  g b = b g 
\eea 
for $ g \in so ( d )$ ( for $d$ large enough). Then the product of $ a$ and $b$ is a map 
\bea 
&& ( a \otimes b ) : ( W_2)^{ \otimes ( m_1 + n_1 )  } \rightarrow ( VW)_2^{ \otimes  ( m_2 + n_2 )  } \cr
&&  ( g \otimes g )  ( a \otimes b ) = ( a \otimes b ) ( g \otimes g ) 
\eea
Following the standard connection between tensor products 
and juxtaposition of diagrams in tensor categories \cite{CP0905,EtingofTC}, this multiplication map 
can be expressed as juxtaposition of the diagrams. This is illustrated in the simple case of 
$ a , b \in H_{ 1,1 }$ below
\bea
&&m:F_{1,1}\otimes F_{1,1}\to F_{2,2}\cr
&&m\left(\begin{gathered}\includegraphics[scale=0.4]{M}\end{gathered}\, ,\,
\begin{gathered}\includegraphics[scale=0.4]{M}\end{gathered}\right)=
\begin{gathered}\includegraphics[scale=0.4]{M}\end{gathered}
\begin{gathered}\includegraphics[scale=0.4]{M}\end{gathered}\label{fig:F11F11prod}
\eea

We will define $\Ust$ as a quotient of $ \cF  $.   This is modelled after the fact 
that $ Uso(d)$ is a quotient of the free algebra generated by $ M_{ ij} $ 
(for $i,j \in \{ 1, 2, \cdots , d \}$) defined by setting to zero the Lie algebra relation 
\bea
M_{ i_1 i_2 } M_{ i_3 i_4 } - M_{ i_3 i_4  } M_{ i_1 i_2 } 
- \delta_{ i_2 i_3 } M_{ i_1 i_4 }  -  \delta_{ i_1 i_4 } M_{i_2 i_3 } + \delta_{ i_1 i_3 } M_{ i_2 i_4 } 
+  \delta_{ i_2 i_4 } M_{ i_1 i_3 }  
\eea
Formally, this setting to zero is done by defining the  left and right ideal 
generated by 
\bea 
C_{ i_1 i_2 i_3 i_4 } = M_{ i_1 i_2 } M_{ i_3 i_4 } - M_{ i_3 i_4  } M_{ i_1 i_2 } 
- \delta_{ i_2 i_3 } M_{ i_1 i_4 }  -  \delta_{ i_1 i_4 } M_{i_2 i_3 } + \delta_{ i_1 i_3 } M_{ i_2 i_4 } 
+  \delta_{ i_2 i_4 } M_{ i_1 i_3 }  
\eea
in the free algebra generated by $M_{ ij}$ and doing the associative algebra quotient by  this ideal (see 
for example \cite{Humphreys} for a discussion of this standard construction of the universal enveloping algebra for general Lie algebras).   A first thought is that we can just quotient the associative diagram algebra $ \cF $ by the left and right ideal 
generated, using the juxtaposition product,  by the following linear combination of diagrams which we can call $C$
\bea
C=
\begin{gathered}\includegraphics[scale=0.4]{M}\end{gathered}
\begin{gathered}\includegraphics[scale=0.4]{M}\end{gathered}
-\begin{gathered}\includegraphics[scale=0.4]{TwistedMs}\end{gathered}
-\begin{gathered}\includegraphics[scale=0.4]{RHSComm1}\end{gathered}
-\begin{gathered}\includegraphics[scale=0.4]{RHSComm2}\end{gathered}
+\begin{gathered}\includegraphics[scale=0.4]{RHSComm3}\end{gathered}
+\begin{gathered}\includegraphics[scale=0.4]{RHSComm4}\end{gathered}\label{fig:CDGI}
\eea
This ideal consists of all elements of the form $A \otimes C \otimes B$, where $A, B$ are arbitrary elements of $\cF $. It turns out that quotienting by this ideal does not set to zero everything that we need to set to zero, when we are working with ACRTDC. At this point it is also useful to note that $C$ can be written as 
\bea\label{Calg}  
M \otimes M ( 1 - \sigma ) - M g_{23} - M g_{ 14} + M g_{ 13} + M g_{ 24}
\eea
which is an operator mapping $ W_2 \otimes W_2$ to $ ( (VW)_2 \otimes (VW)_2 )  \oplus (VW)_2 $. 
$W_2 \otimes W_2$ is a subspace of $ W^{ \otimes 4}$. The permutation 
$\sigma $  is the permutation shown in the second term of (\ref{fig:CDGI}), $ g_{ ij}$ is the paring acting on the $i$'th  and $j$'th copies of $W$.  

The reason it does not suffice to do the associative algebra quotient by the ideal generated 
by $ C$ in $\cF $ is as follows. When we are working in $ Uso(d)$ for general $d$,  we also want to set to zero expressions like 
\bea\label{MstarC} 
M_{ i_5 i_1} C_{ i_1 i_2 i_3 i_4 }
\eea
where an index in $C_{ i_1 i_2 i_3 i_4}$ is contracted with an index in $M_{ pq}$. 
When working in $ Uso(d)$ for any $d$, such expressions are just linear combinations of $ C_{ i_1 , i_2 , i_3 , i_4 }$ so they are included in the ideal generated by $C_{ i_1 , i_2 , i_3 , i_4 }$. 
When we go to the diagram algebra  $ \cF$, the expression (\ref{MstarC}) translates into the diagram
\bea
\begin{gathered}\includegraphics[scale=0.4]{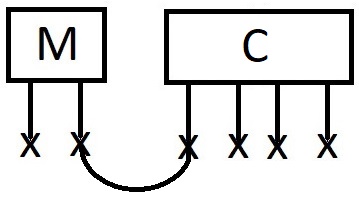}\end{gathered}
\eea
This is not the juxtaposition of $M$ with $C$. It is obtained by taking the juxtaposition, and then adding a downward arc or cap connecting a cross connected to $M$, to a cross connected to $C$. 
We may denote this as $M \star  C$, where $\star $ refers to any of  these more general operations where we juxtapose and then introduce additional arcs.

So what we would like to do in the diagram algebra $ \cF$ is set to zero general elements of the form
$ A \star  C \star  B$, and we would like to show that this setting to zero leaves us with an associative algebra. 
 It turns out that there is  a very general notion of 
quotients, developed in the context of {\it universal algebras},  which makes use of equivalence classes \cite{Wiki:quotalg,Bergman}.   In universal algebras, one studies algebraic structures specified by a 
set, along with a number of n-ary operations. This includes groups, rings, etc. \cite{Bergman}. 
A quotient  of an algebra $ \cA $ is defined by choosing an equivalence relation on the underlying set of $ \cA$ which satisfies a congruence condition between the equivalence relation and the operations defining the algebra. 

Applying this notion to our case, we take two diagrams $ a , b$  in $ \cF $  to be 
equivalent,  denoted $ a\sim b$,  if they are related by an equation of the form 
\bea\label{defEquiv}  
a  = b +   \sum_{i , j } A_i  \star  C \star  B_j 
\eea
where $\star $ is an operation of juxtaposing  and adding some (possibly none) arcs, and $ A_i  , B_j  $ are elements of $ \cF$. When we have no arcs, the star product just reduces to the juxtaposition product denoted by $ \otimes $. It is easy to see that this is reflexive $ a \sim a$, symmetric ($ a \sim b \implies b \sim a$), and 
transitive ( $ a \sim b $, $ b \sim c $ implies that $ a \sim c $). So (\ref{defEquiv}) indeed defines 
an equivalence relation. The congruence condition we need is to show that if $ a_1 \sim b_1 , a_2 \sim b_2 $, then 
$ ( a_1 \otimes  a_2 ) \sim ( b_1 \otimes  b_2)$.  Given  $ a_1 \sim b_1 , a_2 \sim b_2$, this means 
that 
\bea 
&& a_1 = b_1 + \sum_{ i  j } A_i^{(1)}  * C * B^{(1)}_j \cr 
&& a_2 = b_2 + \sum_{ i  j } A_i^{(2)}  * C * B^{(2)}_j 
\eea
which we will abbreviate as 
\bea 
&& a_1 = b_1 + \langle C \rangle_1 \cr 
&& a_2 = b_2 + \langle C \rangle_2 
\eea
It follows that 
\bea 
&& a_1 \otimes  a_2 = ( b_1 + \langle C \rangle_1 )  \otimes  ( b_2 + \langle C \rangle_2 ) \cr 
&& = b_1 \otimes b_2 +   \langle C \rangle_1 \otimes   b_2 + b_1 \otimes  \langle C \rangle_2  + 
     \langle C \rangle_1 \otimes  \langle C \rangle_2 \cr 
  &&    \sim b_1 \otimes  b_2 
\eea
In the last line we have used the fact that $ a_1 \otimes  a_2$ is written as $ b_1 \otimes   b_2$ plus terms 
involving $C$ with operations to the left and right using elements of $\cF$. 
This shows that the multiplication operation on $ \cF $ given by juxtaposition of diagrams is well 
defined on the equivalence classes. Associativity of the product on the equivalence classes follows directly from associativity in $ \cF$. This implies that the multiplication in $ \cF$ descends to an 
associative multiplication on the equivalence classes in $ \cF $ defined by the equivalence relation. 
This quotient algebra will be denoted $\Ust$. The formal variable $ \bfd $ is introduced to 
keep track of loops arising when we do multiplications in the quotient algebra.  

In the discussion above we have replaced the $d(d-1)/2$ generators $M_{ ij}$ by the 
equivariant map $ M$ along with additional equivariant maps such as 
 $ M \otimes M \cdots $, which exist in a large space of equivariant maps defined 
 through a large $d$ limit. This space of equivariant maps admits an associative product
 and allows the definition of an analogous quotient which yields $\Ust$.

In the definition of $ \cF  $ above, we have a bi-grading which gives $ \cF_{ m , n }$. 
The subspace $ \cF_{ 0 ,n } $ (and its image in $\Ust$ after the quotient) is related to Casimirs. 
For example, the quadratic Casimir $ \sum_{ i j } M_{ ij} M_{ ij} $ is related to 
the diagram in equation (\ref{CasDiag}), which is a map from $ \mC = (W_2)^{ \otimes 0 }$ to 
$ (VW)_2^{\otimes 2 }$.  In general the centre of the universal $Uso(d)$, in the large $d$ limit, 
 corresponds to  maps from  $ \mC $ to $ ((VW)_2)^{ \otimes n } $ for general $n$.

To understand how loops arise when we evaluate expressions in $\Ust$, consider  
the quadratic  Casimir operator in $Uso(d)$, which is the sum $\sum_{i,j}M_{ij}M_{ij}$. It acts by successive commutators on the basis  $M_{ kl}$ of the Lie algebra as
\bea\label{Casrel}  
[ M_{ij} , [ M_{ij}, M_{ kl} ] ] = 4 ( 2 - d )   M_{ kl} 
\eea
$4 ( 2-d)$ is the eigenvalue of the Casimir in the adjoint representation. 
When we do this calculation for general $d$, the factor of $d$ arises from evaluating $ \sum_i \delta_{ ii} = d  $  or $ \delta_{ ii} = d $ with summation convention. In the diagrammatic language $d$ comes from loops. 

Let us write out the computation of the above iterated commutator, as a step towards understanding in terms of the construction of $\Ust$ as a quotient of $ \cF $. 
\bea\label{CasOnH}  
&& [ M_{ ij} , [ M_{ ij} , M_{ kl} ] ] = [ M_{ ij } , ( \delta_{ jk} M_{ il} + \delta_{ il} M_{ jk} - \delta_{ jl} M_{ ik } - \delta_{ ik} M_{ jl }  ) ] \cr 
&& = \delta_{ jk }[ M_{ ij} , M_{ il} ] + \delta_{ il} [ M_{ ij} , M_{ jk} ] - \delta_{ jl} [ M_{ ij} , M_{ ik} ] - \delta_{ ik} [ M_{ ij} , M_{ jl} ] \cr 
&& =  \delta_{ jk} ( \delta_{ ji} M_{ il} + \delta_{ il} M_{ ji} - \delta_{ jl} M_{ ii} - \delta_{ ii} M_{ jl}  ) + \delta_{ il} ( \delta_{ jj} M_{ ik} + \delta_{ ik} M_{ jj} - \delta_{ jk} M_{ ij} - \delta_{ ij} M_{ jk} ) \cr 
&&  - \delta_{ jl} ( \delta_{ ji} M_{ ik} + \delta_{ ik} M_{ ji} - \delta_{ jk} M_{ ii} - \delta_{ ii} M_{ jk} )   - \delta_{ ik} ( \delta_{ jj} M_{ il} + \delta_{ il} M_{ jj} - \delta_{ jl} M_{ ij} - \delta_{ ij} M_{ jl} ) \cr 
&& = ( M_{ kl} + M_{ kl} - 0 +  d M_{ kl} ) + ( d M_{ lk} + 0 + M_{ kl} + M_{ kl} ) 
  + ( M_{ kl} + M_{ kl} + 0 + d M_{ lk } )  \cr 
  && + ( -d M_{ kl} + 0 + M_{ kl} + M_{ kl} ) = 4 ( 2 - d ) M_{ kl} 
\eea
To repeat this computation in terms of the quotient in $ \cF$ we have defined, recall that  
the quotient allows us to use the diagrammatic equality (\ref{fig:Mcomm}), or equivalently to set to zero the expression in (\ref{fig:CDGI}). It is useful to express this in terms of the diagram
\bea
\begin{gathered}\includegraphics[scale=0.4]{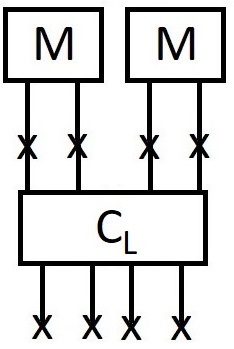}\end{gathered}=
\begin{gathered}\includegraphics[scale=0.4]{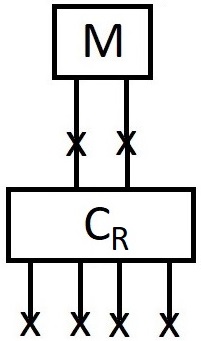}\end{gathered}
\eea
where
\bea\label{CLCRdefs} 
C_L&=&\begin{gathered}\includegraphics[scale=0.4]{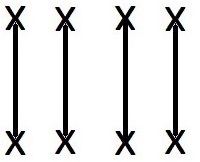}\end{gathered}-
\begin{gathered}\includegraphics[scale=0.4]{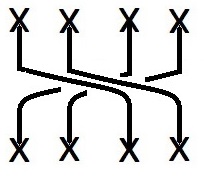}\end{gathered}\cr
C_R&=&
\begin{gathered}\includegraphics[scale=0.4]{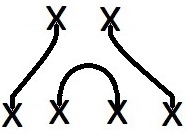}\end{gathered}
+\begin{gathered}\includegraphics[scale=0.4]{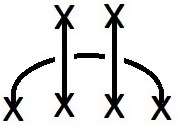}\end{gathered}
-\begin{gathered}\includegraphics[scale=0.4]{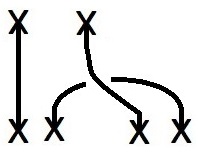}\end{gathered}
-\begin{gathered}\includegraphics[scale=0.4]{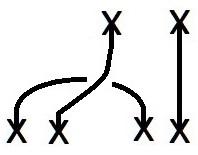}\end{gathered}
\eea
In the quotient we are allowed to use this diagrammatic equality in isolation, or when it comes  juxtaposed with diagrams to the left or right, or further, when it comes juxtaposed with diagrams, along with some down-arcs implementing index contractions (this last operation 
of juxtaposing and contracting has also been described as some star operations used in defining the quotient algebra).

We can now show how (\ref{CasOnH}) is reproduced by a diagrammatic algebra 
calculation in $\Ust$. We know that $ U so(d)$ acts on itself via commutators. Given any elements $a,b,c\in Uso(d)$ the product $ab$ acts on $c$ as 
\bea
ab : c \rightarrow [ a , [ b , c ] ] 
\eea
The decomposition of $ U so(d)$ into irreps, under this commutator action,  is given by the Poincare-Birkhoff-Witt theorem. 
A special case of this decomposition is encoded in (\ref{CasOnH}). This commutator action is translated into 
diagrams to give an action of $\Ust$ on itself.  
We can write out the diagrams relevant to the calculation of (\ref{CasOnH}) as a sum of two terms
\bea
M_{ij}[M_{ij},M_{kl}]&\to&
\begin{gathered}\includegraphics[scale=0.4]{CrossCasimir}\end{gathered}\,\,
\begin{gathered}\includegraphics[scale=0.4]{M}\end{gathered}
-\begin{gathered}\includegraphics[scale=0.4]{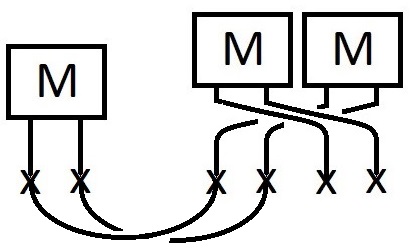}\end{gathered}\cr
&=&\begin{gathered}\includegraphics[scale=0.4]{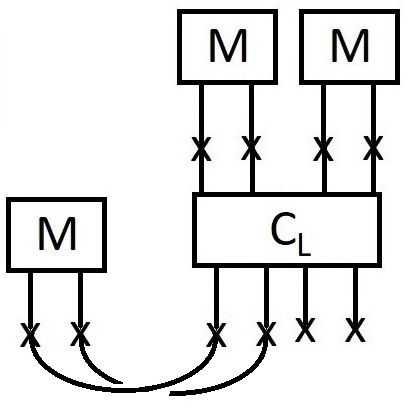}\end{gathered}
=\begin{gathered}\includegraphics[scale=0.4]{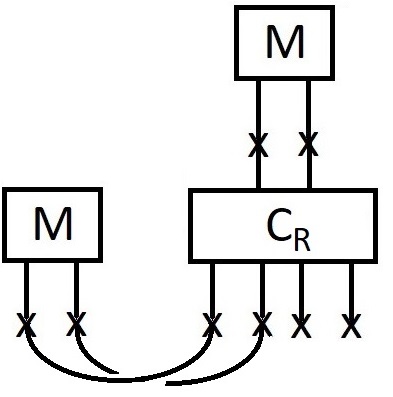}\end{gathered}\cr\cr
[M_{ij},M_{kl}]M_{ij}&\to&
\begin{gathered}\includegraphics[scale=0.4]{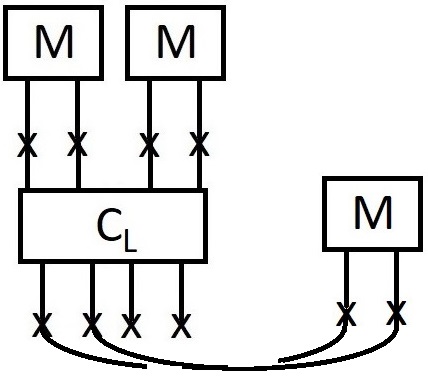}\end{gathered}=
\begin{gathered}\includegraphics[scale=0.4]{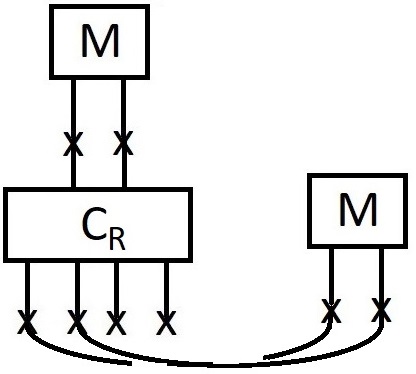}\end{gathered}\cr
&=&\begin{gathered}\includegraphics[scale=0.4]{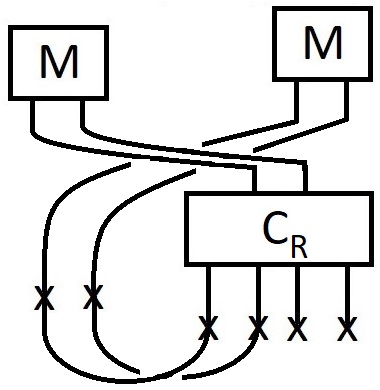}\end{gathered}\label{fig:Adsq2}
\eea
For the first expression, we recognise a diagram which is $ C_L$ composed with two $M$-boxes, along with an $M$-box on the left which is contracted with two lower arcs (an instance of the star-composition
in $\cF$). By the definition of the quotient, we can  use (\ref{fig:CDGI}) to replace $C_L$ with $C_R$ and the two $M$-boxes with a single one. A similar use of (\ref{fig:CDGI}) is applied to the second expression, which is also manipulated into a form which will allow us to write the outer commutator in terms of another application of $C_L$. This is done in the first step of
\bea
[M_{ij},[M_{ij},M_{kl}]]&\to&
\begin{gathered}\includegraphics[scale=0.4]{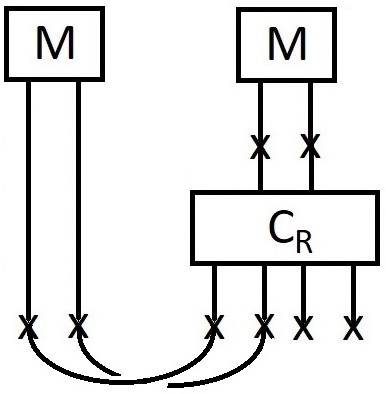}\end{gathered}
-\begin{gathered}\includegraphics[scale=0.4]{MMCRM}\end{gathered}\cr
&=&\begin{gathered}\includegraphics[scale=0.4]{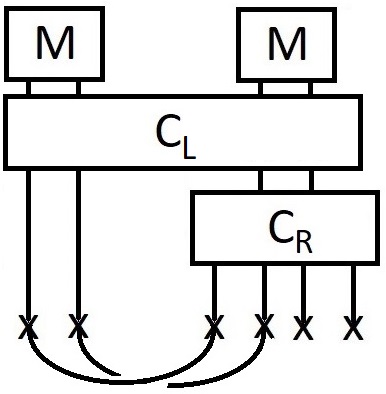}\end{gathered}
=\begin{gathered}\includegraphics[scale=0.4]{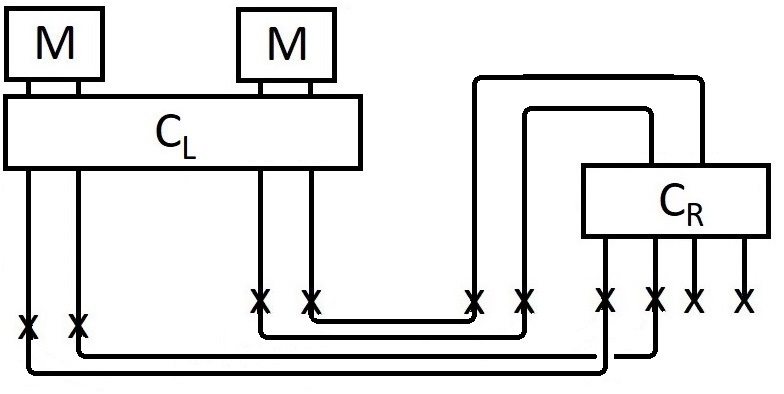}\end{gathered}\cr
&=&\begin{gathered}\includegraphics[scale=0.4]{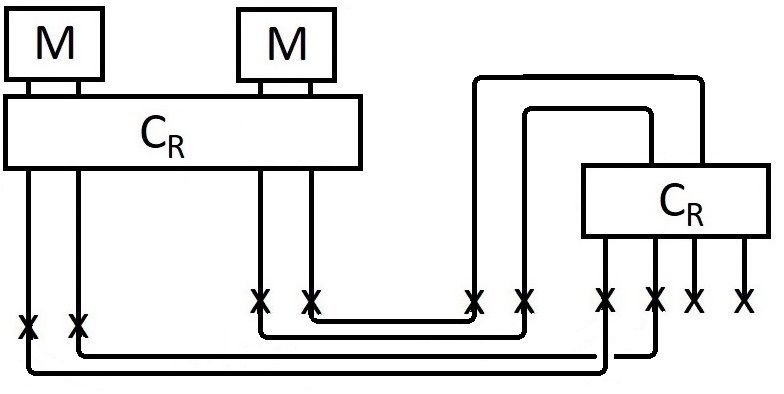}\end{gathered}
=\begin{gathered}\includegraphics[scale=0.4]{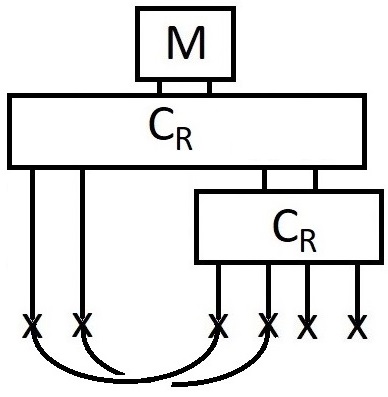}\end{gathered}
\eea 
In the second step, we show that  the diagram can be expressed as a star-composition of a diagram with $C_L$, with another diagram on the right involving six crosses. The diagram on the right is in fact an element of $ Hom_{ so(d)} ( W_2^{ \otimes 3} , \mC )$, i.e. part of the $ \cF_{ 3,0 } $ component of $ \cF$. Recall that $ W_2 = \Lambda^2 ( W )$ : there is anti-symmetry in the first two crosses, because of the contraction with $ C_L$, and anti-symmetry in the next two pairs because of the contraction with $ C_{R}$. 
Now the quotient $\Ust$ allows us to use the relation (\ref{fig:CDGI}) in the presence of star-compositions, so we can replace $ C_L$ with $C_R$ and simplify to get the last diagram in (\ref{fig:Adsq2}).  The next step is to expand out the contractions which make up $ C_L$ and evaluate. In the equation
\bea
&&\begin{gathered}\includegraphics[scale=0.4]{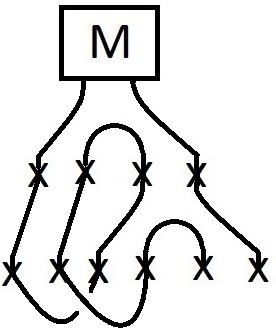}\end{gathered}
+\begin{gathered}\includegraphics[scale=0.4]{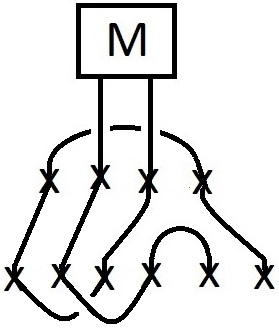}\end{gathered}
-\begin{gathered}\includegraphics[scale=0.4]{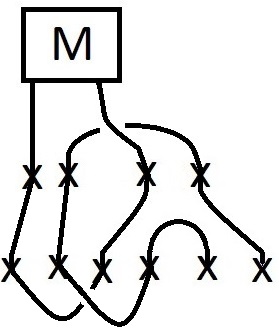}\end{gathered}
-\begin{gathered}\includegraphics[scale=0.4]{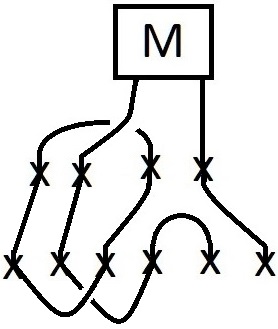}\end{gathered}
\cr
&=&\begin{gathered}\includegraphics[scale=0.4]{M}\end{gathered}+
\begin{gathered}\includegraphics[scale=0.4]{M}\end{gathered}-
\begin{gathered}\includegraphics[scale=0.4]{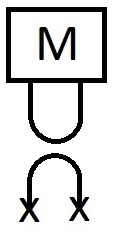}\end{gathered}
-\begin{gathered}\includegraphics[scale=0.4]{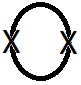}\end{gathered}
\begin{gathered}\includegraphics[scale=0.4]{M}\end{gathered}\cr
&=&2\begin{gathered}\includegraphics[scale=0.4]{M}\end{gathered}
-d\begin{gathered}\includegraphics[scale=0.4]{M}\end{gathered}
\eea
we have exhibited the diagrams corresponding to the first four terms in the third line in (\ref{CasOnH}).  The third term is zero because of the antisymmetry of the $M$-box as in (\ref{fig:antisymmM}). 
In the fourth term, we evaluate the loop as a factor of $d$. 

The diagram algebra $\Ust$ therefore allows us to extend the notion that $ Uso(d)$ 
acts on itself with a Casimir $ 4 ( 2 -d)$, to general $d$.

\section{ Definition of $ \Vst  $ and $\Ust $ action on $ \Vst  $   } \label{sec:DefVstar} 

In Section \ref{sec:DiagsAsEquivs} we have defined a diagram algebra $\Ust$, as a quotient of 
a free associative algebra $ \cF $ generated by diagrams.  The free algebra $\cF$ is 
graded by a pair of integers. The subspace $ \cF_{ 1,1}$ is one-dimensional and is spanned 
by the diagram  in Figure \ref{fig:DiagM}. The $M$-box can be viewed as an unlabelled version of the
generators $M_{ ij}$ of $so(d)$. To make sense of the index-free $M$-box as a mathematical object in $d$-dimensions, we exploited the oscillator realization (\ref{Mosc}) to write 
\bea 
M \in  Hom_{ so(d) }  ( W_2 , (VW)_2 ) 
\eea
The algebra $\cF$ was defined by replacing $W_2,(VW)_2$ by their respective tensor algebras.
We have here used the connection between commutant algebras of $ Uso(d)$ in  tensor spaces and Brauer diagrams.

In order to understand the representation theory of $\Ust$, which will be a diagrammatic analog of 
the representation theory of $Uso(d)$ at large $d$, we will start by interpreting the basic equation 
\bea 
[ M_{ ij} , a^{ \dagger}_{ k} ] = \delta_{ jk} a^{ \dagger}_{ k} - \delta_{ ik} a^{ \dagger}_j 
\eea
which we will also write as 
\bea 
M_{ ij}  \smalltriangleright a^{\dagger}_k = \delta_{ jk} a^{ \dagger}_{ k} - \delta_{ ik} a^{ \dagger}_j 
\eea
which gives the action of $ Uso(d)$ on the $d$-dimensional vector representation. By using 
labelled $M$-box diagrams, and associating to $ a^{\dagger}_k $  a line joining a cross to a circle, the above equation becomes
\bea 
{}_i \begin{gathered}\includegraphics[scale=0.4]{M}\end{gathered}{}_j\,\,\,
\begin{gathered}\includegraphics[scale=0.4]{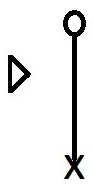}\end{gathered}_k
=\begin{gathered}\includegraphics[scale=0.4]{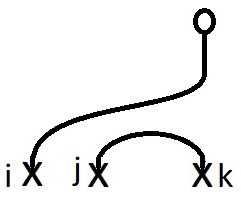}\end{gathered}
-\begin{gathered}\includegraphics[scale=0.4]{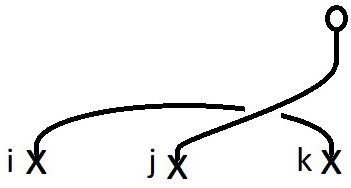}\end{gathered}
\eea
Dropping the labels to get unlabelled diagrams, we get  
\bea
\begin{gathered}\includegraphics[scale=0.4]{M}\end{gathered}\,\,\,
\begin{gathered}\includegraphics[scale=0.4]{MAct}\end{gathered}
&=&\begin{gathered}\includegraphics[scale=0.4]{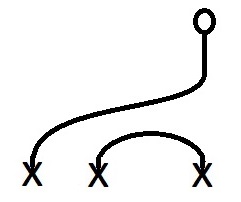}\end{gathered}
-\begin{gathered}\includegraphics[scale=0.4]{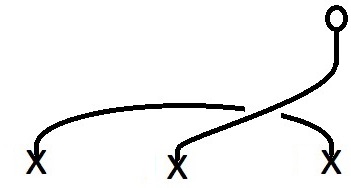}\end{gathered}\cr
&=&\,\,\begin{gathered}\includegraphics[scale=0.4]{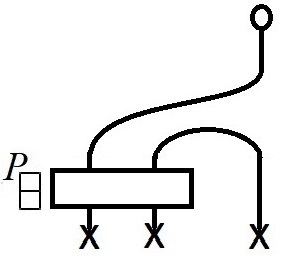}\end{gathered}
\label{fig:DiagMadag}
\eea
where
\bea
P_{\tiny\yng(1,1)}\,\,\begin{gathered}\includegraphics[scale=0.4]{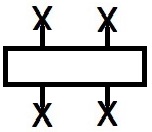}\end{gathered}=
\begin{gathered}\includegraphics[scale=0.4]{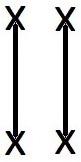}\end{gathered}
-\begin{gathered}\includegraphics[scale=0.4]{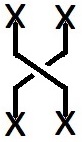}\end{gathered}
\eea
For future computations, it is useful to express the RHS in terms of the linear combination of diagrams representing the 
identity and the swop, which we denoted by $ P_{{}_{\tiny{\yng(1,1)}}}$. 

Following our definition of the unlabelled $M$ as an equivariant map, we will also define
the unlabelled diagram obtained from $a^{\dagger}_k$ as an equivariant map. 
Using the definitions from above, 
\bea 
&& a^{ \dagger}_i = a^{ \dagger} \otimes e_i \in V_+ \otimes W \cr 
&& e_i \in W 
\eea
There is an $so(d)$ equivariant map $ \rho $
\bea
\rho : W \rightarrow ( V_+ \otimes W) 
\eea
We can think of it as the map which attaches $ e_i \in W $ to $ a^{ \dagger} $ to produce $ a^{ \dagger}_i = a^{ \dagger } \otimes e_i $. The map commutes with $so(d)$.

To get a neat description of the general diagrams of the kind encountered in (\ref{fig:DiagMadag}), we 
propose the definition 
\bea 
\Vst = Hom_{ so( \infty ) } ( T ( W_2) \otimes W , V_+  \otimes W ) 
\eea
As in Section \ref{sec:DiagsAsEquivs}, $ W_2$ is the anti-symmetrized subspace of $ W \otimes W$, also denoted $ \Lambda^2 ( W )$. 
The specific diagram in  (\ref{fig:DiagMadag}) is in $Hom_{so(d)}(W_2\otimes W,V_+\otimes W)$. It is a map from the anti-symmetrised tensor product $ W_2 \otimes W $  to $ V_+ \otimes W$, which commutes with $ so(d)$. The significance of the $ d \rightarrow \infty $ limit is that, for general powers of $ W_2$, we will take $so(d)$-equivariant maps for $ d$ large enough, so as to avoid the possibility of $ \epsilon$-contractions. 
\bea 
\Vst = \bigoplus_{ n=0}^{ \infty } Hom_{ so(d)  : d  \gg n  } ( W_2^{ \otimes n } \otimes W  , V_+ \otimes W )  
=\bigoplus_{n=0}^\infty \cV_n
\eea

It is useful to describe how $\cF$ acts on $ \Vst$, and to show that this leads to 
a well-defined action of $\Ust$ on $\Vst$. We expect this will work because the oscillator 
expression for $M_{ ij}$ is consistent with the commutator relation defining $ Uso(d)$.

First let us define an action 
\bea 
\tilde \mu : \cF \otimes \Vst \rightarrow \Vst 
\eea
 Suppose $  a \in \cF_{  m , n  } $ so that 
 \bea  
 a : W_2^{ \otimes m } \rightarrow (VW)_2^{ \otimes n  } 
\eea
And suppose $ \rho \in \cV_{ m' } $, that is 
\bea 
\rho : W_2^{ \otimes m' }  \otimes W  \rightarrow V_+ \otimes W 
\eea
We have 
\bea 
a \otimes \rho : W_2^{ \otimes n } \otimes ( W_2^{ \otimes m' } \otimes W )   \rightarrow (VW)_2^{ \otimes m } \otimes (  V_+ \otimes W  ) 
\eea
To this image we can apply
\bea 
&&   COM^{ \circ m  } = COM \circ COM \circ \cdots \circ COM   \cr 
&& COM^{ \circ m } : (VW)_2^{ \otimes m } \otimes (  V_+ \otimes W  ) \rightarrow V_+ \otimes W 
\eea
The rightmost COM acts on  rightmost $ (VW)_2$ : 
 \bea\label{rightmost}  
COM:  (VW)_2 \otimes ( V_+ \otimes W ) \rightarrow V_+ \otimes W  
\eea
using the expression  (\ref{comdef}) of the oscillator commutation relation as an equivariant map.
For the case $n=1$, the diagrammatic description of the above action is  
\bea
\substack{\underbrace{\begin{gathered}\includegraphics[scale=0.4]{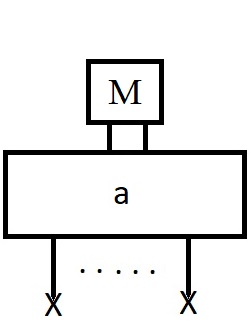}\end{gathered}}\\ m}
\,\, \triangleright\,\,
\substack{\underbrace{\begin{gathered}\includegraphics[scale=0.4]{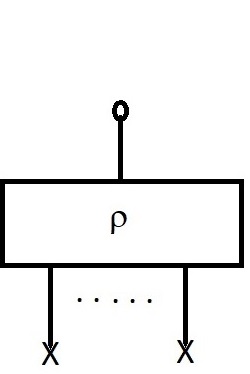}\end{gathered}}\\ m'}
=\begin{gathered}\includegraphics[scale=0.4]{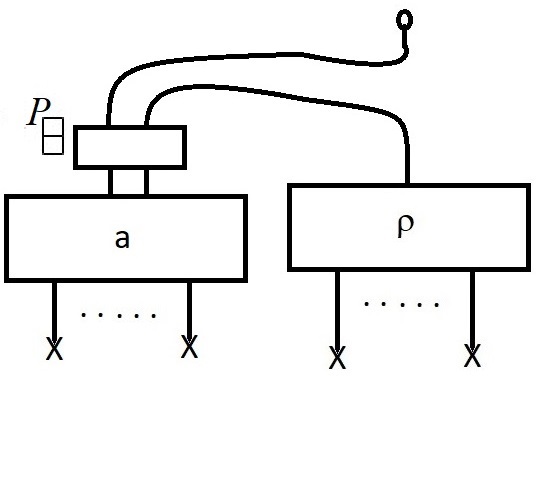}\end{gathered}
\label{fig:a-acting-on-rho}
\eea
which is the appropriate generalization of (\ref{fig:DiagMadag}). 
Then the next COM eats up another $ (VW)_2$ to produce something else in $ V_+ \otimes W$. 
After all the COM have acted, the output is in $ V_+ \otimes W $, with all the $ (VW)_2$ eaten up.   
Hence 
\bea 
COM^{ \circ m } \circ ( a \otimes \rho ) : W_2^{ \otimes n } \otimes W_2^{ \otimes m' }  \otimes W  \rightarrow ( V_+ \otimes W ) 
\eea

Then 
\bea 
COM^{ \circ m  } \circ ( a \otimes \rho )  : ( W_2 )^{ \otimes ( n + m' )  } \otimes W 
\rightarrow V_+ \otimes W  
\eea
We can define 
\bea 
&& COM^{ \star  } = \bigoplus_{ m =0  }^{ \infty }   COM^{ \circ m  } \cr 
&& COM^{ \star  } : \bigoplus_{ m =0}^{ \infty }  ( VW)_{ 2}^{ \otimes m  }  \otimes ( V_+ \otimes W ) \rightarrow ( V_+ \otimes W ) 
\eea
Then we can say 
\bea 
COM^{ \star } \circ ( a \otimes \rho ) :  \left ( \bigoplus_{ n =0}^{ \infty } W_2^{ \otimes n }  \right ) \otimes 
\left ( \bigoplus_{ m' =0}^{ \infty }  W_2^{ \otimes m' } \right )
  \otimes W \rightarrow ( V_+ \otimes W )
  \eea 
  Hence 
  \bea 
  COM^{ \star } \circ ( a \otimes \rho ) :  \left ( \bigoplus_{ n =0}^{ \infty } W_2^{ \otimes n }  \right ) \otimes W \rightarrow V_+ \otimes W 
  \eea
In other words 
\bea 
COM^{ \star  } \circ ( a \otimes \rho ) \in \Vst 
\eea
So we can define $ \tilde \mu$ as 
\bea 
\tilde \mu ( a , \rho ) \equiv COM^{ * } \circ ( a \otimes \rho )    
\eea
This is indeed a map 
\bea 
\tilde \mu  : \Ust  \otimes \Vst \rightarrow \Vst 
\eea
The above actions in terms of equivariant maps are summarised in 
\bea
\substack{n\\ \overbrace{\underbrace{\begin{gathered}\includegraphics[scale=0.4]{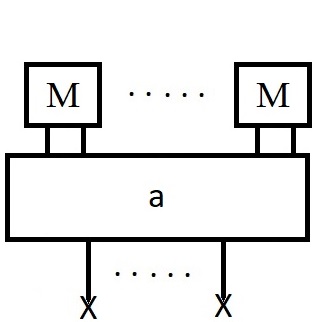}\end{gathered}}}\\ m}
\,\, \triangleright\,\,
\substack{\underbrace{\begin{gathered}\includegraphics[scale=0.4]{rhobox}\end{gathered}}\\ m'}
=\begin{gathered}\includegraphics[scale=0.4]{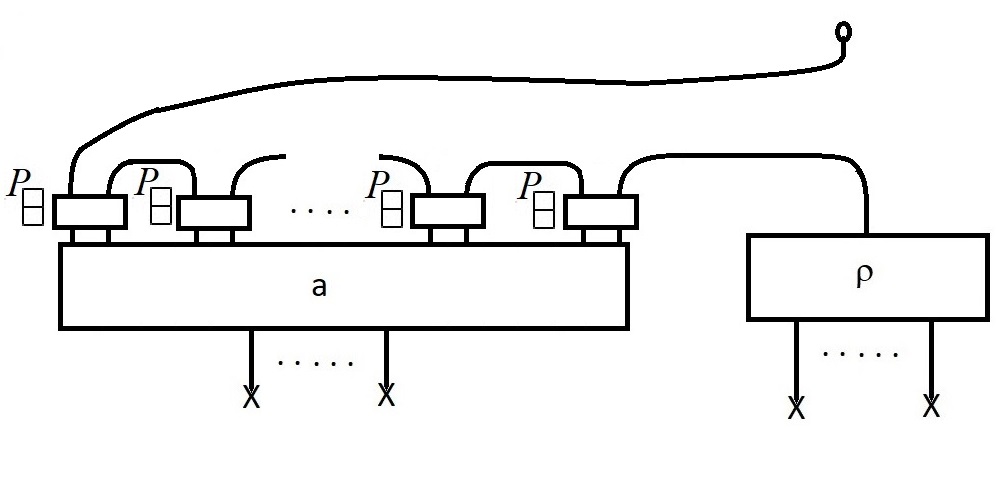}\end{gathered}
=\tilde{\mu}(a,\rho)\in V^*\cr
\label{fig:a-acting-on-rho-two}
\eea

This definition is a translation into equivariant maps of the usual  oscillator  construction of 
$M$ and its action on the fundamental rep. We expect therefore, that this action will 
obey the relations in $\Ust$ which encode, in diagammatic form - or equivalently in terms of equivariant maps - the commutation relations of $Uso(d)$. 
So we expect that 
\bea 
\tilde \mu ( M \otimes M ( 1 - \sigma ) - M g_{23} - M g_{ 14} + M g_{ 13} + M g_{ 24} , \rho  ) 
= 0 \label{expected}
\eea
This is an  equivariant map
 in $ \cF_{ 2,2 } \oplus \cF_{ 2,1} $: it  corresponds to the diagram $C$ in  (\ref{fig:CDGI}).

The $\tilde\mu$ map is defined as
\bea
   \tilde\mu (a,\rho)=COM^{ \star } \circ (a\otimes\rho)
\eea
To recap, the  $\rho$ and $a$ maps are graded by integers $m' , m , n \ge 0$ 
\begin{itemize}
\item $\rho_{m'}: (W_2)^{\otimes m'} \otimes W \to V_+\otimes W$
\item $a_{n,m}:  (W_2 )^{\otimes n}\to (VW)_2^{\otimes m}$
\end{itemize}
and the $COM^{\star } $ is a map 
\begin{itemize} 
\item $COM^{ \star } :T((VW)_2)\otimes (V_+\otimes W)\to V_+\otimes W$
\end{itemize} 
We take $a_{m,n }\in {\rm Hom}_{so(d)}\Big(  W_2^{\otimes n},\big((VW)_2\big)^{\otimes m}\Big)$.
We expect that (\ref{expected}) holds for any choice of $m'$.
Start with $m'=0$. In this case $\rho:W\to V_+\otimes W$ acts as follows
\bea
\rho (e_i)=a^\dagger\otimes e_i
\eea
A simple computation now gives
\bea
&&a\otimes \rho\Big(
(e_j\otimes e_k-e_k\otimes e_j)\otimes (e_l\otimes e_m-e_m\otimes e_l)\otimes e_i\Big)\cr\cr
&&=\Big[
M_{jk}\otimes M_{lm} - M_{lm}\otimes M_{jk}-\delta_{kl}M_{jm}-\delta_{jm}M_{kl}
+\delta_{jl}M_{km}+\delta_{km}M_{jl}\Big]\otimes a^\dagger\otimes e_i\cr\cr
&&
\eea
and then
\bea
&&COM^*\circ a\otimes \rho\Big(
(e_j\otimes e_k-e_k\otimes e_j)\otimes (e_l\otimes e_m-e_m\otimes e_l)\otimes e_i\Big)\cr\cr
&&=1\otimes 1\otimes COM^{\circ 2}\Big[
M_{jk}M_{lm}-M_{lm}M_{jk}\Big]\otimes a^\dagger\otimes e_i\cr\cr
&&+1\otimes 1\otimes COM\Big[-\delta_{kl}M_{jm}-\delta_{jm}M_{kl}+\delta_{jl}M_{km}+\delta_{km}M_{jl}\Big]
\otimes a^\dagger\otimes e_i\cr\cr
&&=(\delta_{kl}\delta_{mi}-\delta_{km}\delta_{li})a^\dagger\otimes e_j
-(\delta_{jl}\delta_{mi}-\delta_{jm}\delta_{li})a^\dagger\otimes e_k\cr\cr
&&\quad -(\delta_{mj}\delta_{ki}-\delta_{km}\delta_{ji})a^\dagger\otimes e_l
-(\delta_{kl}\delta_{ji}-\delta_{jk}\delta_{ki})a^\dagger\otimes e_m\cr\cr
&&\quad
-\delta_{kl}(\delta_{im}a^\dagger\otimes e_j-\delta_{ij} a^\dagger\otimes e_m)
-\delta_{jm}(\delta_{il}a^\dagger\otimes e_k - \delta_{ik}a^\dagger\otimes e_l)\cr\cr
&&+\delta_{jl}(\delta_{im}a^\dagger\otimes e_k-\delta_{ik} a^\dagger\otimes e_m)
+\delta_{km}(\delta_{il}a^\dagger\otimes e_j  -\delta_{ij}a^\dagger\otimes e_l)\cr\cr
&&=0
\eea
This calculation is an expression in terms of equivariant maps, which is therefore a calculation in the diagram algebra $ \cF $ and the diagram space $ \Vst$, of the familiar fact that the 
oscillator expression for $M_{ij}$, when used along with the oscillator commutation relations, obeys 
 the $so(d)$ Lie algebra relations.

It is instructive to display the purely diagrammatic content of this derivation. This is done below

\bea
\begin{gathered}\includegraphics[scale=0.4]{M}\end{gathered}
\begin{gathered}\includegraphics[scale=0.4]{M}\end{gathered}
\triangleright
\begin{gathered}\includegraphics[scale=0.4]{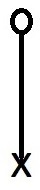}\end{gathered}&=&
\begin{gathered}\includegraphics[scale=0.4]{M}\end{gathered}
\triangleright
\begin{gathered}\includegraphics[scale=0.4]{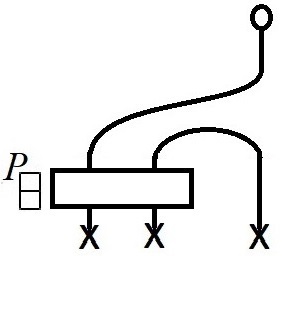}\end{gathered}
=\begin{gathered}\includegraphics[scale=0.4]{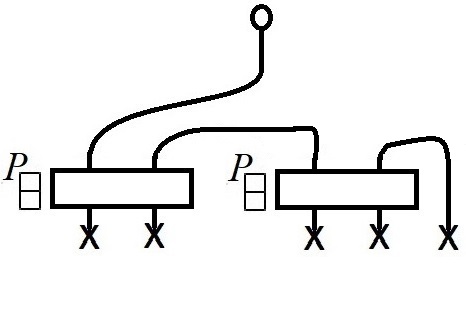}\end{gathered}\cr
\begin{gathered}\includegraphics[scale=0.4]{TwistedMs}\end{gathered}
\triangleright
\begin{gathered}\includegraphics[scale=0.4]{circcross}\end{gathered}&=&
\begin{gathered}\includegraphics[scale=0.4]{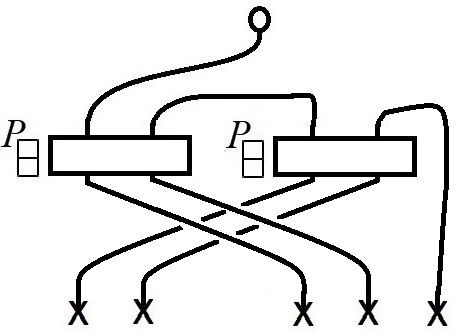}\end{gathered}
\eea

\bea
\left( \begin{gathered}\includegraphics[scale=0.4]{M}\end{gathered}
\begin{gathered}\includegraphics[scale=0.4]{M}\end{gathered}
-\begin{gathered}\includegraphics[scale=0.4]{TwistedMs}\end{gathered}\right)
\triangleright
\begin{gathered}\includegraphics[scale=0.4]{circcross}\end{gathered}=
\begin{gathered}\includegraphics[scale=0.4]{UpActAct}\end{gathered}
-\begin{gathered}\includegraphics[scale=0.4]{UpActActTwisted}\end{gathered}\equiv A_1-A_2
\eea
By definition of the ${\cal F}\to \Ust $ quotient, the above must be equal to
\bea
\begin{gathered}\includegraphics[scale=0.4]{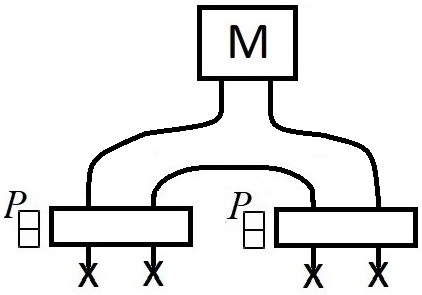}\end{gathered}\triangleright
\begin{gathered}\includegraphics[scale=0.4]{circcross}\end{gathered}&=&
\begin{gathered}\includegraphics[scale=0.4]{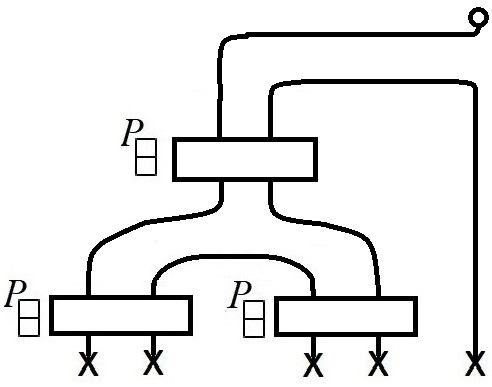}\end{gathered}\cr
&=&\begin{gathered}\includegraphics[scale=0.4]{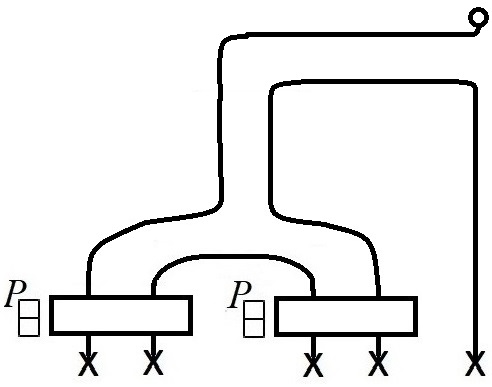}\end{gathered}-
\begin{gathered}\includegraphics[scale=0.4]{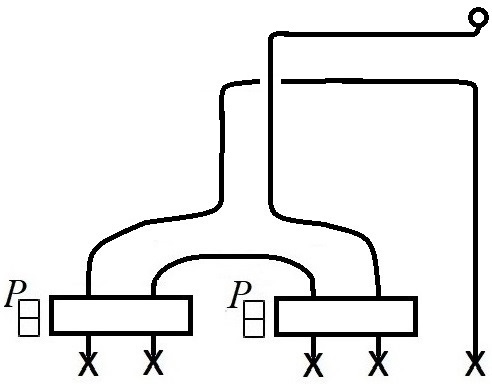}\end{gathered}\equiv B_1-B_2
\eea
A simple diagrammatic manipulation shows that the equivariant maps $ A_2$ and $B_2$ are equal
\bea
A_2=\begin{gathered}\includegraphics[scale=0.4]{UpActActTwisted}\end{gathered}
=\begin{gathered}\includegraphics[scale=0.4]{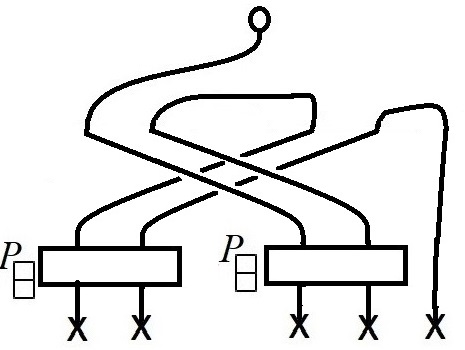}\end{gathered}
=\begin{gathered}\includegraphics[scale=0.4]{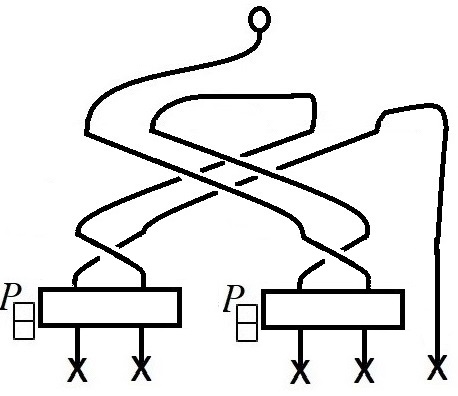}\end{gathered}
=\begin{gathered}\includegraphics[scale=0.4]{ActPPM2}\end{gathered}=B_2\cr
\eea
$A_1=A_2$ is obvious so that we have the equality
\bea
\left( \begin{gathered}\includegraphics[scale=0.4]{M}\end{gathered}
\begin{gathered}\includegraphics[scale=0.4]{M}\end{gathered}
-\begin{gathered}\includegraphics[scale=0.4]{TwistedMs}\end{gathered}\right)\triangleright
\begin{gathered}\includegraphics[scale=0.4]{circcross}\end{gathered}=
\begin{gathered}\includegraphics[scale=0.4]{PPM}\end{gathered}\triangleright
\begin{gathered}\includegraphics[scale=0.4]{circcross}\end{gathered}
\eea

Now consider generic $m'$.
Consider a $\rho$ map defined by 
\bea
  \rho:W_2^{\otimes m'}\otimes W\to V_+\otimes W
\eea
Leave $m'$ unspecified.
Whatever $\rho$ does, the result lives in $V_+\times W$ so that
\bea
  \rho (\alpha)=\sum_i c_i a^\dagger\otimes e_i 
\eea
The argument $\alpha\in W_2^{\otimes m'}\otimes W$ will have $2m'+1$ indices in general, and these would be
inherited by the coefficients $c_i$.
These indices have all been suppressed.
Then,
\bea
&&a\otimes \rho\Big(
(e_j\otimes e_k-e_k\otimes e_j)\otimes (e_l\otimes e_m-e_m\otimes e_l)\otimes \alpha\Big)\cr\cr
&&=\Big[
M_{jk}\otimes M_{lm} - M_{lm}\otimes M_{jk}-\delta_{kl}M_{jm}-\delta_{jm}M_{kl}
+\delta_{jl}M_{km}+\delta_{km}M_{jl}\Big]\otimes \sum_i c_i a^\dagger\otimes e_i\cr\cr
&&
\eea
and then
\bea
&&COM^{ \star} \circ a\otimes \rho\Big(
(e_j\otimes e_k-e_k\otimes e_j)\otimes (e_l\otimes e_m-e_m\otimes e_l)\otimes \alpha\Big)\cr\cr
&&=1\otimes 1\otimes COM^{\circ 2}\Big[
M_{jk}M_{lm}-M_{lm}M_{jk}\Big]\otimes \sum_i c_i \,\, a^\dagger\otimes e_i\cr\cr
&&+1\otimes 1\otimes COM\Big[-\delta_{kl}M_{jm}-\delta_{jm}M_{kl}+\delta_{jl}M_{km}+\delta_{km}M_{jl}
\Big]\otimes \sum_i c_i \,\, a^\dagger\otimes e_i\cr\cr
&&=(\delta_{kl} c_{m}-\delta_{km}c_{l})a^\dagger\otimes e_j
-(\delta_{jl}c_{m}-\delta_{jm}c_{l})a^\dagger\otimes e_k\cr\cr
&&\quad -(\delta_{mj}c_{k}-\delta_{km}c_{j})a^\dagger\otimes e_l
-(\delta_{kl}c_{j}-\delta_{jk}c_{k})a^\dagger\otimes e_m\cr\cr
&&\quad
-\delta_{kl}(c_{m}a^\dagger\otimes e_j-c_{j} a^\dagger\otimes e_m)
-\delta_{jm}(c_{l}a^\dagger\otimes e_k - c_{k}a^\dagger\otimes e_l)\cr\cr
&&+\delta_{jl}(c_{m}a^\dagger\otimes e_k-c_{k} a^\dagger\otimes e_m)
+\delta_{km}(c_{l}a^\dagger\otimes e_j  - c_{j}a^\dagger\otimes e_l)\cr\cr
&&=0
\eea
The diagrammatic version of this more general argument is presented below, using the definitions of
$C_L , C_R$ given in (\ref{CLCRdefs}). 
\vskip2cm 
\bea
\begin{gathered}\includegraphics[scale=0.4]{CL}\end{gathered}\,\,\,\triangleright
&\underbrace{\begin{gathered}\includegraphics[scale=0.4]{rhobox}\end{gathered}}&=
\begin{gathered}\includegraphics[scale=0.4]{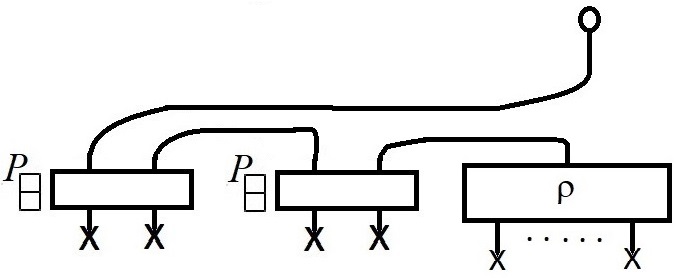}\end{gathered}
-\begin{gathered}\includegraphics[scale=0.4]{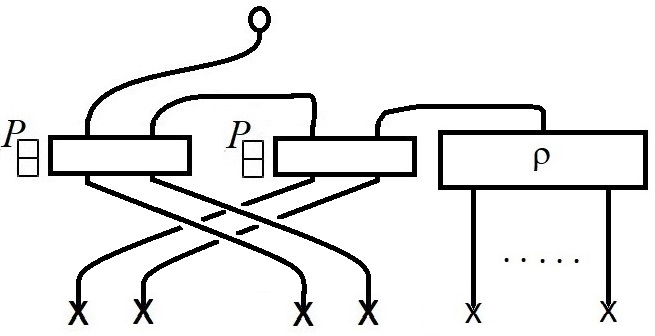}\end{gathered}\cr
&(2m'+1)&
\eea
\bea
\begin{gathered}\includegraphics[scale=0.4]{CR}\end{gathered}\,\,\,\triangleright
&\underbrace{\begin{gathered}\includegraphics[scale=0.4]{rhobox}\end{gathered}}&=
\begin{gathered}\includegraphics[scale=0.4]{PPrho}\end{gathered}
-\begin{gathered}\includegraphics[scale=0.4]{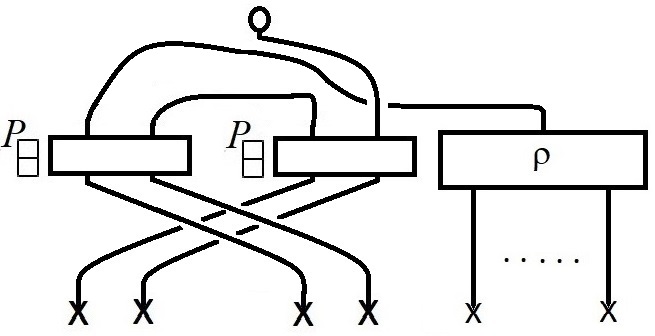}\end{gathered}\cr
&=&\begin{gathered}\includegraphics[scale=0.4]{PPrho}\end{gathered}
-\begin{gathered}\includegraphics[scale=0.4]{UpActActTwistedrho}\end{gathered}
\eea

\subsection{ $\Ust $ action on $ \Vst \otimes \Vst $ } \label{sec:HVVdecomp}

We have described the action of $\Ust$ on $ \Vst$ in the previous section. 
Now we will extend to $\Vst \otimes \Vst$ and show that the action commutes with the permutation 
$ \sigma $ and  the contraction $ C $.  We will define a map $ \tilde \mu : \cF \otimes ( \Vst \otimes \Vst ) \rightarrow \Vst \otimes \Vst $ which gives a well-defined 
map $ \mu :\Ust\otimes ( \Vst \otimes \Vst ) \rightarrow \Vst \otimes \Vst $.  In other words 
\bea 
 \mu ( a , \rho_1 \otimes \rho_2 )&&  \in ( \Vst \otimes \Vst )  \hbox{ for } a \in \Ust
  ~,~ \rho_1 \in \Vst ~ , ~ \rho_2 \in \Vst 
\eea

Recall that $\Ust$ is a quotient of a graded algebra $ \cF $ 
\bea 
\cF = \bigoplus_{ n , m } \cF_{ n , m } 
\eea
with 
\bea 
&& \cF_{ n , m } = Hom_{ so(d) : d ~large } (W_2^{ \otimes n }  ,  (VW)_2^{ m } ) 
\eea

\begin{figure}[ht]%
\begin{center}
\includegraphics[width=1.0\columnwidth]{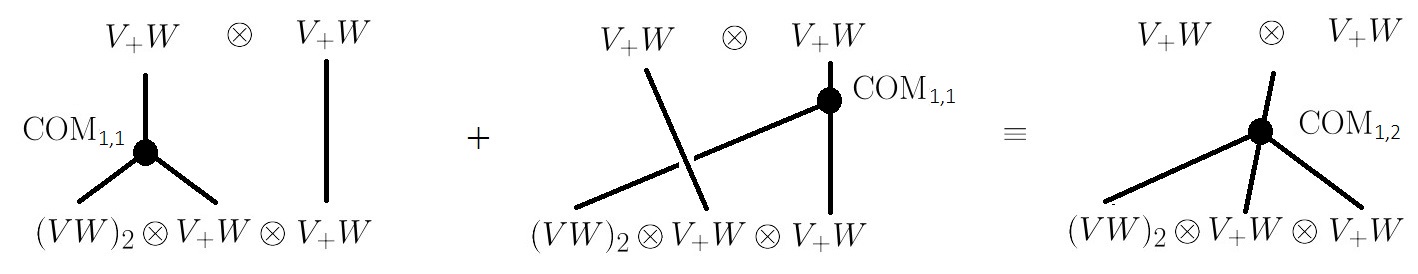}%
\caption{Definition of $ COM_{1,2} $}%
\label{fig:defcom2}%
\end{center}
\end{figure}

In  the discussion of  the action of $\Ust$ on $\Vst$, we have  a map $ \hbox{ COM }  $ 
\bea 
 \hbox{ COM }   : (VW)_2 \otimes (V_+W) \rightarrow ( V_+W)  
\eea
in equation (\ref{rightmost}). We will rewrite this map 
as 
\bea 
 \hbox{ COM}_{1,1}     : (VW)_2 \otimes (V_+W) \rightarrow ( V_+W) 
\eea
 The $ ( 1,1)$ refers to the fact that we have 
one $(VW)_2$ (in the associated diagram, one $M$-box) on the left and one $ V_+W$ on the right.  
Now we define maps 
\bea 
\hbox{ COM}_{ 1,  2} : ( VW)_2 \otimes ( V_+ W ) \otimes (V_+ W ) \rightarrow ( V_+W) \otimes (V_+W) 
\eea
with one copy of $ (VW)_2$ and two copies of $ V_+W$. 
A formula for $\hbox{COM}_{ 1 , 2} $ in terms of $ \hbox{COM}_{1,1}  $ comes from the usual formula 
\bea 
[ M_{ ij} , a^{ \dagger}_{ k_1 } \otimes a^{ \dagger}_{ k_2}  ] 
 = [ M_{ ij} , a^{ \dagger}_{ k_1} ] \otimes a^{ \dagger}_{ k_2} 
   + a^{ \dagger}_{ k_1} \otimes [ M_{ ij} , a^{ \dagger}_{ k_2}  ]
\eea
The formula is 
\bea 
\hbox{COM}_{1 ,2 } &=& ( \hbox{COM}_{1,1}  )_{ \sc{ ( VW)_2 \otimes ( V_+ W ) }  }  \otimes 1_{ \sc { V_+W }  } \cr
 &+& ( 1_{ \sc{ V_+W }  } \otimes (\hbox{COM}_{1,1}  )_{  \sc{ ( VW)_2 \otimes ( V_+ W )   } } )
  \circ  ( (\sigma )_{ \sc{  ( VW)_2 \otimes ( V_+ W ) }   } \otimes 1_{ \sc{ V_+W} }  )
\eea
This construction of $ \hbox{ COM}_{ 1,2}$   is shown in terms of   diagrams in Figure \ref{fig:defcom2}.

\begin{figure}[ht]%
\begin{center}
\includegraphics[width=1.0\columnwidth]{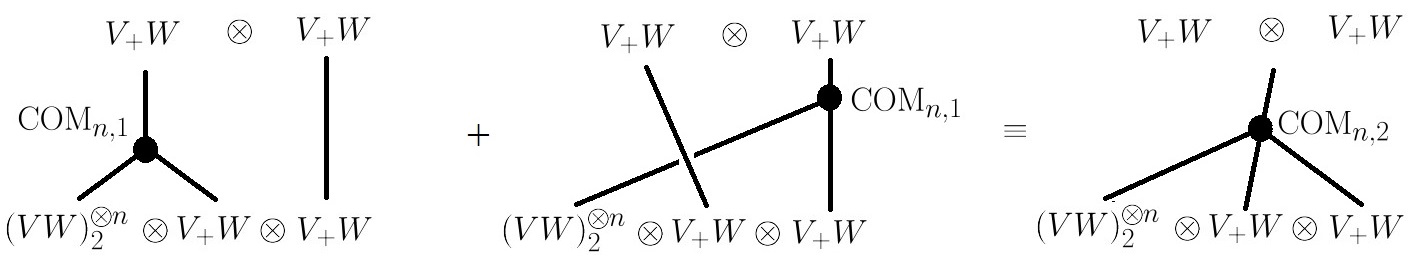}%
\caption{Definition of $COM_{n,2}$}%
\label{fig:COMn2}%
\end{center}
\end{figure}

Along similar lines, we can define 
\bea 
COM_{ n,1 }  : ( VW)_2^{ \otimes n } \otimes ( V_+W ) \rightarrow V_+W
\eea
This comes from the fact that $ M_{ i_1 j_1} M_{ i_2 j_2} \cdots M_{ i_n j_n } $ 
act on an oscillator $ a^{ \dagger}_k  $, by successive commutators,  as 
\bea 
[ M_{ i_1 j_1} , [ M_{ i_2 j_2} , \cdots [ M_{ i_n j_n } , a^{ \dagger}_k ] \cdots ]  
\eea
\bea 
&& COM_{ n , 1 } =  ( COM_{ 1,1} )_{ \sc{  (VW)_2 \otimes (V_+W) }  } )  \circ  \cdots \circ (1_{ \sc {  (VW)_2^{ \otimes n-2}   }}  \otimes  ( COM_{ 1,1} )_{ \sc{  (VW)_2 \otimes (V_+W) }  } ) \cr 
&&   \circ ( 1_{ \sc {  (VW)_2^{ \otimes n-1}   }}  \otimes ( COM_{ 1,1} )_{ \sc{  (VW)_2 \otimes (V_+W) }  }  ) 
\eea
We can then  define 
\bea 
COM_{ n , 2 } : (VW)_2^{ \otimes n }  \otimes (V_+ W)^{ \otimes 2} \rightarrow (V_+W)^{ \otimes 2 } 
\eea
using $ COM_{ n,1}$ in the same way that $ COM_{2,1}$ is built from $COM_{1,1}$
in Figure \ref{fig:defcom2}. The analogous figure here is Figure \ref{fig:COMn2}. 

\begin{figure}[h]%
\begin{center}
\includegraphics[width=0.75\columnwidth]{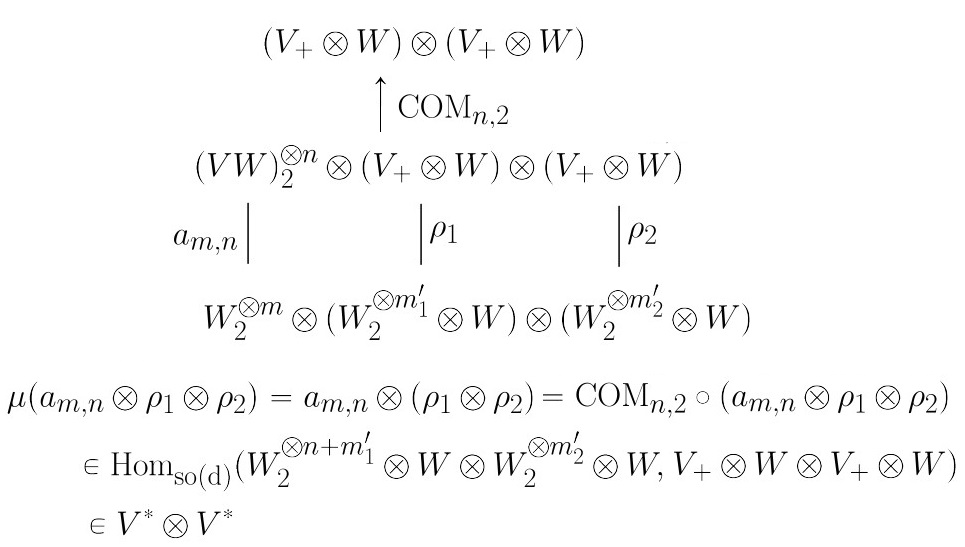}%
\caption{Definition of $\Ust$ action on $\Vst \otimes \Vst $   }%
\label{fig:HonVV}%
\end{center}
\end{figure}

 A direct sum over all $n$ gives us 
\bea 
&& COM_{ * , 2 } = \bigoplus_{ n } COM_{ n , 2 } \cr 
&& COM_{ * , 2 } : 
\bigoplus_n ( VW)_2^{ \otimes n } \otimes V_+ \otimes W \otimes V_+ \otimes  W
  \rightarrow V_+ \otimes W \otimes V_+ \otimes W 
\eea
Now we will use this to define the map 
\bea 
\tilde \mu : \cF\otimes ( \Vst \otimes \Vst ) \rightarrow ( \Vst \otimes \Vst ) 
\eea
This will  give  $ ( \Vst)^{ \otimes 2 }  $ the structure of  a representation of
 $ \cF$, and will lead to a well-defined map 
\bea 
\tilde \mu : \cF\otimes ( \Vst \otimes \Vst ) \rightarrow ( \Vst \otimes \Vst ) 
\eea
The last step should follow along  the same lines as in the discussion of the action of $ \cF $ 
on $\Vst$ descending to the action of $\Ust$, explained earlier, which essentially uses the 
fact that the oscillator construction we are using gives representations of $ Uso(d)$ and 
the quotient from $ \cF $ to $\Ust$ is implementing the $ Uso(d)$ commutation relation in the
diagrammatic setting. 
To elaborate on the construction of $ \tilde \mu $, take 
\bea 
&&  a_{ m  , n  } \in Hom_{ so(d) : d ~ large } ( (W)_2^{ \otimes m } , (VW)_2^{ \otimes n } ) 
  \in \cF  \cr 
&&  v_m  \in W_2^{ \otimes m } \cr 
&&  a_{ m  , n  } ( v_m   ) \in  ( VW)_2^{ \otimes n } 
\eea
specify $ m_1' , m_2'$ for $ \rho_1 , \rho_2 \in \Vst$: 
\bea 
&& \rho_1 : (W_2)^{ \otimes m_1'} \otimes W \rightarrow (V_+ \otimes W) \cr 
&& \rho_2 : (W_2)^{ \otimes m_2'} \otimes W \rightarrow (V_+ \otimes W)
\eea
The formula for $ \tilde \mu $ is 
\bea 
 \tilde \mu  ( a_{ m , n } ( v_m) \otimes \rho_1 \otimes \rho_2 ) )  
&&= COM_{ n  , 2 } ( a_{ m,n } ( v_m ) \otimes \rho_1 \otimes \rho_2\cr 
&& \in Hom_{ so (d)  ~ : ~ d  ~ large } ( W_2^{ \otimes m } \otimes W_2^{ \otimes m'} \otimes W \otimes W_2^{ \otimes m_2'} \otimes W , V_+ \otimes W ) \cr 
&& 
\eea
This is illustrated with   the picture in Figure \ref{fig:HonVV}

The action of $ \cF $ on $ \Vst \otimes \Vst$ we have defined in generality above, have simple 
diagrammatic expressions which can be worked out from the above definitions. 
For example if $ a \in \cF_{ 1,1} $ and $ \rho_1 , \rho_2\in Hom ( W , V_+ \otimes W ) $, i.e. $ m =1 , n=1, m_1' = 0 , m_2' = 0$ in the above discussion, the action above is given by the diagram below

\bea
\begin{gathered}\includegraphics[scale=0.4]{M}\end{gathered}\,\,
\triangleright\,\,
\begin{gathered}\includegraphics[scale=0.4]{circcross}\end{gathered}\,\,\,
\begin{gathered}\includegraphics[scale=0.4]{circcross}\end{gathered}=
\begin{gathered}\includegraphics[scale=0.4]{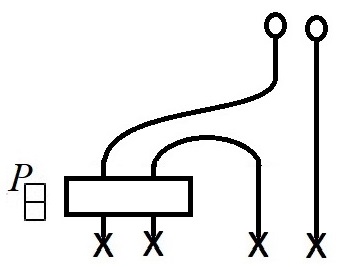}\end{gathered}+
\begin{gathered}\includegraphics[scale=0.4]{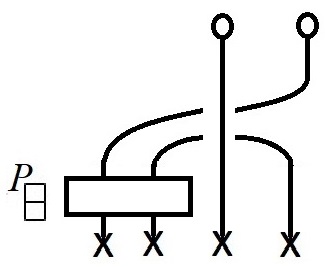}\end{gathered}\label{fig:HonVVSD1}
\eea

Keeping $ a \in M_{ 1,1}$ and considering general $m_1' , m_2'$, the action of the diagram algebra on 
the diagram module is shown below

\bea
\begin{gathered}\includegraphics[scale=0.4]{M}\end{gathered}\,\,\triangleright
\,\,
&\underbrace{\begin{gathered}\includegraphics[scale=0.4]{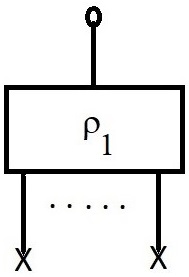}\end{gathered}}&
\underbrace{\begin{gathered}\includegraphics[scale=0.4]{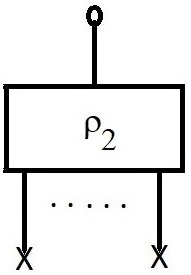}\end{gathered}}
=\begin{gathered}\includegraphics[scale=0.4]{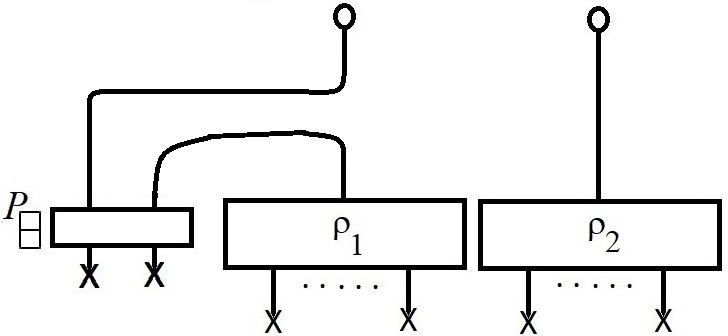}\end{gathered}
+\begin{gathered}\includegraphics[scale=0.4]{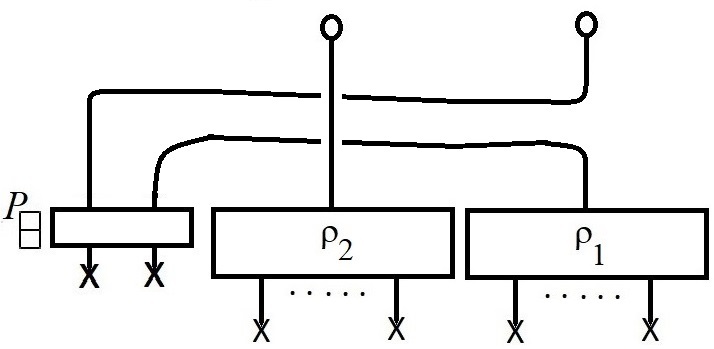}\end{gathered}\cr
&2m_1'+1&\quad 2m_2'+1\cr
&&\qquad\qquad\quad
=\sum_{\alpha\in S_2}\,\,\,
\begin{gathered}\includegraphics[scale=0.4]{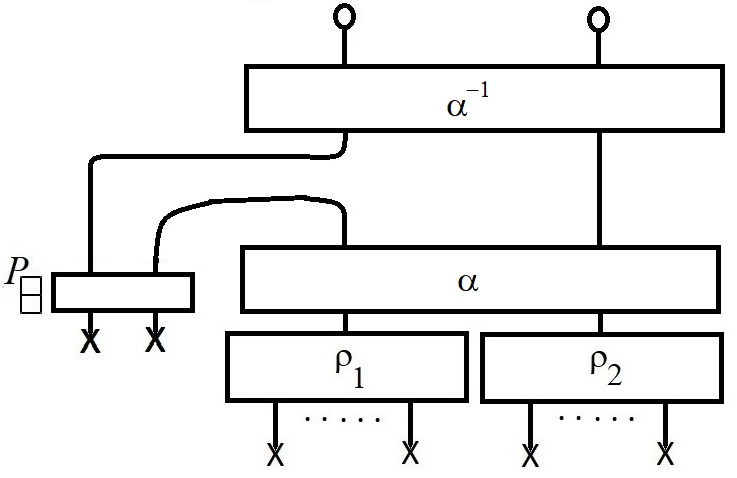}\end{gathered}
\eea
The last expression in the above equation admits a simple generalization to  general $ ( V^*)^{ \otimes n }$ by replacing the sum over permutations $ \alpha $  taking values in $ \{ () , (1,2) , ( 1,3) , \cdots , (1,n) \}$ where $()$ is the trivial permutation, $ (1,2) $ is the $(1,2)$ swop etc.

\subsection{ Linear operators commuting with the action of $ \Ust $  on $ \Vst \otimes \Vst$  } 
\label{sec:ObsComm}

Important information about the structure of a representation $ \cW $ of an associative algebra $ \cA $, specifically its decomposition into irreducibles, is contained in the sub-algebra of $ End ( \cW ) $ (space of linear maps from $\cW $ to $ \cW $ ) which commutes with $\cA$ \cite{GW98,ProcesiBook,FulHar,RamDissChap1}. This is the commutant of $ \cA $ in $ End ( \cA )$.  For example if $ \cA $ is the enveloping algebra of $ gl(N)$, $V_N$ is the fundamental representation of $ gl(N)$ and $V_N^{ \otimes n }$ is the $n$-fold tensor product,  then the commutant of 
$ \cA $ is $ \mC ( S_n)$, the group algebra of the symmetric group $ S_n$. As explained, we have constructed $\Ust$, employing a few key connections to $Uso(d)$, with the aim that it will  recover the representation theory of $ Uso(d)$ at large $d$. $V^*$ is the analog of the fundamental representation of $ V_d$ of $so(d)$. As a vector space over $ \mC$, $V^*$ is infinite dimensional. The commutant of $ Uso(d)$ in $V_d^{ \otimes 2 }$ is spanned by the identity operator, along with the permutation $ \sigma $ of the two factors along with the contraction operator $C$. In  this section we show that $ \sigma , C$, defined on $  ( V^* )^{ \otimes 2 }$ commute with the action of $\Ust$. We also find that there are additional additional operators, closely related to $ \sigma , C$ which commute with $\Ust$. We will develop the implications of these observations for  the decomposition of $ ( V^{ * } )^{ \otimes 2}$ into irreps in Section \ref{sec:DecomHonVV}.

Working with $ Uso(d)$ for generic $d$, and using a basis $e_i$ for $V_d$, we have for the action on $ V_d \otimes V_d$, the expressions 
\bea 
M_{ ij} e_k \otimes e_l = M_{ ij} e_k \otimes e_{ l } + e_k \otimes M_{ ij} e_l  \cr 
M_{ ij} e_k = \delta_{ jk} e_i - \delta_{ ik} e_j 
\eea
while the swop $ \sigma $ acts as 
\bea 
\sigma  ( e_i \otimes e_j ) = e_j \otimes e_k 
\eea
and the contraction $C$ acts as 
\bea 
C ( e_i \otimes e_j ) = \delta_{ ij} \sum_{ p } e_p \otimes e_p 
\eea
We know that $M_{ ij}$ commutes with $ \sigma $ and $C$. The commutation of $ M_{ ij}$ with $ \sigma $ 
leads us to expect that  the element $M \in \bU_{ 1,1} = \cF_{ 1,1} $ commutes with 
$ \sigma $ when acting on $V^* \otimes V^*$.  This commutation is demonstrated directly in terms of 
the diagrammatic action of the diagram algebra $\Ust $ on $ V^* \otimes V^*$ in
(\ref{fig:MsigCommVV1}), (\ref{fig:MsigCommVV2}) and (\ref{fig:MsigCommVV3}). 
\bea
\sigma\triangleright\,\,
\substack{\underbrace{\begin{gathered}\includegraphics[scale=0.4]{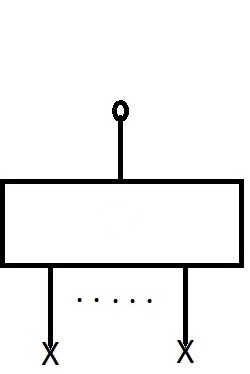}\end{gathered}}\\ 2m_1+1}\,\,\,\,
\substack{\underbrace{\begin{gathered}\includegraphics[scale=0.4]{circbox}\end{gathered}}\\ 2m_2+1}
=\begin{gathered}\includegraphics[scale=0.4]{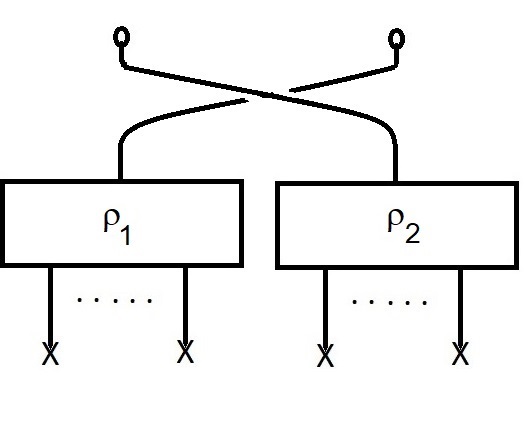}\end{gathered}\label{fig:sigonVV}
\label{fig:MsigCommVV1}
\eea

\bea
\begin{gathered}\includegraphics[scale=0.4]{M}\end{gathered}\,\cdot\,\sigma\,\triangleright\,
\!\!\!\!\!\!\!\!\!\!\!\!\!\!\!\!\!\!\!\!\!\!\!\!
&\underbrace{\begin{gathered}\includegraphics[scale=0.4]{rho1box}\end{gathered}}&
\!\!\!\!\!\!\!\!\!\!\!\!\!\!\!\!\!\!\!\!\!
\underbrace{\begin{gathered}\includegraphics[scale=0.4]{rho2box}\end{gathered}}
=
\begin{gathered}\includegraphics[scale=0.4]{M}\end{gathered}\,\triangleright\,\,
\begin{gathered}\includegraphics[scale=0.4]{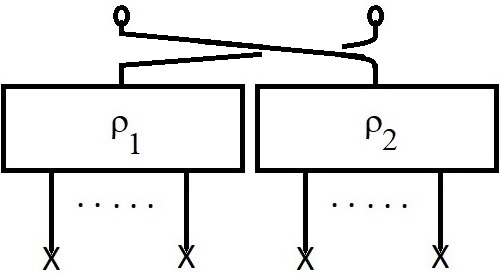}\end{gathered}\cr
&2m_1+1& 
\!\!\!\!\!\!\!\!\!\!\!\!\!\!\!\!\!\!\!\!
2m_2+1\cr\cr
=&\begin{gathered}\includegraphics[scale=0.4]{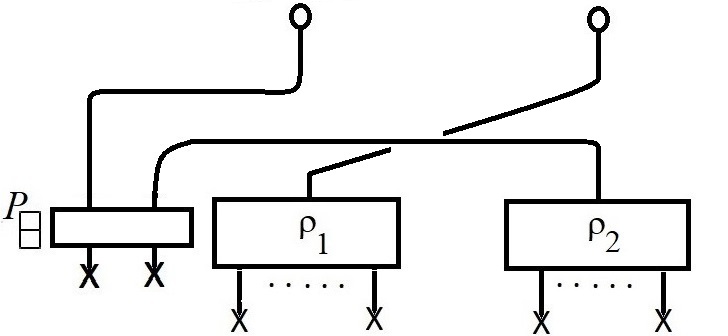}\end{gathered}&
+\,\,\,\begin{gathered}\includegraphics[scale=0.4]{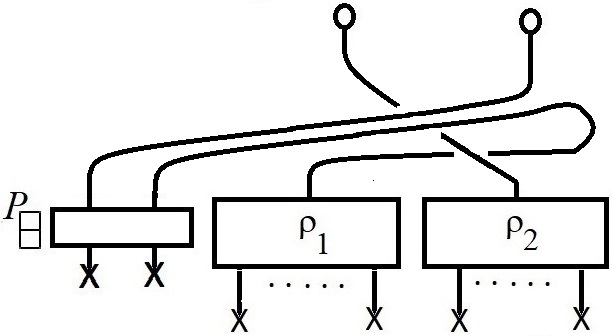}\end{gathered}
=A_1+A_2\cr
&&\label{fig:MsigCommVV2}
\eea

\bea
\sigma\,\cdot\,
\begin{gathered}\includegraphics[scale=0.4]{M}\end{gathered}\,\triangleright\,
&\underbrace{\begin{gathered}\includegraphics[scale=0.4]{rho1box}\end{gathered}}&
\underbrace{\begin{gathered}\includegraphics[scale=0.4]{rho2box}\end{gathered}}
=
\sigma\,\, \triangleright\,\,\left(\begin{gathered}\includegraphics[scale=0.4]{Prho1}\end{gathered}
+\begin{gathered}\includegraphics[scale=0.4]{Prho2}\end{gathered}
\right)\cr
&2m_1+1& \quad
2m_2+1\cr\cr
&&=\begin{gathered}\includegraphics[scale=0.4]{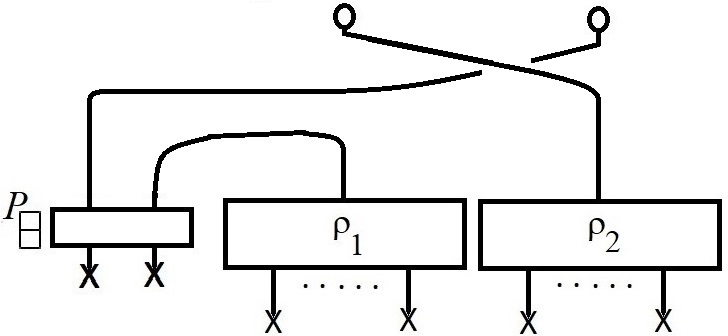}\end{gathered}
+\begin{gathered}\includegraphics[scale=0.4]{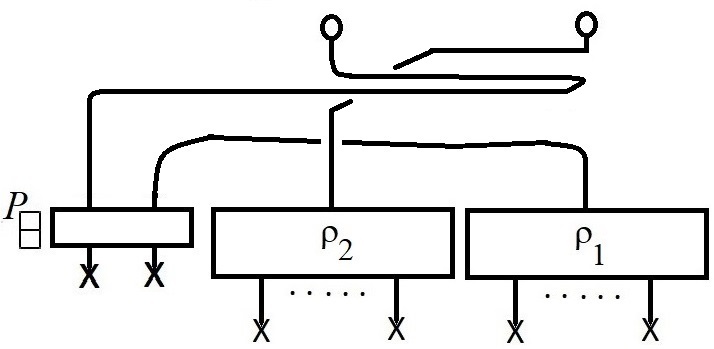}\end{gathered}\cr
&&=\begin{gathered}\includegraphics[scale=0.4]{Prho1twist}\end{gathered}
+\begin{gathered}\includegraphics[scale=0.4]{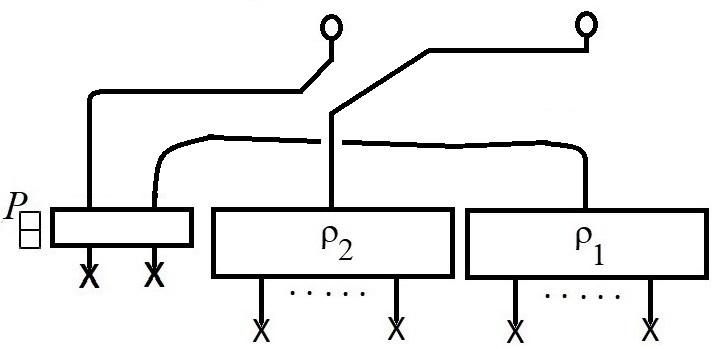}\end{gathered}=B_1+B_2\cr
&&\label{fig:MsigCommVV3}
\eea
Clearly
\bea
A_1=B_1\qquad A_2=B_2
\eea

As we have explained these index-free diagrams are associated with 
equivariant maps. It is instructive to write down the proof, using these equivariant maps, 
that the action of any element in 
$\Ust$ on $ V^* \otimes V^* $ commutes with the action of $ \sigma $. 

We will take   $ \rho_1 \otimes \rho_2 \in \Vst \otimes \Vst$. 
\bea 
&& \rho_1 \otimes \rho_2 : ( T (W_2)  \otimes W ) \otimes ( T ( W_2 ) \otimes W )  \rightarrow V_+W \otimes V_+W 
\eea
where we have abbreviated $ V_+ \otimes W $ as $ V_+W$. The permutation $ \sigma  $ is the swop 
of the states in the two factors of $W$
\bea 
\sigma = \sigma_{ W \otimes W } : V_+W \otimes V_+W \rightarrow V_+W \otimes V_+W 
\eea
which is indicated in the subscript on the RHS. 
\bea  
&& \mu ( \sigma , \rho_1 \otimes \rho_2 ) = \sigma_{ W \otimes W } \circ  ( \rho_1 \otimes \rho_2) \cr 
&& \mu ( \sigma , \rho_1 \otimes \rho_2 ) : T (W_2) \otimes T ( W_2 ) \rightarrow V_+W \otimes V_+W  \cr 
&& \mu ( \sigma , \rho_1 \otimes \rho_2 )  \in \Vst \otimes \Vst 
\eea 
For $ a \in \Ust  = Hom_{so(d) } ( T ( W_2) , T ( (VW)_2)  ) $, we want to show 
\bea 
\mu ( a , \mu ( \sigma , \rho_1 \otimes \rho_2 ) ) = 
\mu ( \sigma  , \mu ( a , \rho_1 \otimes \rho_2 ) ) 
\eea
Given any $ v_3 \in T ( ( W_2) ) $, $ a ( v_3 ) \in T ( (VW)_2 ) $
\bea 
&& \mu ( a ( v_3 ) , \rho_1 ( v_1 ) ) = COM_{*,1} ( a ( v_3) \otimes \rho_1 ( v_1 ) ) \cr 
&& \mu ( a ( v_3 ) , \rho_1 ( v_1 ) \otimes  \rho_2 ( v_2) ) 
= COM_{ \star , 2 } ( a ( v_3 ) \otimes \rho_1 ( v_1 ) \otimes \rho_2 ( v_2 ) ) \cr 
&& = COM_{\star ,1} ( a (v_3 ) , \rho_1 ( v_1 ) ) \otimes \rho_2 ( v_2 ) 
     + \rho_1 ( v_1) \otimes COM_{ \star ,1} ( a (v_3 ) \otimes \rho_2 ( v_1) ) \cr 
     && 
\eea
Apply $ a ( v_3) $ first and then $ \sigma $ 
\bea\label{athensig} 
&& \mu ( \sigma , \mu ( a ( v_3 ) , \rho_1 ( v_1) \otimes \rho_2 ( v_2 ) )  \cr 
&& = \mu ( \sigma , COM_{ *,1 } ( a ( v_3 ) \otimes \rho_1 ( v_1 )  )  \otimes \rho_2 ( v_2 )  ) + 
          \mu ( \sigma ,         \rho_1 ( v_1 ) \otimes COM_{*,1} ( a ( v_3 ) , \rho_2 ( v_2 ) ) ) \cr  
&& =  \rho_2 ( v_2 ) \otimes COM_{ \star ,1 } ( a ( v_3 ) \otimes \rho_1 ( v_1 )  ) 
       +  COM_{\star ,1} ( a ( v_3 ) , \rho_2 ( v_2 ) ) \otimes \rho_1 ( v_1 ) \cr  
       && 
\eea
Now apply $ \sigma $ first, then $ a ( v_3)$ 
\bea\label{sigthena}  
&& \mu ( a ( v_3 ) , \mu ( \sigma , \rho_1 ( v_1 ) \otimes \rho_2 ( v_2 ) )  
= COM_{ \star ,2 } ( a ( v_3 ) , \rho_2 ( v_2 ) \otimes \rho_1 ( v_1) ) \cr 
&& = COM_{\star ,1} ( a ( v_3 ) \otimes \rho_2 ( v_2 ) ) \otimes \rho_1 ( v_1 ) 
  + \rho_2 ( v_2 ) \otimes COM_{ \star ,1} ( a (v_3) \otimes \rho_1 ( v_1 ) ) \cr 
  && 
\eea

The outcomes in (\ref{athensig}) and (\ref{sigthena}) are equal. So we have proved that the action of $\Ust$ on $ \Vst \otimes \Vst$  commutes with $ \sigma $ on $ \Vst \otimes \Vst$.

We will now show that the action of $\Ust$ on $ \Vst \otimes \Vst  $ commutes with  the contraction operator. 
Let us first show that the action of $M \in \Ust$ on $V^{\star }  \otimes \Vst$ commutes with 
the contraction and permutation operators $C $. 
The $M$ box acting on $ \Vst \otimes \Vst$ followed by the action of $C$ gives zero. 
In fact it is easy to  show this for any $ a \in \cF_{ n , 1 }$ the $ n=1$ case corresponds to the $M$-box.  
The diagrammatic manipulation in the equations below demonstrate that $ C \cdot a $ (which is the composition of $C$ and $a$) acting on $ V^{\star } \otimes \Vst $ gives zero  
\bea
\underbrace{\begin{gathered}\includegraphics[scale=0.4]{rho1box}\end{gathered}}
&\underbrace{\begin{gathered}\includegraphics[scale=0.4]{rho2box}\end{gathered}}&=
\underbrace{\begin{gathered}\includegraphics[scale=0.4]{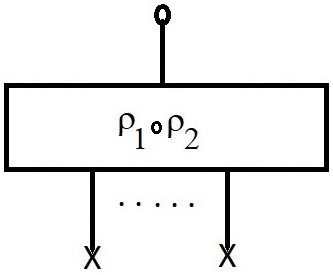}\end{gathered}}\cr
2m_1+1 &2m_2+1&\quad 2m_1+2m_2+2
\eea

\bea
&\underbrace{\begin{gathered}\includegraphics[scale=0.4]{abox}\end{gathered}}&\,\,
\triangleright\,\, \begin{gathered}\includegraphics[scale=0.4]{rho1circrho2box}\end{gathered}
=\sum_{\alpha\in S_2}
\begin{gathered}\includegraphics[scale=0.4]{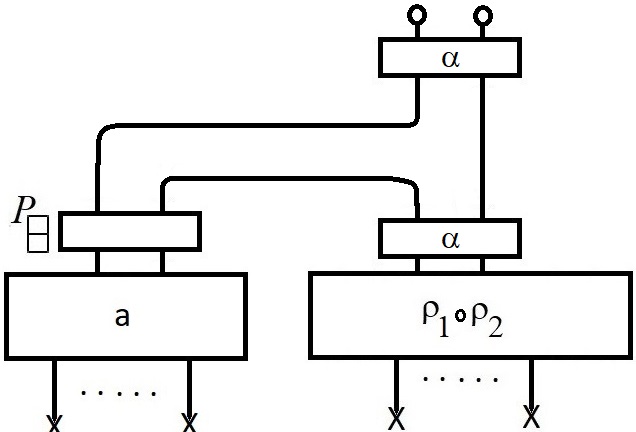}\end{gathered}\cr
&m&
\eea

\bea
C\cdot \begin{gathered}\includegraphics[scale=0.4]{abox}\end{gathered}\,\,
\triangleright\,\, \begin{gathered}\includegraphics[scale=0.4]{rho1circrho2box}\end{gathered}
&=&\sum_{\alpha\in S_2}
\begin{gathered}\includegraphics[scale=0.4]{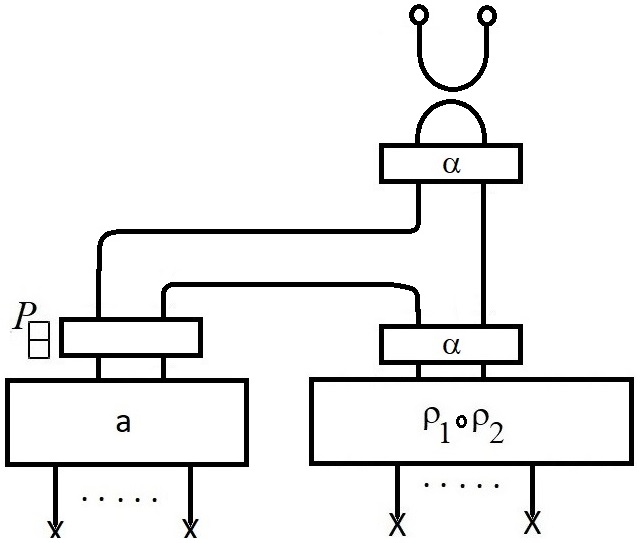}\end{gathered}\cr
&=&\sum_{\alpha\in S_2}
\begin{gathered}\includegraphics[scale=0.4]{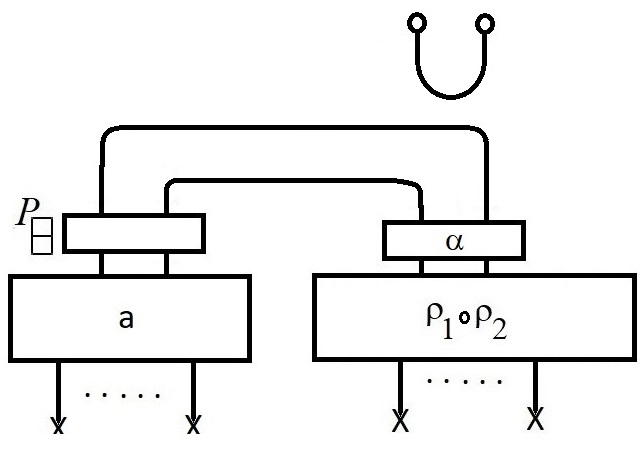}\end{gathered}\cr
&=&
\begin{gathered}\includegraphics[scale=0.4]{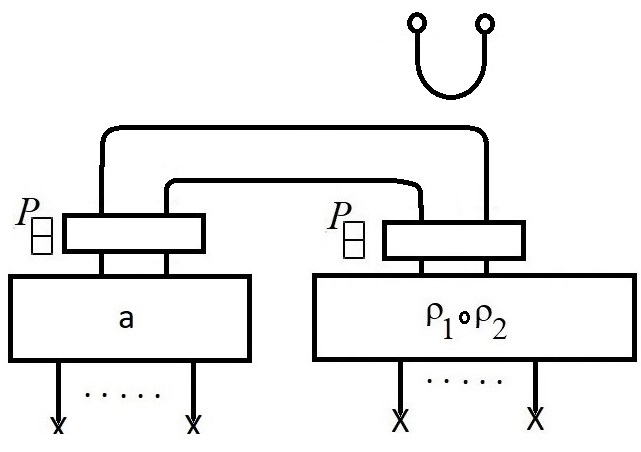}\end{gathered}=0
\eea
where we have used the fact that
\bea
\begin{gathered}\includegraphics[scale=0.4]{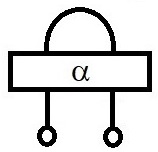}\end{gathered}=
\begin{gathered}\includegraphics[scale=0.4]{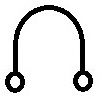}\end{gathered}
\eea
Likewise $  a \cdot C =0$ as shown below.
\bea
\begin{gathered}\includegraphics[scale=0.4]{abox}\end{gathered}\cdot C\triangleright
\,\,\begin{gathered}\includegraphics[scale=0.4]{rho1circrho2box}\end{gathered}&=&
\begin{gathered}\includegraphics[scale=0.4]{abox}\end{gathered}\triangleright
\begin{gathered}\includegraphics[scale=0.4]{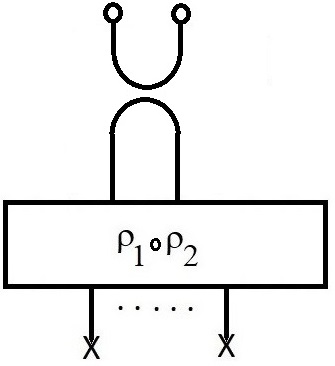}\end{gathered}\cr
&=&\sum_{\alpha\in S_2}
\begin{gathered}\includegraphics[scale=0.4]{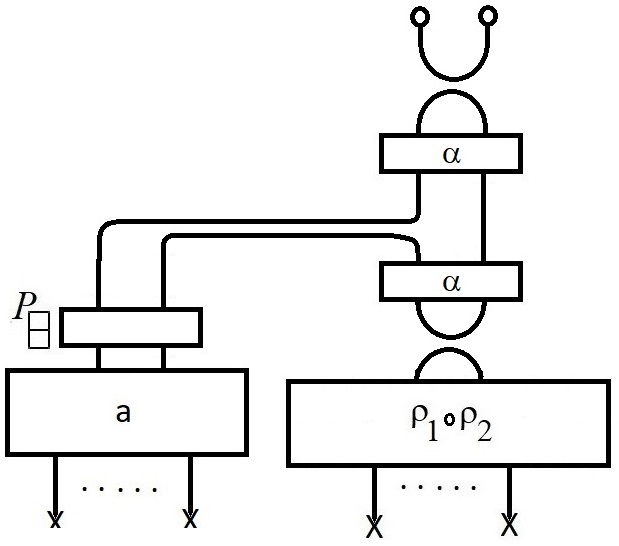}\end{gathered}\cr
&=&
\begin{gathered}\includegraphics[scale=0.4]{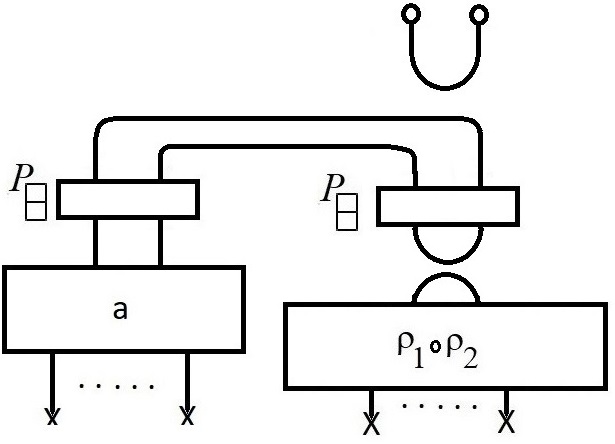}\end{gathered}=0
\eea

These arguments can be generalized to $ a_{ n , m } \in \cF_{ n , m } $ for general $ n , m $.
Diagrammatically, $a_{n,m}$ is represented as
\bea
&&\quad n\text{ M-boxes}\cr
&&\underbrace{\overbrace{\begin{gathered}\includegraphics[scale=0.4]{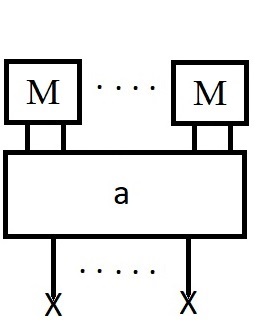}\end{gathered}}}
\text{ for }a\in{\cal F}_{n,m}\cr
&&\quad 2m
\eea
The arguments are purely diagrammatic and are given  in (\ref{fig:CAmnVV0}) and (\ref{fig:AmnCVV0}). 
The equation (\ref{fig:CAmnVV0}) shows that the action of $ C \cdot a_{ m , n } =0$ on a general 
$ \rho_1 \otimes \rho_2 \in \Vst \otimes \Vst $. 
The equation (\ref{fig:AmnCVV0}) shows that $ a_{ m , n } \cdot C = 0$. 
Taken together they imply $ [ a_{ m  , n } , C ] = 0$. 
\bea
C\cdot\begin{gathered}\includegraphics[scale=0.4]{aboxes}\end{gathered}\triangleright
\begin{gathered}\includegraphics[scale=0.4]{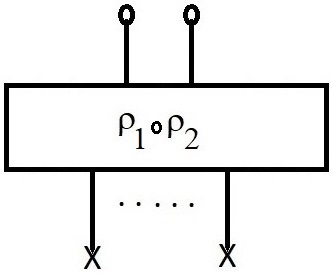}\end{gathered}
=\sum_{\alpha_1,\cdots,\alpha_n\in S_2}
\begin{gathered}\includegraphics[scale=0.4]{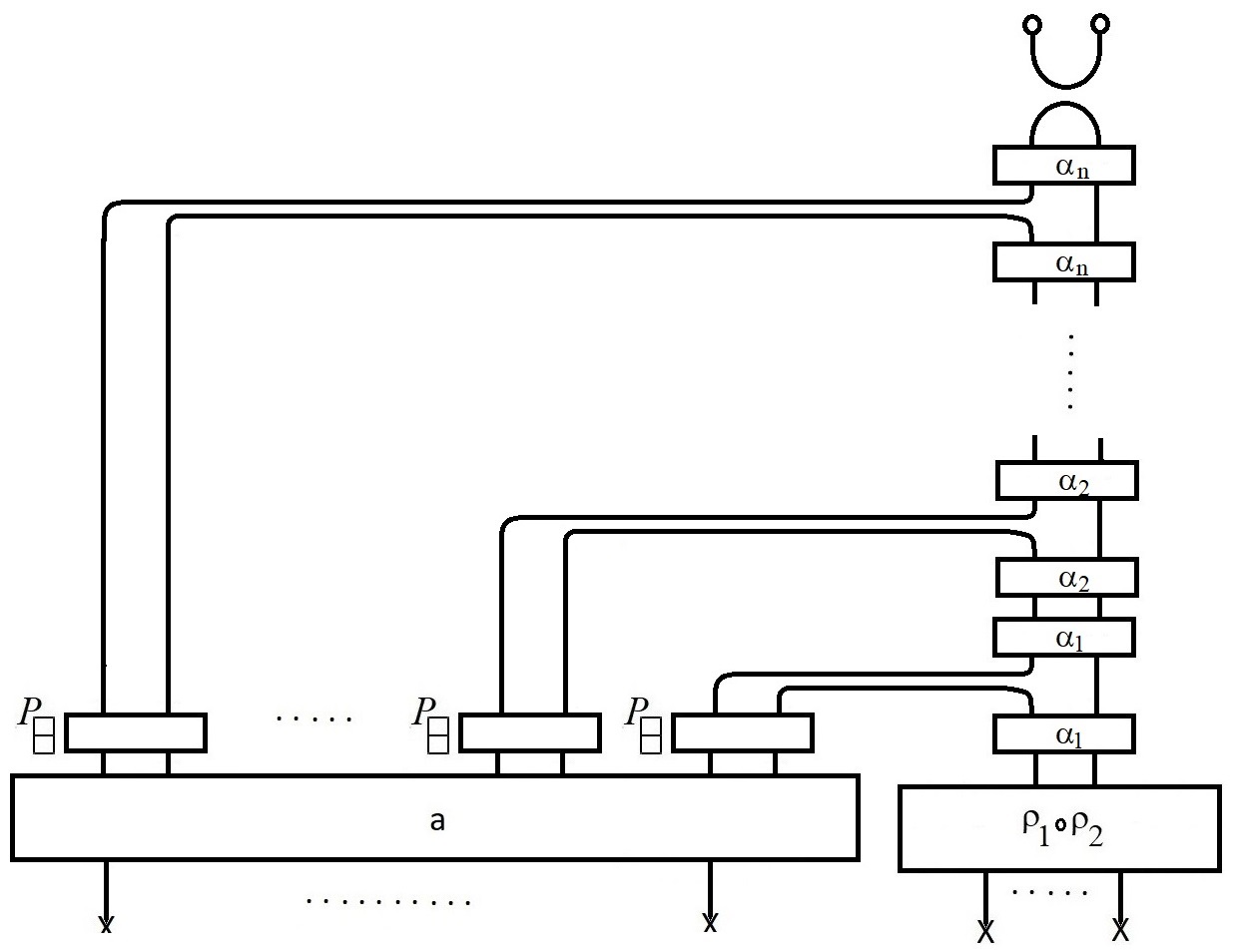}\end{gathered}\cr
=\sum_{\alpha_1,\cdots,\alpha_n\in S_2}
\begin{gathered}\includegraphics[scale=0.4]{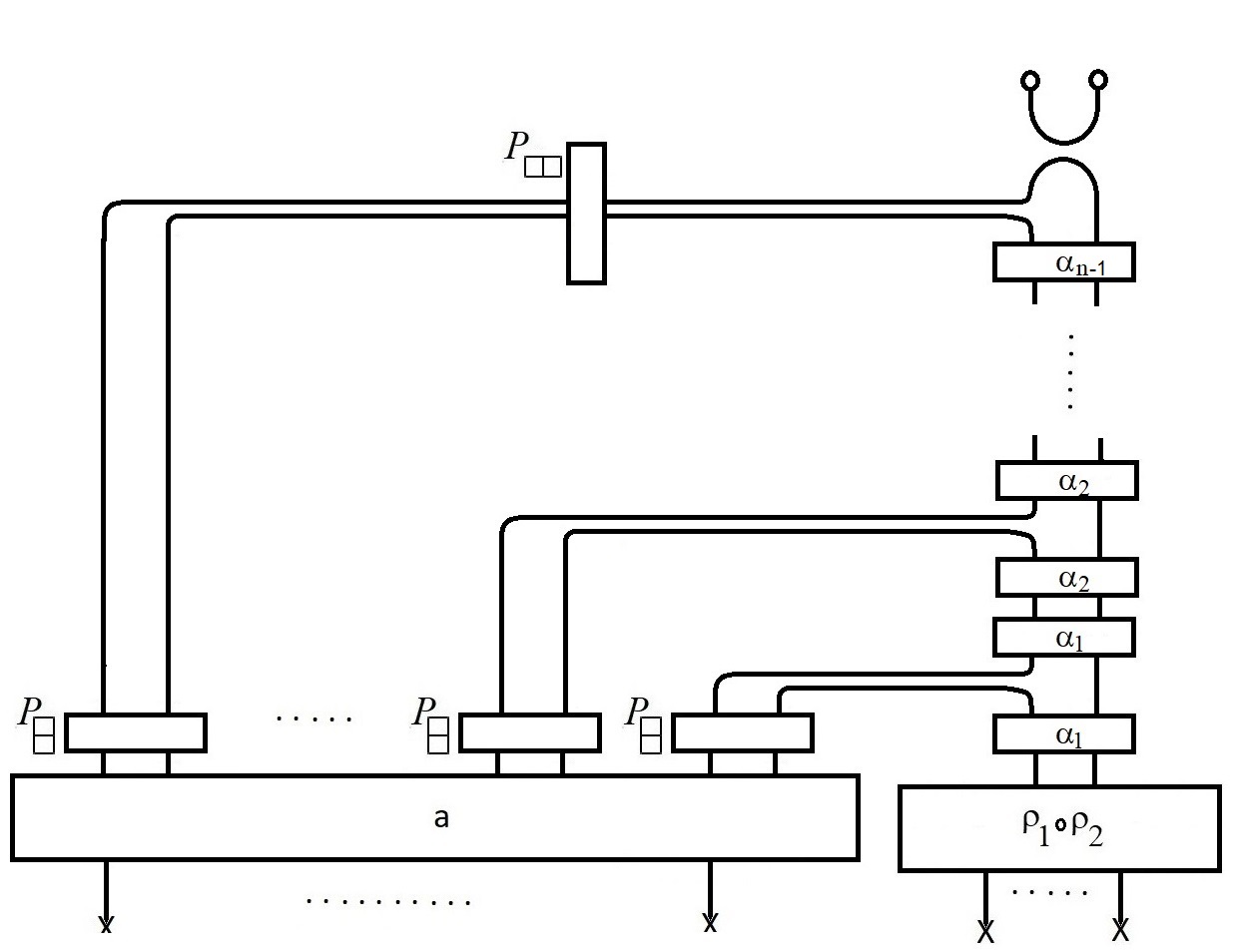}\end{gathered}=0\cr
\label{fig:CAmnVV0}
\eea
\bea
\begin{gathered}\includegraphics[scale=0.4]{aboxes}\end{gathered}\cdot C \triangleright
\begin{gathered}\includegraphics[scale=0.4]{rho1circcircrho2box}\end{gathered}
=\begin{gathered}\includegraphics[scale=0.4]{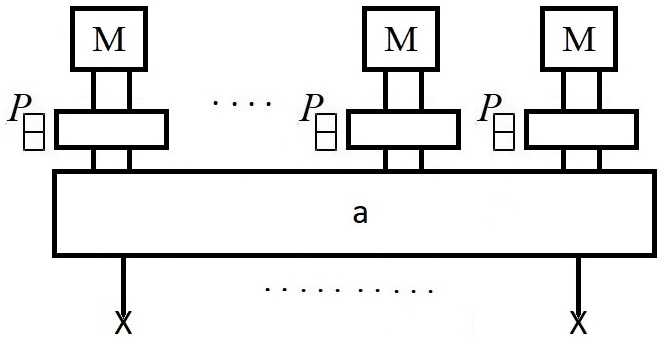}\end{gathered}\,\,\triangleright\,\,
\begin{gathered}\includegraphics[scale=0.4]{rho1circrho2C}\end{gathered}\cr
=\sum_{\alpha_1,\cdots,\alpha_n\in S_2}
\begin{gathered}\includegraphics[scale=0.4]{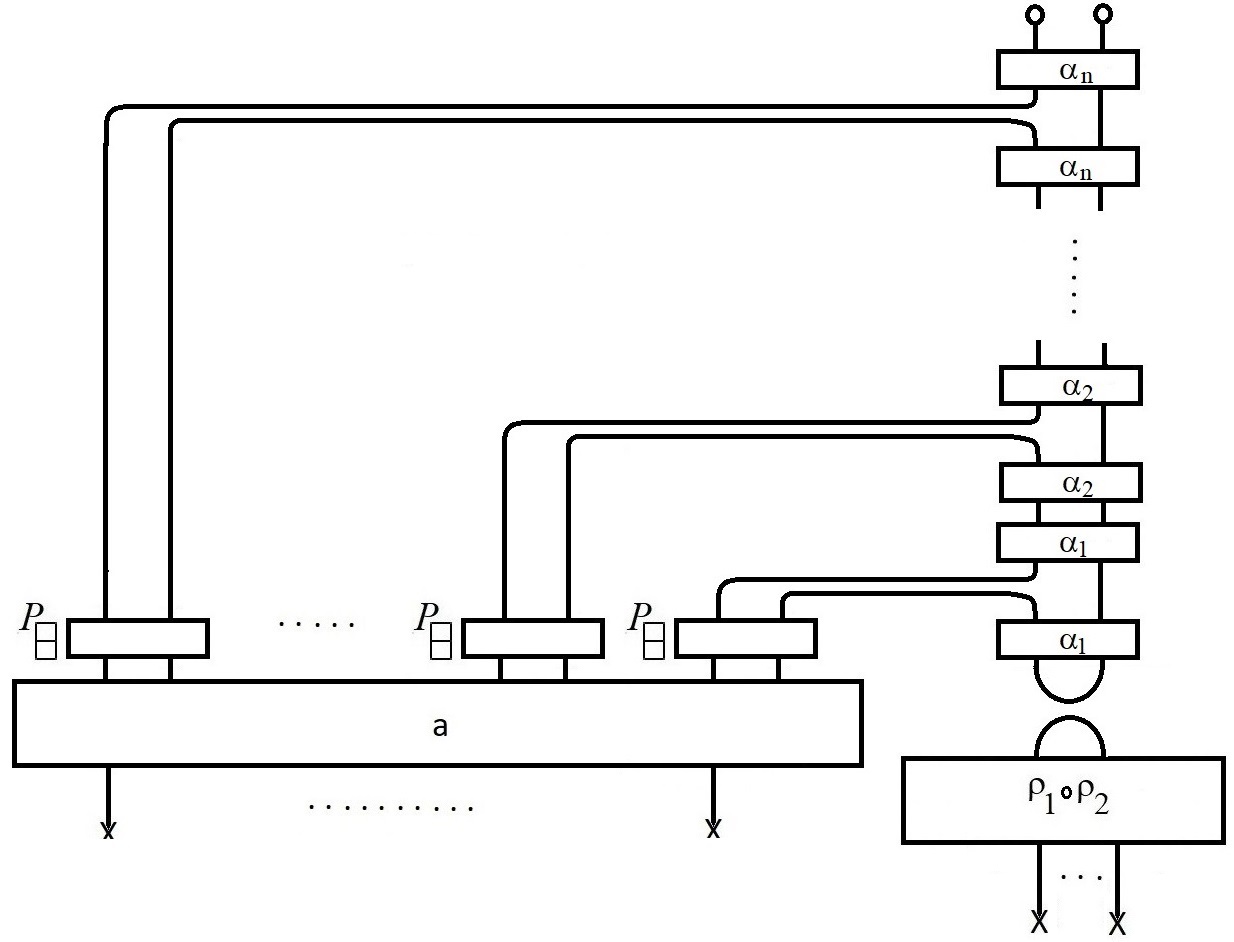}\end{gathered}\cr
=\sum_{\alpha_2,\cdots,\alpha_n\in S_2}
\begin{gathered}\includegraphics[scale=0.4]{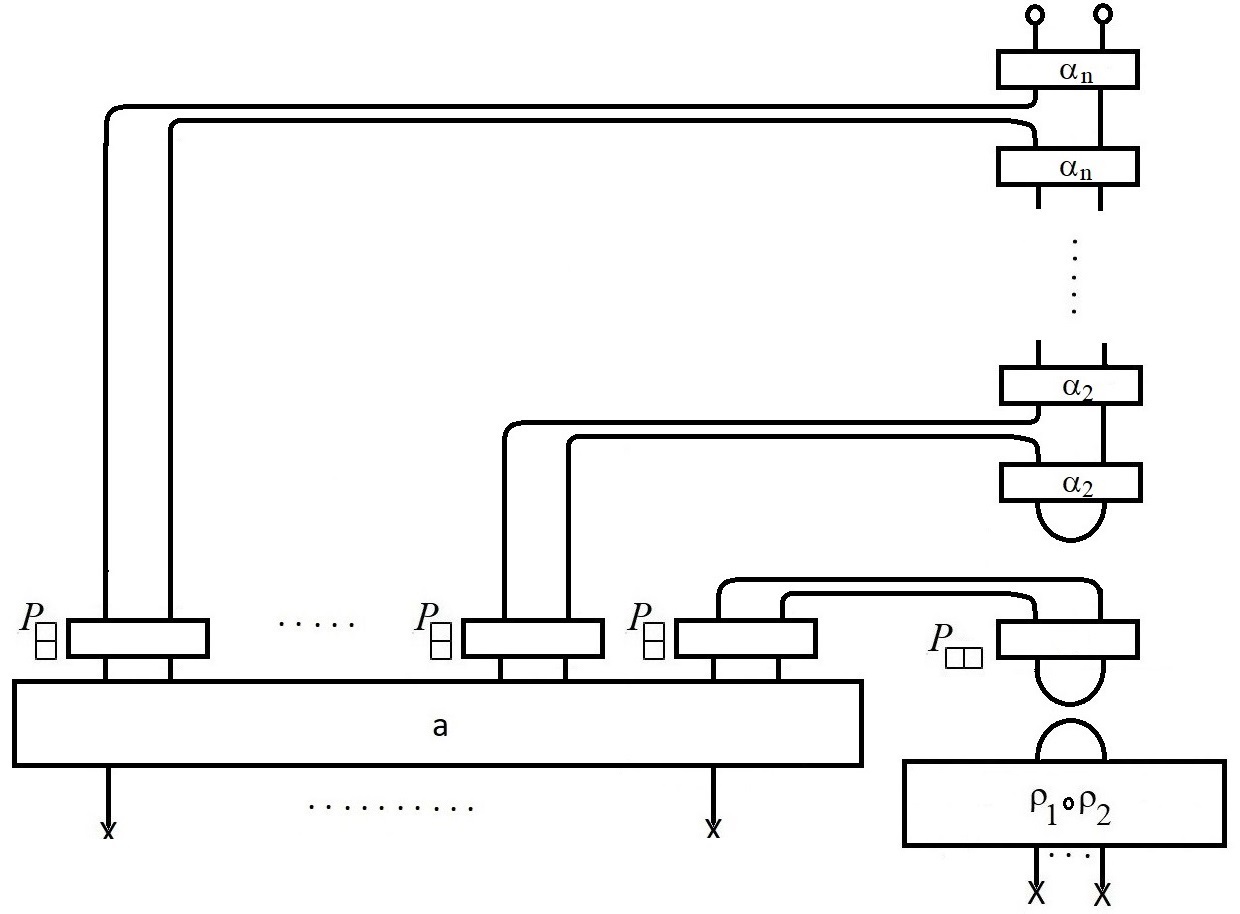}\end{gathered}=0
\label{fig:AmnCVV0}
\eea
The conclusion is that since $C, \sigma $ commute with $\Ust$, it follows that 
the images of the three projectors are invariant subspaces under the action of 
$\Ust$. By specialising  $ \bfd $ to any real number $d$, we have a generalization of 
the $so(d)$ decomposition of $V_d \otimes V_d$ to generic $d$.

Our argument proves that $\sigma$ and $C$ commute with $\Ust$ on $ \Vst \otimes \Vst $. 
We have not shown that anything commuting with $\Ust$ is generated by $ \sigma , C$. 
In fact, it is easy to show that operators $ a_{ m , 0 } \sigma $ and $ a_{ m , 0 } C$, for 
$ a_{ m, 0 } \in \cF_{ m , 0 }$,  also commute with 
general $a_{ m , n } \in \cF $. Scaling $ \sigma $ (or $ C$) means that we are acting with $ \sigma$  ( or $ C$) and then tensoring from the left with some diagrams involving $2m$ crosses going to the vacuum. 
Acting subsequently with a general $a$ just involves operations on the open circles in the diagram, which are unaffected by the $a_{ m , 0 }$. Thus all the above commutation arguments can be carried over. 
It is tempting to conjecture that these operators from  $ B_d(n)$ scaled by $ a \in \cF_{ \star , 0 }$
 are in fact the complete commutant of $\Ust$ acting on $ \Vst  \otimes \Vst$. This suggests we should be thinking about 
$\Ust$ as a module over the ring $ \cF_{ * , 0 } $. In other words we should treat these as scalars.  
Schur-Weyl duality in the more general setting of rings over modules is discussed in \cite{Cruz-SWrings}.

A remark is in order. The above discussion shows the special role that is played by elements of the ring $ \cF_{ * , 0 } $. In achieving the best definition of $\Ust$, which has the cleanest connections (these connections will be expressed as conjectures in section \ref{sec:TensRepsHd})  to the representation theory of $ Uso(d)$, there is another place where we may tweak the definition of $ { \bf U_{ \star  } }$ and indeed of the parent algebra $ \cF_{ \star }$. A variation of $ \cF $ imposes the condition that any element of the form $ a \otimes b  \otimes c $ where $ b \in \cF_{ \star , 0 }$  should be identified with 
$b\otimes a\otimes c$. In other words, we should treat any $b \in \cF_{ \star , 0 }$ as a scalar in the sense that it commutes with other sub-diagrams in $ \cF $. In pictures, we are imposing relations of the following form
\bea\label{Fst0asScal}
\begin{gathered}\includegraphics[scale=0.3]{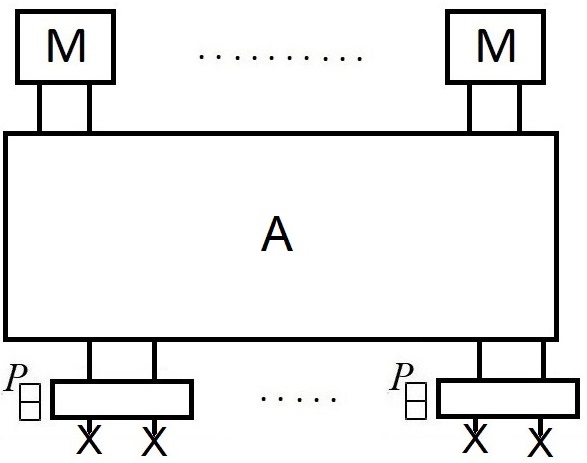}\end{gathered}\,\,\,
\begin{gathered}\includegraphics[scale=0.3]{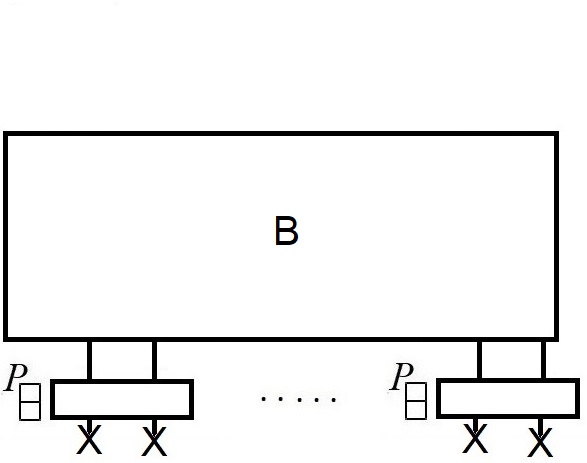}\end{gathered}
=\begin{gathered}\includegraphics[scale=0.3]{PdotsPB}\end{gathered}\,\,\,
\begin{gathered}\includegraphics[scale=0.3]{PdotsPAMM}\end{gathered}
\eea

\subsection{ Decomposing $ \Vst \otimes \Vst $ into orthogonal subspaces  invariant under $ \Ust$  } 
\label{sec:DecomHonVV}

Recall that we have 
\bea 
\Vst = Hom_{ so(\infty )}  ( T ( \Lambda^{ 2 } ( W ) ) \otimes W \rightarrow V_+ \otimes W ) 
\eea
We have shown that the  $ \Ust $ action on  $ \Vst \otimes \Vst $  commutes with 
$  \sigma , C$. This implies that there is a map from  $\Vst \otimes \Vst$ to orthogonal 
subspaces analogous to the three irreps which appear in decomposing $ V_d \otimes V_d$ for generic $d$, in terms of $so(d)$ irreps. There is a subtlety   in this analogy which we will subsequently discuss, stemming from the fact that $ \sigma , C$ do not, when treating $ \Ust , \Vst$ as vector spaces over $\mC$, generate the full commutant of $ \Ust$.

We have three orthogonal projectors in the Brauer algebra $ B_d (2)$
\bea\label{3proj}  
&& P_{ [2] } = \left (   { 1 \over 2 } ( 1 + \sigma ) - { C \over d } \right ) \cr 
&& P_0 =  { C \over d } \cr 
&& P_{ [1^2] } = { 1 \over 2 } ( 1 - \sigma ) 
\eea
obeying 
\bea 
P_i P_j & = &  \delta_{ ij} P_j \hbox { for } i \in \{ [2] , 0 , [1^2] \} \cr 
P_{ [2] } + P_{ 0 } + P_{ [1^2]}  & =  & 1 
\eea
Multiplication of the projectors is done by vertically stacking  the diagrams,  which show that 
\bea 
C^2 & = &  d C \cr 
\sigma C & = &  C \sigma = C 
\eea
This is illustrated using diagrams as follows
\bea
C=\begin{gathered}\includegraphics[scale=0.4]{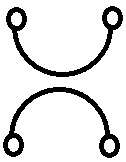}\end{gathered}\qquad
C^2 &=&\begin{gathered}\includegraphics[scale=0.4]{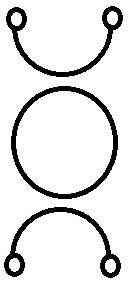}\end{gathered}
=d\begin{gathered}\includegraphics[scale=0.4]{C}\end{gathered}
=dC\cr
C\cdot\sigma =\begin{gathered}\includegraphics[scale=0.4]{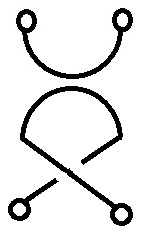}\end{gathered}
=\begin{gathered}\includegraphics[scale=0.4]{C}\end{gathered}=C\qquad
\sigma\cdot C&=&\begin{gathered}\includegraphics[scale=0.4]{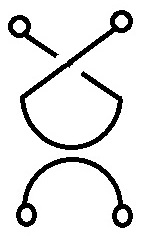}\end{gathered}
=\begin{gathered}\includegraphics[scale=0.4]{C}\end{gathered}=C
\eea
These relations form part of the standard relations of the Brauer algebra which is the commutant of $so(d)$ 
in $ V_d \otimes V_d$ ( for $ d > 2 $ ).

We have seen in Section \ref{sec:ObsComm} that
$ \sigma , C $ commute with the action of $\Ust$ on $\Vst \otimes \Vst$. 
Thus these projectors commute with  $\Ust$. The images of these projectors are invariant under the action of $\Ust$.   

Thus we have well-defined actions of the three projectors on the diagrams $D \in \Vst \otimes \Vst$. 
\bea
D = ( P_0 + P_{ [2]] } + P_{ [1^2] } ) D = ( P_0 D )  + ( P_{ [1^2]} D ) + ( P_{ [2] } D )  
\eea
The image of these three projectors are subspaces 
\bea 
\Vst_0   & = &  P_{ 0 } ( \Vst \otimes \Vst )  \cr 
V_{[1^2]}^* & = &  P_{ [1^2] }  ( \Vst \otimes \Vst ) \cr 
V_{ [ 2]] }^* & = & P_{ [2] } ( \Vst \otimes \Vst )
\eea

We would like to say that these are orthogonal subspaces. 
i.e. we want to define an inner product  $ \langle * , * \rangle $ for the diagrams in $ \Vst \otimes \Vst$  with respect to which the projectors are hermitian so that we can  write 
\bea 
\langle P_i D_1 , P_j D_2 \rangle = \langle D_1 , P_i P_j D_2 \rangle = 0 \hbox{ unless } i = j 
\eea

Given two diagrams $A , B$, we define the inner product to be zero if the number of incoming crosses 
in the two diagrams is different. If the number is the same, 
then we use the operation defined in (\ref{fig:InnerProdVVdef}). This involves inverting one of the diagrams stacking it below the first diagram, and joining the open circles. This will result in a
number of loops; the diagram is evaluated as $ d^{ L } $ where $L$ is the number of loops. 
\bea\label{InnerProdAB} 
\left\langle 
\begin{gathered}\includegraphics[scale=0.4]{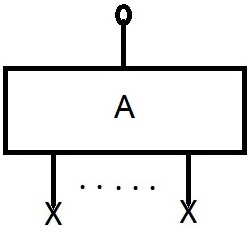}\end{gathered},
\begin{gathered}\includegraphics[scale=0.4]{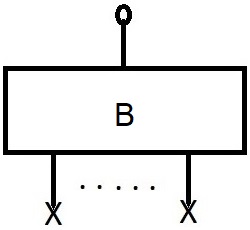}\end{gathered}
\right\rangle
=\begin{gathered}\includegraphics[scale=0.4]{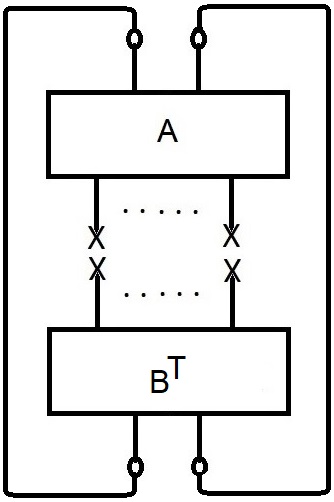}\end{gathered}
\equiv {\rm Tr}(AB^T)\label{fig:InnerProdVVdef}
\eea
We have shown this for $ \Vst \otimes \Vst$, but the same idea applies to any tensor power of $\Vst$. 
Some  inner products in $\Vst$ are illustrated in the equation below 
\bea
\left\langle\begin{gathered}\includegraphics[scale=0.4]{circcross}\end{gathered}\, ,\,
\begin{gathered}\includegraphics[scale=0.4]{circcross}\end{gathered}\right\rangle
&=&\begin{gathered}\includegraphics[scale=0.4]{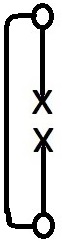}\end{gathered}=d\qquad\qquad
\left\langle\begin{gathered}\includegraphics[scale=0.4]{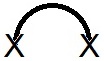}\end{gathered}
\begin{gathered}\includegraphics[scale=0.4]{circcross}\end{gathered}\, ,\,
\begin{gathered}\includegraphics[scale=0.4]{Cap}\end{gathered}
\begin{gathered}\includegraphics[scale=0.4]{circcross}\end{gathered}\right\rangle
=\begin{gathered}\includegraphics[scale=0.4]{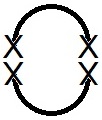}\end{gathered}\,\,
\begin{gathered}\includegraphics[scale=0.4]{innercirccrosscirccross}\end{gathered}=d^2\cr
\left\langle\begin{gathered}\includegraphics[scale=0.4]{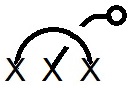}\end{gathered}\, ,\,
\begin{gathered}\includegraphics[scale=0.4]{CapCircTangle}\end{gathered}\right\rangle
&=&\begin{gathered}\includegraphics[scale=0.4]{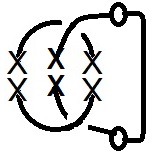}\end{gathered}=d^2\qquad\qquad
\left\langle\begin{gathered}\includegraphics[scale=0.4]{Cap}\end{gathered}
\begin{gathered}\includegraphics[scale=0.4]{circcross}\end{gathered}\, ,\,
\begin{gathered}\includegraphics[scale=0.4]{CapCircTangle}\end{gathered}\right\rangle
=\begin{gathered}\includegraphics[scale=0.4]{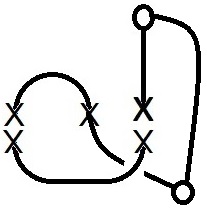}\end{gathered}
=d\cr
\Big\langle\begin{gathered}\includegraphics[scale=0.4]{Cap}\end{gathered}
\begin{gathered}\includegraphics[scale=0.4]{circcross}\end{gathered}+
\begin{gathered}\includegraphics[scale=0.4]{CapCircTangle}\end{gathered}&,&
\begin{gathered}\includegraphics[scale=0.4]{Cap}\end{gathered}
\begin{gathered}\includegraphics[scale=0.4]{circcross}\end{gathered}+
\begin{gathered}\includegraphics[scale=0.4]{CapCircTangle}\end{gathered}\Big\rangle
=2d^2-2d=2d(d-1)
\eea
This construction of an inner product is very similar to a standard inner product 
in Brauer algebras.

From the diagrams it is easy to see that
\bea 
\langle A , B \rangle = \langle B , A \rangle 
\eea
The counting of loops is unchanged by inverting the diagram for the inner product. 
Also note that $ \sigma^T = \sigma , C^T = C$. 
With these the desired hermiticity of the innerproduct follows and we have  
a map from  $ \Vst \otimes \Vst $ to $\Ust$ invariant subspaces
\bea 
 V_0^{ \star }  \oplus V_{ [2] }^{ \star }  \oplus V_{ [1^2] }^{ \star }  
\eea
corresponding to the three projectors in equation (\ref{3proj}). 

We have explained that there are additional commuting operators  of the form $ a C $ and $a \sigma $ 
for $ a \in \cF_{ \star , 0 }$. In the  light of the double commutant theorem,
the existence of these additional commuting operators means that the orthogonal subspaces 
of $ \Vst  \otimes \Vst$ constructed above are not in fact irreducible as representations 
of $ \Ust $ over $ \mC$. An important question is whether  these additional commuting operators generate
 the full commutant of $ \Ust $ in $ \Vst \otimes \Vst$. If that is the case, it would make sense to develop a treatment of $ \cF$ as a module-algebra over the ring $ \cF_{ \star , 0}$, and define an analogous quotient $ \Ust' $  to impose the diagrammatic version of the $ Uso(d)$ relations. 
 This would also mean treating $ \cF_{ \star , 0 } $ as scalars in the definition of  the inner product.  
 We would then  conjecture that if we  take the tensor product $\Vst\otimes \Vst$ over $\cF_{*,0}$, 
then we get exactly 3 irreps (simple modules) labelled by the symmetric, the anti-symmetric and the trace, just as we  do for $Uso(d)$ acting on $V_d \otimes V_d$ at large $d$.  We leave a more precise discussion of this point for the future.

\subsection{ Connection to Brauer category diagrams  }\label{BrauerCat} 

The diagrams in $ \cF_{ m , n } \subset \cF $ have $ 2m $ incoming crosses and $n$ $M$-boxes. 
In the definition of the quotient $ \Ust$ of $ \cF $, and the subsequent construction of the representations $ \Vst^{ \otimes n }$, a key role is played by the anti-symmetriser $ P_{\tiny\yng(1,1)}$. 
In fact if, in any diagram $ a \in \cF$,   we  get rid of the $M$-boxes, replace them with these projectors ending on pairs of crosses, we do not lose any information. So in this simplified picture, we would make the replacements in equation (\ref{fig:JustPs}) below. 
\bea
\begin{gathered}\includegraphics[scale=0.4]{M}\end{gathered}\qquad &\rightarrow&\qquad
\begin{gathered}\includegraphics[scale=0.4]{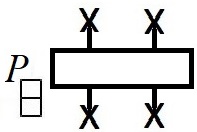}\end{gathered}\cr
\begin{gathered}\includegraphics[scale=0.4]{M}\end{gathered}
\begin{gathered}\includegraphics[scale=0.4]{M}\end{gathered}\qquad &\rightarrow&\qquad
\begin{gathered}\includegraphics[scale=0.4]{ProjBox}\end{gathered}
\begin{gathered}\includegraphics[scale=0.4]{ProjBox}\end{gathered}\cr
\begin{gathered}\includegraphics[scale=0.4]{TwistedMs}\end{gathered}\qquad &\rightarrow&\qquad
\begin{gathered}\includegraphics[scale=0.4]{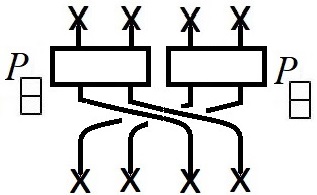}\end{gathered}\cr
\begin{gathered}\includegraphics[scale=0.4]{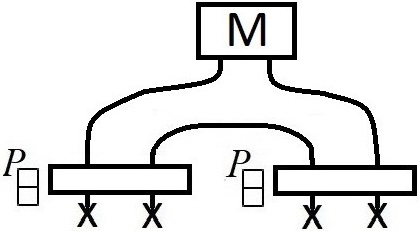}\end{gathered}\qquad &\rightarrow&\qquad
\begin{gathered}\includegraphics[scale=0.4]{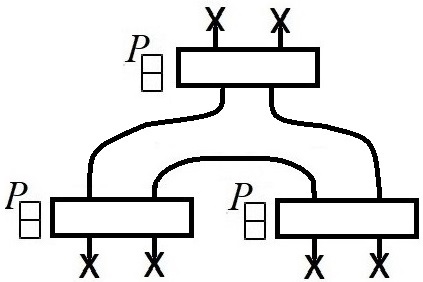}\end{gathered}\label{fig:JustPs}
\eea
Now we just have diagrams having a lower rung of $ 2m $ crosses and an upper rung of $ 2n $ crosses, along with a number of lines joining these crosses. 
These are Brauer diagrams in  the Brauer category \cite{LZ1207}.  

In the Brauer category,  there are two products : the tensor product - or diagrammatically 
a horizontal juxtaposition and the vertical composition or concatenation product. 
Brauer algebras  $B_d(n) $ which form the commutant of $ so(d)$ in tensor products $V_d^{ \otimes n }$ (let us keep $ d > n $ for the simplest statements) use the vertical product. The algebras $ \cF , \Ust$ we have defined here use the horizontal product, along with the star products, which we defined using juxtaposition followed by contractions. It would be interesting to study  the full structure and interplay of all the products : horizontal, vertical and star products.

\section{ Conjectures on tensor  representations of  $\Ust $ }\label{sec:TensRepsHd}

Based on our discussion of the action of $ \Ust $ on $ \Vst $ and  $ \Vst \otimes \Vst$ in Section \ref{sec:DecomHonVV}, we present here some conjectures for the action of $ \Ust $ 
 in $ \Vst^{ \otimes n }$, and for the action of $ \Ust $ on $ \Ust$ by commutators. 
The construction of $ \Ust$ employed a large $d$ limit, so it is reasonable to expect 
a simple relation to the large $d$ limits (also called stable limits) of $ Uso(d)$ representation 
theory of $ V_d^{ \otimes n }$. The decomposition into irreducible representations of the 
action of $ Uso(d)$ on itself by commutators is related, by the Poincare-Birkhoff-Witt theorem \cite{Humphreys}, 
to the decomposition of symmetric powers $ Sym^n ( V_{ [1^2] } )$ where $V_{[1^2]} $ is the anti-symmetric part of $ V_d \otimes V_d$ and $ Sym^n$ is the projection to the $S_n$ symmetric part of the $n$-fold tensor product. Likewise, for the decomposition of $ \Ust $ under commutator action by $ \Ust$, it is reasonable to expect a link to the large $d$ limit of $ Sym^n ( V_{ [1^2] } ) $. The discussion of 
the commutant in Section \ref{sec:ObsComm}  suggests some care is required with the treatment of the $ \cF_{ \star  , 0}$ subspace of $ \cF $, which we elaborate on below.

\noindent 
{\bf Conjecture for  $\Ust $ action on  $(\Vst)^{\otimes n}$}

We will denote by  $ B_d ( n )$, the Brauer algebra on $ n$ strands with loop parameter $d$ \cite{Brauer}.  We defined an action of  $ B_d (2)$ on $\Vst \otimes \Vst $ and showed that the generators of 
$B_d(2)$ - the swop of the tensor factors and the contraction operation - 
commute with the action of $ \Ust $ on $ \Vst \otimes \Vst $.
There is straightforward generalization to a definition of $ B_d ( n )$ on $  ( \Vst )^{ \otimes n }$.
Our first conjecture is that 

\noindent 
{\bf Conjecture 1} The action of $ \Ust $ on $ ( \Vst )^{ \otimes n } $ commutes with 
the action of $ B_d (n)$. 

The proof will be a straightforward generalization of the  diagrammatic argument at $ n=2$. 
A corollary  is  that  ${\bold V}^{ \star\,\otimes n} $ at generic $d$
admits $ \Ust$-equivariant maps to orthogonal direct sums  of spaces, invariant under $\Ust$ action, and  
in 1-1 correspondence with Brauer algebra projectors corresponding to Young diagrams which 
appear in $ V_{d}^{ \otimes n }$ for $d > 2n$.  These projectors were discussed for $ n=2$ in Section \ref{sec:DecomHonVV}. 

\noindent 
{\bf Conjecture 2a}  $ ( \Vst )^{ \otimes n } $ can be decomposed into orthogonal subspaces, invariant under the action of $ \Ust $ by using projectors in $B_d (n)$. 

This should also be a  straightforward generalization of the discussion we gave for $n=2$. For explicit construction of projectors, we can  use results on characters of Brauer algebras \cite{Ram95}. 

As we saw in the discussion of the $n=2$ case, it is easy to construct additional operators in the endomorphism algebra of $ \Vst$ of the form $ a A $ where $ A \in B_d ( n )$ and $ a \in \cF_{ \star , 0 }$.  Because of the double commutant theorem, we therefore do not expect that the above orthogonal  invariant subspaces are irreducible representations of $ \Ust$. 
These elements $ aA$  are linearly independent of $A$ if we think of $ \cF $, and the quotient $\cU_{ \star } $  as  algebras over $ \mC$ and $ \Vst$ as a vector space over $ \mC$. On the other hand, we may treat $ \cF$ and $ \Ust $  as  algebras over the ring $ \cF_{ \star , 0}$, in which case we will refer to 
them as $ \cF' , \Ust' $. From a physical point of view, this is highly sensible, since the diagrams in $ \cF_{ \star , 0 }$ are linear combinations of Kronecker deltas : the diagrams in $ \cF_{ m , 0 }$ are transitions from  $ ( \Lambda^2 ( W ) )^{ m } $ to the ground field $ \mC$.   To define $ \cF'$, it  makes sense to start from $ \cF$ and impose relations of the form $ A_1 \otimes a  = a \otimes A_1  $ for $ a \in \cF_{ \star , 0 }$, i.e. to  treat $ a$ as commuting scalars (see equation (\ref{Fst0asScal})).
Forming the quotient by the subspace generated by the 
commutator diagram  will define $ \Ust' $. When we define the tensor product of $\Vst$ we should treat elements of $\cF_{ \star ,0}$ as scalars : 
\bea 
( a v_1 )  \otimes v_2 = v_1 \otimes ( a v_2 ) 
\eea
More formally, we should treat $ \Ust'$ as a module-algebra over the ring $ \cF_{ \star ,0 }$, $ \Vst $ as a module over  the ring. The definition of the inner product on $ \cF$ would also treat $ a \in \cF_{ \star , 0 }$ as scalars i.e 
\bea 
\langle a_1 B_1 , a_2 B_2 \rangle = {\bar a_1} a_2 \langle B_1 , B_2 \rangle    
\eea
where $ a_1 , a_2 $ are complex linear combinations of diagrams in $ \cF_{ \star ,0 }$ 
and $B_1 , B_2$ are general complex linear combinations of diagrams in $ \cF$. $ {\bar a_1} $ is obtained  by complex-conjugating the complex coefficients in $ a_1$. 

\noindent 
{\bf Conjecture 2b} The full commutant of the action of 
$  \Ust'  $  on $ ( \Vst )^{ \otimes n } $, when loops are evaluated to $d$ ( for $ d > 2n $) 
 is $ B_d (n)$. 

A corollary of this conjecture is that the sub-modules of $ ( \Vst )^{ \otimes n }  $
labeled by $so(d)$ Young diagrams appearing in $ V_d^{ \otimes n }$ for $ d > 2n $ are irreducible 
representations (simple modules) of $ \Ust'$. 

Finally, note that this is a useful statement as a consequence of the fact that the representation theory of $SO(d)$ has
nice stability properties.
The tensor product $V_d^{\otimes 2k}$ decomposes into a direct sum of irreps labeled by every possible Young
diagram with $2q$ boxes for $q=0,1,2,...,k$. The multiplicity of a given irrep $\Lambda\vdash 2q$ is the symmetric 
group dimension of $\Lambda$ times the number of ways of making $k-q$ pairs from $2k$ objects.
This rule is stable for $so(d)$ with $d\ge 4k$.
The tensor product $V_d^{\otimes 2k+1}$ decomposes into a direct sum of irreps labeled by every possible Young 
diagram with $2q+1$ boxes for $q=0,1,2,...,k$. The multiplicity of a given irrep $\Lambda\vdash 2q+1$ is the symmetric 
group dimension of $\Lambda$ times the number of ways of making $k-q$ pairs from $2k+1$ objects.
This rule is stable for $so(d)$ with $d\ge 4k+2$.
These rules can be reproduced from explicit computations using characters at low $d,k$ and are related to the representation theory of Brauer algebras $B_d(n)$ for $ n = 2k$ or $ n = 2k+1$ \cite{KT87,Ram95}.

\noindent
{ \bf Remarks on $ d \le 2n $ } 

In this regime, there are two cases to consider : integer $d$ and non-integer $d$. 
In the above conjectures we have proposed relations between the tensor representations $ \Vst^{ \otimes n } $ of  $ \Ust$, with loops evaluated at $d$, 
and the representation $ V_d^{ \otimes n }$  of $ Uso(d)$, at $ d > 2n$. 
To make contact between the representation $ \Vst^{ \otimes n} $ of  $ \Ust $ in the $ d \le 2n$ 
 regime and 
$Uso(d)$,  we expect there should be 
 finite $d$ diagrammatic quotients of the $ \Vst^{ \otimes n } $ representations of $ \Ust$. These finite $d$ quotients would involve  finite $d$  projectors acting on the $ 2m + 2n$ crosses  
associated with the  $ \cF_{ n ,m  }$ subspace of $ \cF$ in the Brauer category picture described in Section \ref{BrauerCat}. These finite $d$ projectors would be a 
sum of Young projectors for Young diagrams having $2m + 2n $ boxes and no  more than $d$ rows.
These should lead to well-defined finite $d$ versions of $ \Ust$. 
Likewise finite $d$ version of $\Vst $ are defined using finite $d$ projectors acting the 
$ 2m+2$ crosses associated with the $\cV_{ m }$ subspace of $ \Vst$.   This regime of $ d \le  2n $ is somewhat more subtle: it  involves significant differences between $ so(d)$ and $ o(d)$ centralizers (see for example \cite{Grood}). These differences should be reflected in variations of possible quotients of $ \Ust$. We leave a more precise discussion of the relevant quotients of  $ \Ust , \Vst $ which makes contact 
with tensor representations of $ Uso(d)$  and $ o(d)$ for $ d \le  2n $ for the future.

\noindent 
{\bf Conjecture 3 :  $ \Ust $ action on $ \Ust  $ : } We can define an action of $ \Ust $ on itself by commutators. $ \Ust$ will decompose into irreps labelled by Young diagrams. The subspace of $ \Ust$ corresponding to $ \cF_{ m , n}$ with $ n \le k$ decomposes into Young diagrams in the same way that 
\bea 
\bigoplus_{ l  =0}^{ k  } Sym^l  ( \Lambda^2 ( V_d ) ) 
\eea
decomposes into irreps of $ Uso(d)$ for $ d > 2k $.

\noindent 
{ \bf Remarks on non-unitarity}

Based on recent discussions of  non-unitarity in the Wilson-Fischer fixed point CFTs \cite{HRV1512}
we expect that $ ( \Vst )^{ \otimes n } $ will have states of negative norm under a natural inner product. 
Using the inner product defined in (\ref{InnerProdAB}), and caculating the norm  for a state in $ \Vst^{ \otimes n } $ constructed from the anti-symmetric projector 
\bea 
P_{[1^n]} = { 1 \over n! } \sum_{ \sigma \in S_n } (-1)^{ \sigma }  \sigma 
\eea 
with permutation $ \sigma $ represented as a winding of strands in the diagram, the inner product is calculated as follows
\bea
&\underbrace{\begin{gathered}\includegraphics[scale=0.4]{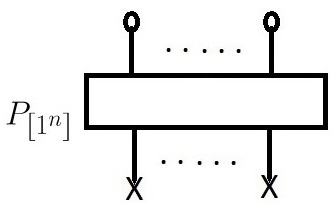}\end{gathered}}&\in (V^*)^{\otimes\, n}\cr
&n&
\eea
\bea
\left\langle \begin{gathered}\includegraphics[scale=0.4]{boxP1n}\end{gathered}\, ,\,
\begin{gathered}\includegraphics[scale=0.4]{boxP1n}\end{gathered}\right\rangle
&=&
\begin{gathered}\includegraphics[scale=0.4]{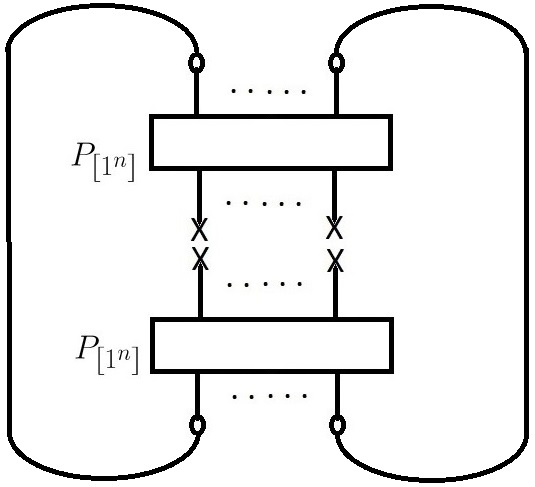}\end{gathered}\cr
&=&
\begin{gathered}\includegraphics[scale=0.4]{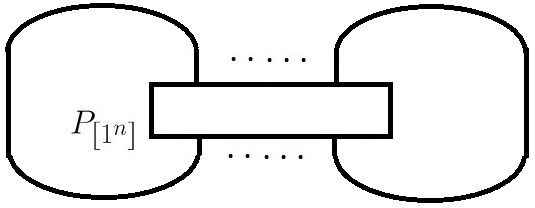}\end{gathered}\cr
&=&{ ( d ) ( d -1 ) \cdots ( d -n+1) \over n! } 
\eea
This is negative for $ d < n-1$.

\section{ An algebra $\Ustt $ extending $ Uso(d,2)$ to generic $d$ }\label{sec:UsoStar}

Along similar lines to the definition of $ \Ust $ we have proposed to generalize  $ Uso(d)$ representation theory for general $d$, we sketch the definition  $\Ustt$ appropriate for generalizing $ Uso(d,2)$ beyond integer $d$. This gives a proposed  framework for discussing the states of Wilson-Fischer CFT away from integer dimensions as representations of  conformal symmetry beyond integer dimensions, and in particular 
gives a representation theoretic meaning to the  diagrammatic computation of the properties of the stress tensor state we gave in Section \ref{sec:ACRTDC}. 

Corresponding to the generators $ M_{ \mu \nu} , D , K_{ \mu} ,  P_{ \nu}$ of the conformal algebra,  there are boxes for $ M , D , K , P $ as shown below 
\bea
\begin{gathered}\includegraphics[scale=0.4]{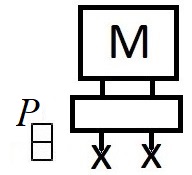}\end{gathered}\qquad
\begin{gathered}\includegraphics[scale=0.4]{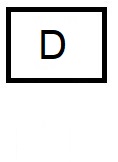}\end{gathered}\qquad
\begin{gathered}\includegraphics[scale=0.4]{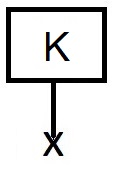}\end{gathered}\qquad
\begin{gathered}\includegraphics[scale=0.4]{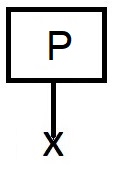}\end{gathered}\label{fig:Us2PA}
\eea
Each diagram is made of two rungs - as in Brauer algebras. The lower rung consists of either $0,1$ or $2$ crosses - zero in 
the case of $D$, one in the case of $P,K$ and two in the case of $M$. We have made the anti-symmetry of $ M$ manifest by
inserting a projector $ P_{ [1^2] } = { 1 \over 2 } ( 1 - \sigma ) $. 

Commutation relations are diagrammatic: they involve setting to zero the linear combinations of diagrams shown below
\bea
\begin{gathered}\includegraphics[scale=0.4]{P}\end{gathered}
\begin{gathered}\includegraphics[scale=0.4]{P}\end{gathered}
&-&\begin{gathered}\includegraphics[scale=0.4]{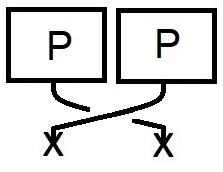}\end{gathered}\qquad\qquad\qquad\qquad
\begin{gathered}\includegraphics[scale=0.4]{K}\end{gathered}
\begin{gathered}\includegraphics[scale=0.4]{K}\end{gathered}
-\begin{gathered}\includegraphics[scale=0.4]{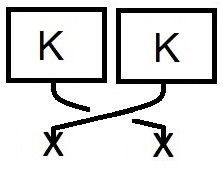}\end{gathered}\cr
\begin{gathered}\includegraphics[scale=0.4]{D}\end{gathered}
\begin{gathered}\includegraphics[scale=0.4]{P}\end{gathered}
&-&\begin{gathered}\includegraphics[scale=0.4]{P}\end{gathered}
\begin{gathered}\includegraphics[scale=0.4]{D}\end{gathered}
-\begin{gathered}\includegraphics[scale=0.4]{P}\end{gathered}\qquad\qquad
\begin{gathered}\includegraphics[scale=0.4]{D}\end{gathered}
\begin{gathered}\includegraphics[scale=0.4]{K}\end{gathered}
-\begin{gathered}\includegraphics[scale=0.4]{K}\end{gathered}
\begin{gathered}\includegraphics[scale=0.4]{D}\end{gathered}
+\begin{gathered}\includegraphics[scale=0.4]{K}\end{gathered}\cr
\begin{gathered}\includegraphics[scale=0.4]{projM}\end{gathered}
\begin{gathered}\includegraphics[scale=0.4]{P}\end{gathered}
&-&\begin{gathered}\includegraphics[scale=0.4]{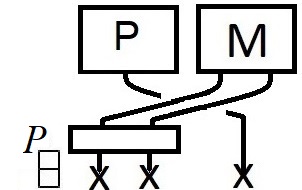}\end{gathered}
-\begin{gathered}\includegraphics[scale=0.4]{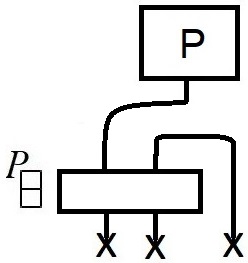}\end{gathered}\cr
\begin{gathered}\includegraphics[scale=0.4]{projM}\end{gathered}
\begin{gathered}\includegraphics[scale=0.4]{K}\end{gathered}
&-&\begin{gathered}\includegraphics[scale=0.4]{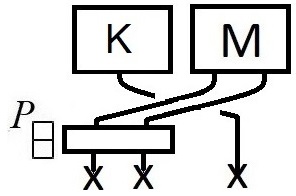}\end{gathered}
-\begin{gathered}\includegraphics[scale=0.4]{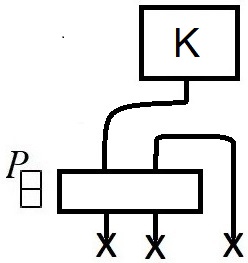}\end{gathered}\cr
\begin{gathered}\includegraphics[scale=0.4]{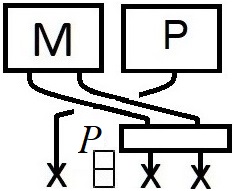}\end{gathered}
&-&\begin{gathered}\includegraphics[scale=0.4]{P}\end{gathered}
\begin{gathered}\includegraphics[scale=0.4]{projM}\end{gathered}
-\begin{gathered}\includegraphics[scale=0.4]{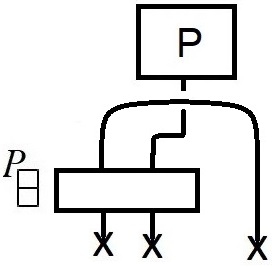}\end{gathered}\cr
\begin{gathered}\includegraphics[scale=0.4]{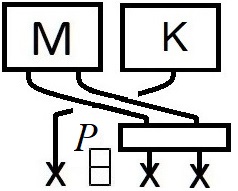}\end{gathered}
&-&\begin{gathered}\includegraphics[scale=0.4]{K}\end{gathered}
\begin{gathered}\includegraphics[scale=0.4]{projM}\end{gathered}
-\begin{gathered}\includegraphics[scale=0.4]{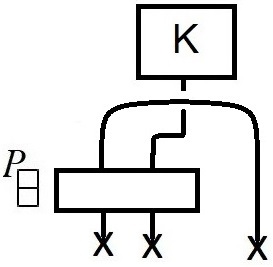}\end{gathered}\cr
\begin{gathered}\includegraphics[scale=0.4]{M}\end{gathered}\,\,\,
\begin{gathered}\includegraphics[scale=0.4]{D}\end{gathered}
&-&\begin{gathered}\includegraphics[scale=0.4]{D}\end{gathered}\,\,\,
\begin{gathered}\includegraphics[scale=0.4]{M}\end{gathered}\cr
\begin{gathered}\includegraphics[scale=0.4]{K}\end{gathered}
\begin{gathered}\includegraphics[scale=0.4]{P}\end{gathered}
&-&\begin{gathered}\includegraphics[scale=0.4]{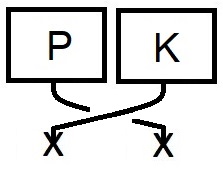}\end{gathered}
-2\,\,\begin{gathered}\includegraphics[scale=0.4]{projM}\end{gathered}
+2\,\,\begin{gathered}\includegraphics[scale=0.4]{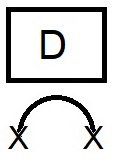}\end{gathered}\cr
\begin{gathered}\includegraphics[scale=0.4]{projM}\end{gathered}
\begin{gathered}\includegraphics[scale=0.4]{projM}\end{gathered}
&-&\begin{gathered}\includegraphics[scale=0.4]{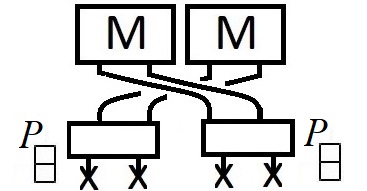}\end{gathered}
-\begin{gathered}\includegraphics[scale=0.4]{Mprojproj}\end{gathered}\label{fig:CUsod2}
\eea
These diagrams do not involve indices and have a meaning in terms of equivariant maps between tensor spaces as  discussed earlier in the context of $\Ust$. As  observed in that discussion, an easy way to understand the diagrams 
is to attach indices and read off the usual commutation relations of $Uso(d,2)$. For example, setting to zero the 
first line expresses $ P_{ \mu_1} P_{ \mu_2} = P_{ \mu_2 } P_{ \mu_1} $. The third and fourth lines express the 
commutators 
\bea 
&&   [  M_{ \mu \nu }  , P_{ \alpha } ] = \delta_{ \nu \alpha } P_{ \mu} - \delta_{ \mu \alpha } P_{ \nu }\cr  
&&   [ M_{ \mu \nu } , K_{ \alpha } ] = \delta_{ \nu \alpha } K_{ \mu}  - \delta_{ \mu \alpha } K_{ \nu }  
\eea
The reason we have two  linear combinations of diagrams  for each of  this pair of equations is that 
when we interpret these diagrams in terms of equivariant maps involving $ W , \Lambda^2 ( W )$, 
we need to take into account that $ W \otimes \Lambda^2 ( W )$ and $ \Lambda^2 ( W ) \otimes W$
are distinct subspaces of the tensor algebras we will be considering. 


We would like (as in the analogous discussion of  of $ \Ust $)  to 
give a general description of a space of diagrams, containing and generalizing 
the diagrams in (\ref{fig:Us2PA}) or (\ref{fig:CUsod2}), expressed in terms of 
$so(d)$ equivariant maps (in a large $d$ limit)  between  appropriate tensor algebras, such that 
the tensor product operation gives an associative algebra structure to the space of diagrams. 
Introduce one-dimensional vector spaces over $ \mC$ : 
\bea 
&& V_{ K } = { \rm Span }  \{ K \}  \cr 
&& V_P = { \rm Span }  \{ P \} \cr 
&& V_D = { \rm Span }  \{  D \}  \cr 
&& V_{ M } =  { \rm Span }   \{ M \} 
\eea
Let $ W$ be the $d$-dimensional vector  irrep of $ so(d)$. 
Define 
\bea 
&& V_{ K}^w  = V_K \otimes W \cr 
&& V_{ P}^w = V_P \otimes W \cr 
&& V_{ D}^w = V_D  \cr 
&& V_{ M}^w = V_M \otimes \Lambda^{ 2 } ( W )  = V_M \otimes W_2  
\eea
Observe that the space of $ so(d)$ equivariant maps $ Hom_{ so(d)} ( W , V_K^w )$ is spanned by 
\bea 
\delta : e_{\mu }  \rightarrow \delta_{ \mu \nu  } K \otimes e_{ \nu }  
\eea
which can be associated with the  diagram for $ K$  in Figure \ref{fig:Us2PA} : the cross on the lower rung is 
associated with $e_{\mu } $, the box on the upper rung with $ K \otimes e_\nu  $, 
the line with the $\delta $ map. 
Similarly $ Hom_{ so(d)} ( W , V_P^w )$  corresponds to  the diagram for $P$. 
In the case $Hom_{ so(d)} ( \mC , V_D^w = V_D )$, the lower rung has no crosses since it corresponds to the vacuum which is associated with  the ground field in standard diagrammatic representations of tensor categories. The map is just 
\bea 
a \rightarrow a D \in V_D   \hbox{ for } a \in \mC 
\eea 
The association of $ Hom ( \Lambda^2 ( W ) , V_M^w ) $  to the diagram is as described in the discussion of $ U_{ \star} $. 

To have a structure which contains all the diagrams, define 
\bea 
&& \cL = V_M^{ w} \oplus V_{P}^w \oplus V_{K}^{w} \oplus V_D^{ w} \cr 
&& \cS = \mC \oplus  W \oplus W_2 
\eea
The space  $ Hom_{ so(d)} ( \cS , \cL ) $ is spanned by  exactly four diagrams, which are the ones 
shown in (\ref{fig:Us2PA}),  associated with the generating diagrams for $ K , P , D , M$. 

The natural proposal for a general space of diagrams is 
\bea 
&& \cF^{  (2)  }  = Hom_{ so(\infty )}  ( T ( \cS ) , T ( \cL  )  \cr 
&& = \bigoplus_{ m , n  =0}^{ \infty } Hom_{ so(d) : d ~  large } (   \cS^{ \otimes m } , \cL^{ \otimes n } )   \cr 
&& =  \bigoplus_{ m , n } \cF^{(2)}_{ m , n }    
\eea
$ \cF^{(2)}_{ 0 , 0 } = \mC$. 
$ \cF^{ (2)  }_{ 0 , 1} $ is one-dimensional. It is the map which takes $ a \in \mC $ to $ a D \in V_D^w $. 
$ \cF^{ (2)}_{ 1,1} $ is four-dimensional, as we discussed.

Equation (\ref{fig:CUsod2}) gives the   linear combinations 
of diagrams which should be  set to zero. 
The first two express the commutativity of $P$'s and $K$'s. 
Note that we have two relations  coming from $ [ M, K ] \sim K $ commutator and two relations 
from $ [ M , P  ] \sim P$. This is because $ W \otimes W_2 $ and $ W_2 \otimes W $ are distinct subspaces of $ \cS \otimes \cS$.  
$ \cF^{ (2)} $ has a product given by the tensor product operation : if $ a \in \cF^{(2)}_{ m_1  , n_1 } , 
b \in \cF_{ m_2 , n_2 } $ then $ a \otimes b \in \cF_{ m_1 + m_2 , n_1  + n_2 }^{ (2)} $ is the tensor product of  the equivariant maps. This corresponds to the juxtaposition of the diagrams for 
 $ a $ and $ b$. $ \cF^{(2)}$ contains additional composition operations consisting of juxtaposition followed by  contractions. We referred to the analogous operations as  star products
 in the discussion of $ \cF $ and $ { \bf U_{ \star } }$. As in that discussion, the quotienting operation 
 involves setting to zero all elements in $ \cF^{(2)} $ of the form $ a * C * b$ where $C$ is any of the linear combinations in  (\ref{fig:CUsod2}) and $ a, b \in \cF^{(2)}$  are being composed with $C$ using any of the star operations.  $ { \bf U ( \star , 2 ) } =  Uso( \star  , 2 )$ is defined as  this quotient of $ \cF^{(2)}$.

We now describe the construction of representations of  ${ \bf U_{ \star , 2 }  }$ which correspond to 
scalar primaries of $ U so(d,2)$. These representations are spaces of $ so(d)$ equivariant maps, which have diagrammatic representations using standard facts about $so(d)$ invariant theory, which can be described as
\bea
&& \widetilde V^{\star ,2}_{ \delta } = Hom_{ so(\infty )} ( Sym ( W ) , Sym  ( V_P^w ) ) \cr 
&&   = \bigoplus_{ m , n  = 0}^{ \infty }   Hom_{ so(d):  ~ d ~ large } ( Sym^m ( W )  ,  Sym^n (V_P^w )) \cr 
&& \equiv \bigoplus_{ m ,n } V^{ \star , 2 }_{ m , n } 
\eea
In this definition $ Sym (W )$ and $ Sym ( V_P^w )$ are the symmetric
 algebras over $ \mC $ of $ W$ and $ V_P^w$; $Sym^m ( W )$, $Sym^n ( V_P^w) $ are the $S_m$ and $S_n$ symmetric subspaces respectively of $ W^{ \otimes m } $ and $ ( V_P^w)^{ \otimes n } $.  
Large $d$ means $ d >  m + n $ so that the homomorphisms are constructed from $ \delta $ contractions only and not $ \epsilon $ contractions. 
The subspace  $ V^{ \star , 2 }_{ 0 , 0 } $ is  just the ground field $ \mC$.

The  space $\widetilde V^{\star ,2}_{ \delta } $ is made a representation of $ { \bf U_{ \star , 2 }  }  $ by specifying that $D$ acts on $ a \in  V^{ \star , 2 }_{ 0 , 0 } $ as: 
\bea 
D  ( a ) = \delta ~  a  ~ \in  V^{ \star , 2 }_{ 0 , 0 }
\eea
The usual  actions of $ K_{ \mu } , M_{ \mu \nu } , P_{ \alpha } , D $ on states of the form 
$ P_{ \beta_1 } P_{ \beta_2 } \cdots P_{ \beta_k } | \delta  \rangle $ can be translated into diagrams 
which give the action of $ \Ustt $ on $ \widetilde V^{\star ,2}_{ \delta }  $. 
For example the actions 
\bea
&& P_{ \mu } ( P_{ \nu} | \delta \rangle  ) = P_{ \mu } P_{ \nu } | \delta \rangle  \cr 
&& K_{ \mu } ( P_{ \nu } | \delta \rangle  ) =  [  K_{ \mu } , P_{ \nu } ] | \delta \rangle 
 = 2 M_{ \mu \nu } - 2 \delta_{ \mu \nu } D ) | \delta \rangle  = - 2 \delta \delta_{ \mu \nu }
   | \delta \rangle   \cr 
&& M_{ \mu \nu } ( P_{ \alpha } | \delta \rangle  ) = (  \delta_{ \nu \alpha } P_{ \mu } - \delta_{ \mu \alpha } P_{ \nu } ) | \delta \rangle 
\eea
translate into the diagrammatic actions given below
\bea
\begin{gathered}\includegraphics[scale=0.4]{P}\end{gathered}\,\,\triangleright\,\,\left(
\begin{gathered}\includegraphics[scale=0.4]{P}\end{gathered}\right)&=&
\begin{gathered}\includegraphics[scale=0.4]{P}\end{gathered}\,\,
\begin{gathered}\includegraphics[scale=0.4]{P}\end{gathered}\cr
\begin{gathered}\includegraphics[scale=0.4]{K}\end{gathered}\,\,\triangleright\,\,\left(
\begin{gathered}\includegraphics[scale=0.4]{P}\end{gathered}\right)&=&\,-\,2\,\,\delta\,\,\,
\begin{gathered}\includegraphics[scale=0.4]{Cap}\end{gathered}\cr
\begin{gathered}\includegraphics[scale=0.4]{projM}\end{gathered}\,\,\triangleright\,\,
\left(\begin{gathered}\includegraphics[scale=0.4]{P}\end{gathered}\right) &=&\,\,
\begin{gathered}\includegraphics[scale=0.4]{ProjPTwist}\end{gathered}
\eea

\section{ Summary and Outlook  } \label{sec:discussion}

In \cite{CFT4TFT2} we gave a complete description of the correlators of general composite operators at separated space-time points in free scalar field theory in terms of an $so(4,2)$ equivariant algebra, where the underlying state space  
is the direct sum of symmetric powers of $ V = V_+ \oplus V_-$ and $ V_+, V_-$ are two dual irreducible representations of $ so(4,2)$. $V_+$ is spanned by the states obtained from  general derivatives of the scalar field using the state-operator correspondence. The algebra satisfies a non-degeneracy condition, which allows us to define a genus-restricted two-dimensional topological field theory. 
We have addressed  two challenges faced by the programme of extending CFT4/TFT2 to interacting conformal  field theories such as $ \cN =4$ SYM and the Wilson-Fischer fixed point near four dimensions. 
A deformed co-product in Section \ref{sec:defcoprod}  for $ so(4,2)$  was defined which allows us to describe the states corresponding to composite operators, in the presence of non-additive anomalous dimensions. To make sense of tensor algebra in general dimensions e.g. $ d = 4 - \epsilon $, we defined an algebra $ \Ust $  (Section \ref{sec:diagalg}) and its tensor representations $ \Vst^{ \otimes n }$, which is abstracted from $ Uso(d)$ at general $d$ and tensor products of the vector representation $V_d^{ \otimes n } $.  We motivated and presented some conjectures for the decomposition of $ \Vst^{ \otimes n }$ 
in terms of irreducible representations of $ \Ust$. We then sketched the definition of a diagram algebra $ \Ustt$, which allows generalization of $Uso(d,2)$ to general $d$, and the construction of its diagrammatic  representations corresponding  to scalar primary fields.

Some future goals in the programme of developing  perturbative-CFT4/TFT2 are 

\begin{itemize} 

\item Describe the state space of the Wilson-Fischer theory, describing  the structure which replaces the direct sum of symmetrised tensor products  $ Sym^n ( V_+)$ which corresponds to local operators in free scalar theories. We have made some preliminary remarks about the role of indecomposable representations in connection with multiplet-recombination in Section \ref{sec:indecomp} and in \cite{II1512}. We have found here that the indecomposables also play a role in the formulation of the deformed co-product. 

\item Given the results of \cite{RT1505} which demonstrate that algebraic perspectives based on the quantum equation of motion constrain CFT data of anomalous dimensions, it is reasonable to expect that the characterisation  of  aspects of the conformal equivariance properties 
in interacting theories we have given here, will contribute to the programme of constraining CFT data
algebraically. 

\item Extend the description of  general primary fields in free CFT in terms of a system of linear equations for multi-variable polynomials \cite{DRRR170506}, associated ring structures and algorithms for constructing primaries \cite{HLMM1706,DR1806} to the case of  perturbatively interacting CFTs. Useful formulae and algorithms for constructing primary fields can be used to approach the next element in CFT data - the calculation of OPE coefficients. 

\end{itemize}

\begin{center} 
{ \bf Acknowledgements}
\end{center} 
This work of RdMK  is supported by the South African Research Chairs
Initiative of the Department of Science and Technology and National Research Foundation
as well as funds received from the National Institute for Theoretical Physics (NITheP).
SR is supported by the STFC consolidated grant ST/P000754/1 `` String Theory, Gauge Theory \& Duality” and  a Visiting Professorship at the University of the Witwatersrand, funded by a Simons Foundation grant (509116)  awarded to the Mandelstam Institute for Theoretical Physics. We are grateful for useful conversations on the subject of this paper to Matt Buican, Igor Frenkel, Antal Jevicki,  Shahn Majid, Joao Rodrigues,  Bill Spence, Gabriele Travaglini, Congkao Wen, Chris White.

\end{document}